\documentclass[twocolumn]{aastex631}
\usepackage[flushleft]{threeparttable}
\usepackage{lmodern}

\begin{document}

\title{A Near-Infrared Spectral Library of Very Young Brown Dwarfs and Planetary-Mass Objects in the Orion Nebula Cluster}

\author[0000-0003-0192-6887]{Elena Manjavacas}
\affiliation{AURA for the European Space Agency (ESA), ESA Office, Space Telescope Science Institute, 3700 San Martin Drive, Baltimore, MD, 21218 USA}
\affiliation{Department of Physics and Astronomy, Johns Hopkins University, Baltimore, MD 21218, USA}
\correspondingauthor{Elena Manjavacas}
\email{emanjavacas@stsci.edu}

\author[0000-0002-5581-2896]{Mario Gennaro}
\affiliation{Space Telescope Science Institute, 3700 San Martin Dr., Baltimore, MD 21218, USA}

\author[0000-0003-0192-6887]{Sarah Betti}
\affiliation{Space Telescope Science Institute, 3700 San Martin Dr., Baltimore, MD 21218, USA}

\author[0000-0003-3818-408X]{Laurent Pueyo}
\affiliation{Space Telescope Science Institute, 3700 San Martin Dr., Baltimore, MD 21218, USA}
\affiliation{Department of Physics and Astronomy, Johns Hopkins University, Baltimore, MD 21218, USA}

\author[0000-0001-6396-8439]{William O. Balmer}
\affiliation{Department of Physics and Astronomy, Johns Hopkins University, Baltimore, MD 21218, USA}
\affiliation{Space Telescope Science Institute, 3700 San Martin Dr., Baltimore, MD 21218, USA}

\author[0000-0002-9573-3199]{Massimo Robberto}
\affiliation{Department of Physics and Astronomy, Johns Hopkins University, Baltimore, MD 21218, USA}
\affiliation{Space Telescope Science Institute, 3700 San Martin Dr., Baltimore, MD 21218, USA}



\begin{abstract}

Age-benchmark {brown dwarfs' and planetary-mass objects' spectroscopy is} key to characterize substellar evolution. In this paper we present the {JHK} {medium resolution (R$\sim$3000)} spectra of {25} 7--76~$\mathrm{M_{Jup}}$ (spectral types L3.0-M6.0) brown dwarfs and planetary-mass objects in the Orion Nebula Cluster obtained with MOSFIRE installed at the W. M. Keck\,I telescope. We obtained the spectral types of the targets in our sample using template brown dwarf and {planetary-mass objects' spectra}. We confirmed their extreme youth  {($<$5~Myr)} and {membership to the cluster} using spectral indices, and the diversity of their spectra even for targets with similar spectral types. {Six} of our targets presented Pa-$\beta$ and Bra-$\gamma$ emission lines, suggesting the existence of accreting protoplanetary disks to objects with masses as low as 7~$\mathrm{M_{Jup}}$. After analyzing the emission lines of those objects, and measuring their accretion rates, we compared them to those of stars, brown dwarfs and planetary-mass objects, confirming that planetary-mass young objects deplete their disks quickly at young ages. Finally, we illustrate the spectral evolution of a 7--10~$\mathrm{M_{Jup}}$ planetary-mass object through its life from 1--3~Myr to 200~Myr old using one of our latest spectra type targets, and other targets from the literature with older age, but similar estimated masses.
{The spectra are publicly available for the community's use as data behind the figures.}

\end{abstract}



\section{Introduction} \label{sec:intro}

Brown dwarfs are substellar objects (13~$\mathrm{M_{Jup}}<$M$<75~\mathrm{M_{Jup}}$) unable to sustain hydrogen burning as stars do. Brown dwarfs cool down and contract as they age \citep{Burrows, Baraffe}, never reaching the main sequence. Thus, it is extremely challenging to constrain the ages and masses of brown dwarfs.
As brown dwarfs age, the decrease in effective temperature produces a shift in spectral types from the spectral type they are initially born with (M, L, T or Y) to later spectral types. Simultaneously, {brown dwarfs contract, producing} an increase in surface gravity \citep{Burrows, Baraffe}, which also affects some brown dwarfs' spectral characteristics. Understanding how surface gravity affects the spectral characteristics of brown dwarfs of any given spectral type is key to estimating their ages.  

Characterizing the spectra of brown dwarfs members of open clusters, associations, young moving groups or companions to stars for which we have an estimation of their age, has been a priority since brown dwarfs were discovered (e.g. \citealt{Gorlova2003, McGovern, Cruz2007, Allers2007, Allers2013, Bonnefoy2014a, Manjavacas2014, Martin2017, Manjavacas2020}). These age-benchmark brown dwarfs are key spectral templates to further characterize other brown dwarfs with no age estimation, and also planetary-mass objects, or even directly-imaged exoplanets.

After more than two decades of community effort, {it has been concluded} that young brown dwarfs usually show weaker K\,I and Na\,I alkali lines (e.g. \citealt{Gorlova2003, McGovern, Martin2017}) in the optical and in the near-infrared spectra in comparison to their older counterparts. In addition, FeH absorptions {features} in the $J$- and $H$-bands are also less prominent for low-gravity dwarfs \citep{Allers2013, Lodieu2018}, producing the characteristic triangular $H$-band {shapes} in young brown dwarfs. Very red near-infrared colors were believed to be an additional indication of youth (\citealt{Marocco2014, Faherty2016, Liu2016}, among others), however, some of those red-brown dwarfs did also show other spectroscopic signposts of low-gravity atmospheres in their spectra, like weak alkali lines \citep{Marocco2014}. \cite{Vos2017} discovered that brown dwarfs that show redder colors usually are observed {equator-on (i.e. rotational axis aligned)}, suggesting a different cloud distribution from the equator to the poles. \cite{Suarez2023} further confirmed that L-dwarfs observed equator-on show a deeper silicate feature at 10~$\mu$m confirming that L-dwarfs show higher cloud coverage in their equatorial latitudes. Thus, red colors alone are not necessarily an indication of low surface gravity in brown dwarfs \citep{Manjavacas2020}.

The most reliable age-benchmark brown dwarfs and planetary-mass objects are those that are members of young open clusters or stellar associations. However, due to their relatively far distances, the crowding of open cluster's fields {and possible contamination from faint but old background sources}, finding brown dwarfs of very young ages $<$5~Myr and with spectral types {later than} L0 is relatively challenging. Only a few spectroscopic surveys have particularly targeted brown dwarfs and planetary-mass object members of very young ($<$5~Myr) open clusters {and young associations}, due to instrumental limitations, and the challenge of finding such low-mass objects {since} they are embedded in their star-forming cloud. Those surveys have targeted the $\sim$0.3~Myr $\rho$-Ophiuci Star Forming region (e.g. \citealt{Martin2017}), the $\sim$2~Myr Chamaeleon star-forming region (e.g. \citealt{Luhman2007, Manjavacas2020, Kubiak2021}), the Taurus $\sim$2~Myr Star-Forming region (e.g. \citealt{Martin2017}), the 1--3~Myr Orion Nebula Cluster (e.g. \citealt{Weights2009, Caballero2019, Luhman2024}), and the 5-10~Myr Upper-Scorpious Association (e.g \citealt{Lodieu, Martin2017, Lodieu2018, Luhman2018, Manjavacas2020}), among others. {However, none of them had medium-resolution spectra with full near-infrared coverage between 1.0 and 2.5~$\mu$m, and they did not reach the brown dwarf-planetary-mass {boundary}}. {Here} we aim to {start filling} this gap.  {To do so, we obtained high signal-to-noise, near-infrared medium-resolution spectra brown dwarf and planetary-mass members {of the Orion Nebula Cluster (ONC)} with the MOSFIRE instrument at the W. M. Keck Observatory. Our sample was carefully selected by \cite{Robberto2020}, based on HST photometry. The { mass estimates} of our targets lie between 76~$\mathrm{M_{Jup}}$ and 7~$\mathrm{M_{Jup}}$ (spectral types M6.0-L3.0), {with an estimated age between 1 and 3~Myr \citep{Jeffries2011}}.}

 This paper is structured as follows: in Section \ref{sec:targets}, we describe our target selection. In Section \ref{sec:observations} we describe our observations with MOSFIRE at the Keck\,I telescope. In Section \ref{sec:data_reduction} we explain the data reduction for the MOSFIRE spectra. In Section \ref{sec:data_analysis} we describe the data analysis carried out in our medium-resolution near-infrared spectra. In Section \ref{sec:characteristics_ONC_sample} we confirmed the youth of our sample. In Section \ref{sec:evol_planetary_mass_objects} we compare the near-infrared spectra of one of our planetary-mass ONC members with near-infrared spectra of other later spectral type planetary-mass objects with similar mass to illustrate the spectral evolution with time of a planetary-mass object. In Section \ref{sec:discussion} we present our discussion of the results obtained in this paper. Finally, in Section \ref{sec:conclusions} we summarize our conclusions. {The spectra are publicly available for the community's use as data behind the figures.}

\section{{Target Selection}}\label{sec:targets}


Our targets were selected using the catalog provided in \cite{Robberto2020} {which found} 742 sources with a negative color index in the F130N and F139N (F130N-F139N) filters of the Wide Field Camera 3 (WFC3) instrument onboard the Hubble Space Telescope (HST). These {colors} suggest the presence of the $\mathrm{H_{2}O}$ absorption at 1.4~$\mu$m, and can therefore be classified as bona fide ONC members {(and not background contaminants}) with late M and early L spectral types, with an estimated $\mathrm{T_{eff}}$ below 2850~K assuming a cluster age of 1~Myr. 

We selected a subsample of planetary-mass objects, brown dwarfs, and stars located in the outskirts of the ONC {for our Keck/MOSFIRE observations}. 
{Because of the brightness (high background), the crowding, and dust extinction of the nebula are higher near the core, JWST/NIRspec (GTO program 1228, P.I. Alves de Oliveira) is more suited for sources near the core}. 
In total, we targeted {59 low-mass stars and substellar object photometric candidates with magnitudes between 15.0 and 18.9} in the F130N filter, corresponding to objects of temperatures between 3150~K and 1600~K, and masses between 3~$\mathrm{M_{Jup}}$ and 167~$\mathrm{M_{Jup}}$. Brown dwarfs with masses estimated $<$76~$\mathrm{M_{Jup}}$ had magnitudes fainter than 15.8 in the $J$-band. In total, {46 of the 59 objects were substellar object candidates in our sample}. 

{\cite{Robberto2020} selected these brown dwarf candidates using the BT-Settl models isochrones over the color-magnitude diagram of all their candidates, assuming an age of 1~Myr for the ONC. However, these isochrones did not reproduce the distribution of the substellar objects below 45~$\mathrm{M_{Jup}}$ in {a F130N-F149M} color-magnitude diagram (see Fig. 7 in \citealt{Robberto2020}), probably due to the underprediction of the depth of the water band for low-mass substellar objects by the BT-Settl models. Thus, they {adjusted} the isochrones using the position of the objects with masses $<$45~$\mathrm{M_{Jup}}$ in the color-magnitude diagram (see their Fig. 7) deriving an empirical isochrone. They dereddened their candidates in the color-magnitude diagram to derive a rough estimate of effective temperatures and masses, but they warned that significant uncertainties should be expected in these derived parameters. {The published extinction values for those objects are based on the empirically-corrected isochrones}. }

We were able to extract only {34} of the 46 substellar object candidates observed with MOSFIRE in the three nights due probably to low SNR or to a slight misalignment of the position of slits. The extracted spectra are shown in Fig. \ref{fig:JHK-high_mass} through \ref{fig:JHK-planetary_mass} in Appendix \ref{app:all_spectra}. {Nine} of those objects were removed from the sample because they were found to be stellar-mass cluster members or too noisy to classify (objects 3382, 363, 367, 365, 1436, 884, 3300, 1339 and 334), probably indicating that the empirical isochrone adjustment in \cite{Robberto2020} could be improved. We do not discuss these sources in the present paper. We end up with a final sample of 25 objects, shown in Table \ref{table:parameters}, {but for completeness, we include all the spectra from Fig. \ref{fig:JHK-high_mass} through \ref{fig:JHK-planetary_mass} as data behind the figures}.

\begin{table*}
	\small
	\caption{Fundamental parameters of our sample as derived by \cite{Robberto2020} and signal-to-noise of our MOSFIRE spectra.}  
	\label{table:parameters}
	\begin{center}
		\begin{tabular}{lllllllll}
			\hline
			\hline 
				
        Target \# &  RA (deg )&  DEC (deg)&  $\mathrm{Mag_{F130M}}$ & $\mathrm{Mag_{F139N}}$ & $\mathrm{A_{V}}$ & $\mathrm{T_{eff}}$ (K) & Mass ($\mathrm{M_{Jup}}$) & {SNR}\\ 
        \hline
        262 & 83.77654681 & -5.48144079 & 18.3 & 18.7 & 0.0 & 2102 & 7.4 & {10.7}\\ 
        1549 & 83.75503092 & -5.42337210 & 17.6 & 18.1 & 0.0 & 2081 & 7.1 & {22.1} \\ 
        473 & 83.83997125 & -5.45587588 & 17.9 & 18.3 & 0.0 & 2141 & 8.1 & {23.8} \\ 
        335 & 83.76552335 & -5.44101921 & 18.3 & 18.7 & 2.3 & 2179 & 8.8 & {23.8}\\ 
        442 & 83.84664424 & -5.50015704 & 17.3 & 17.8 & 0.8 & 2276 & 11.2 & {36.9} \\ 
        471 & 83.83463471 & -5.45185125 & 17.6 & 18.0 & 1.7 & 2265 & 10.9 & {22.5}\\ 
        379 & 83.82282938 & -5.47746217 & 18.0 & 18.3 & 3.9 & 2308 & 12.2 & {13.3} \\ 
        1386 & 83.76937999 & -5.43057666 & 17.0 & 17.4 & 0.7 & 2328 & 12.9 & {8.3}\\ 
        3155 & 83.86960570 & -5.32384671 & 17.2 & 17.6 & 1.6 & 2343 & 13.4 & {30.0} \\ 
        3251 & 83.83086549 & -5.28534493 & 17.2 & 17.6 & 1.6 & 2333 & 13.1 & {34.4} \\ 
        3307 & 83.85833662 & -5.32031533 & 17.3 & 17.7 & 1.7 & 2336 & 13.2 & {28.2}\\ 
        3311 & 83.85787387 & -5.28772208 & 18.4 & 18.7 & 5.8 & 2322 & 12.6 & {13.0} \\ 
        3256 & 83.81190488 & -5.29820963 & 18.0 & 18.3 & 4.6 & 2345 & 13.5 & {18.0} \\ 
        1572 & 83.76288214 & -5.42676104 & 16.5 & 16.9 & 0.0 & 2386 & 15.2 & {53.0} \\ 
        333 & 83.78237448 & -5.45448142 & 16.8 & 17.2 & 1.2 & 2402 & 15.9 & {30.3} \\ 
        3253 & 83.82259091 & -5.28678238 & 17.4 & 17.7 & 4.4 & 2434 & 17.9 & {28.6} \\ 
        3385 & 83.81244898 & -5.28040512 & 16.5 & 16.9 & 1.5 & 2466 & 19.8 & {57.8} \\ 
        833 & 83.78670799 & -5.51599523 & 15.7 & 16.0 & 1.0 & 2649 & 35.1 & {42.2} \\ 
        366 & 83.83709122 & -5.50135968 & 15.5 & 15.8 & 0.4 & 2675 & 36.9 & {135.9} \\ 
        372 & 83.83379331 & -5.48769436 & 16.7 & 16.9 & 5.6 & 2720 & 40.1 & {56.8} \\ 
        1376 & 83.77932948 & -5.41683643 & 16.0 & 16.2 & 3.1 & 2732 & 40.9 & {8.3} \\ 
        255 & 83.77999388 & -5.48822882 & 15.5 & 15.7 & 3.0 & 2761 & 48.9 & {109.3}\\ 
        3238 & 83.82993951 & -5.31090854 & 15.4 & 15.6 & 3.0 & 2763 & 50.2 & {123.7} \\ 
        469 & 83.82622430 & -5.46586584 & 15.7 & 15.7 & 5.2 & 2846 & 71.0 & {145.7} \\ 
        3241 & 83.80120292 & -5.29723920 & 15.8 & 15.9 & 6.5 & 2852 & 76.2 & {117.4}\\ 
        \hline
				
		\end{tabular}
	\end{center}
	
\end{table*}

\section{Observations} \label{sec:observations}

\begin{figure*}
    \centering
    \includegraphics[width=0.9\linewidth]{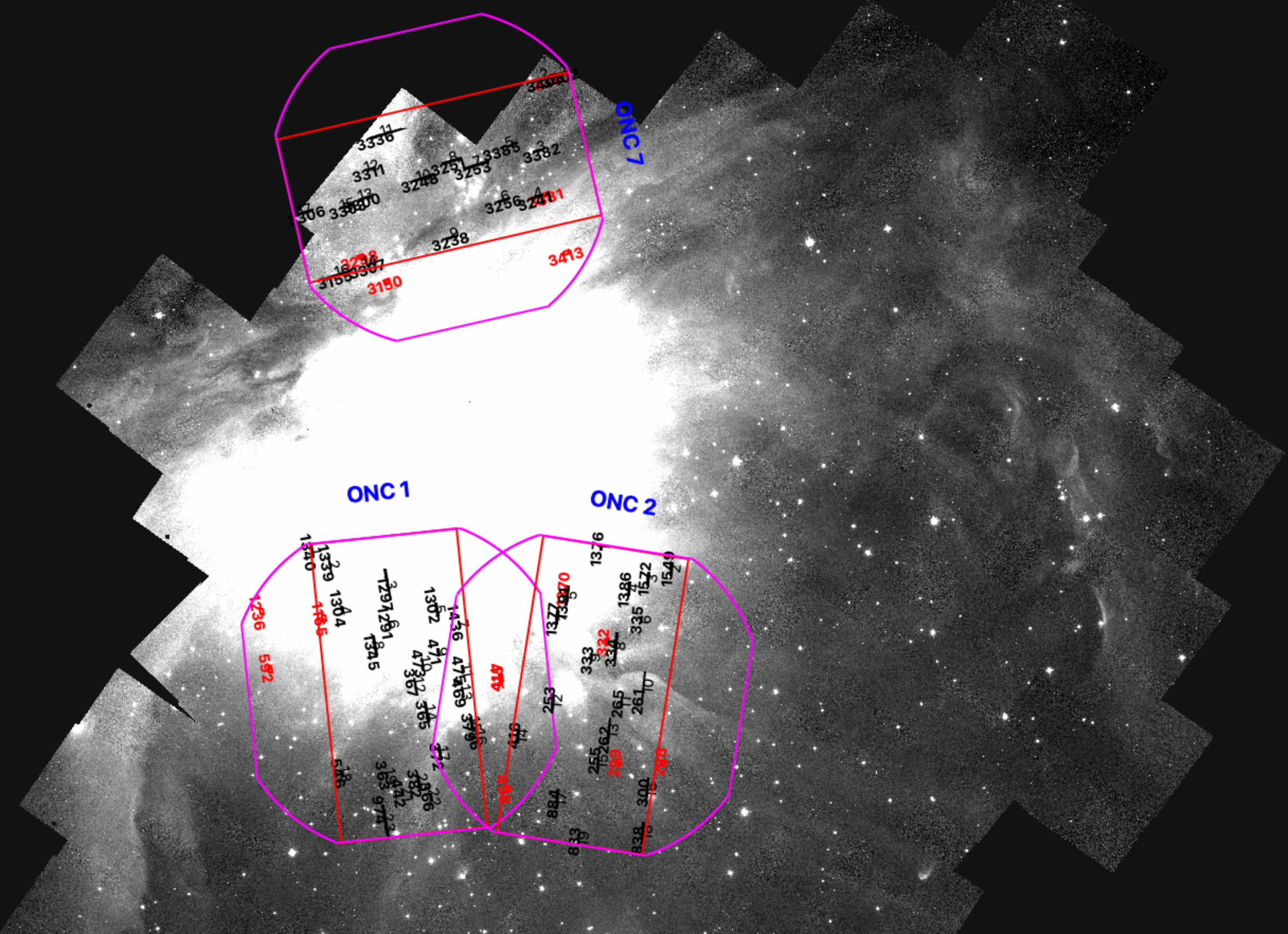}
    \caption{Region files (purple) with the three fields observed (red squares) with MOSFIRE showing the positions of the objects overplotted over the ONC HST/WFC3 image from \cite{Robberto2020}. The objects in black indicate our science targets, and the objects in red are stars used for aligning the MOSFIRE mask.}
    \label{fig:all_ONC_masks}
\end{figure*}

\begin{figure}
    \centering
    \includegraphics[width=0.99\linewidth]{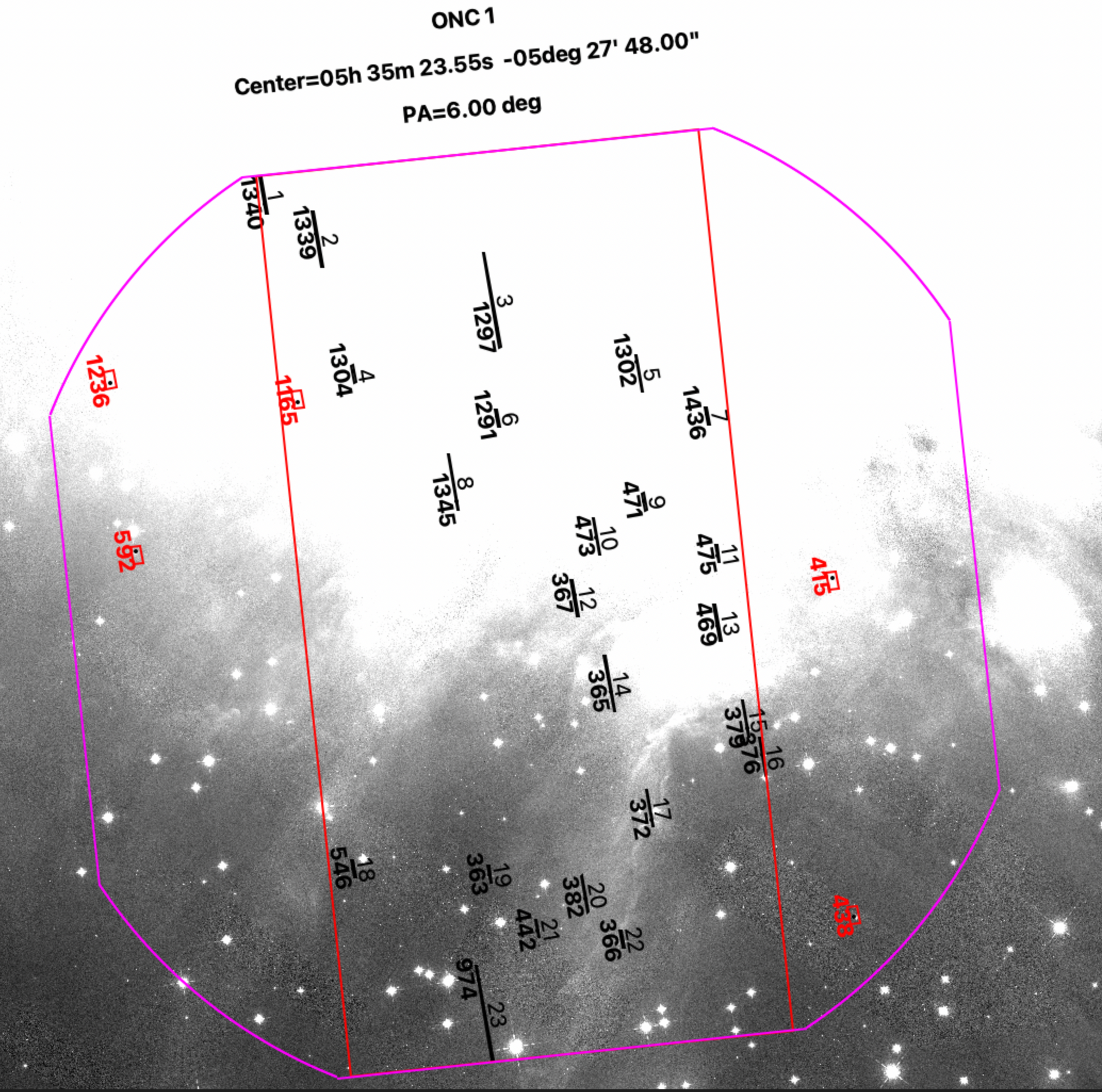}
    \caption{Region file for the first field observed on the 2021 October 27 in the ONC. The objects in black indicate our science targets, and the objects in red are stars used for aligning the MOSFIRE mask.}
    \label{fig:ONC_1}
\end{figure}

\begin{figure}
    \centering
    \includegraphics[width=0.99\linewidth]{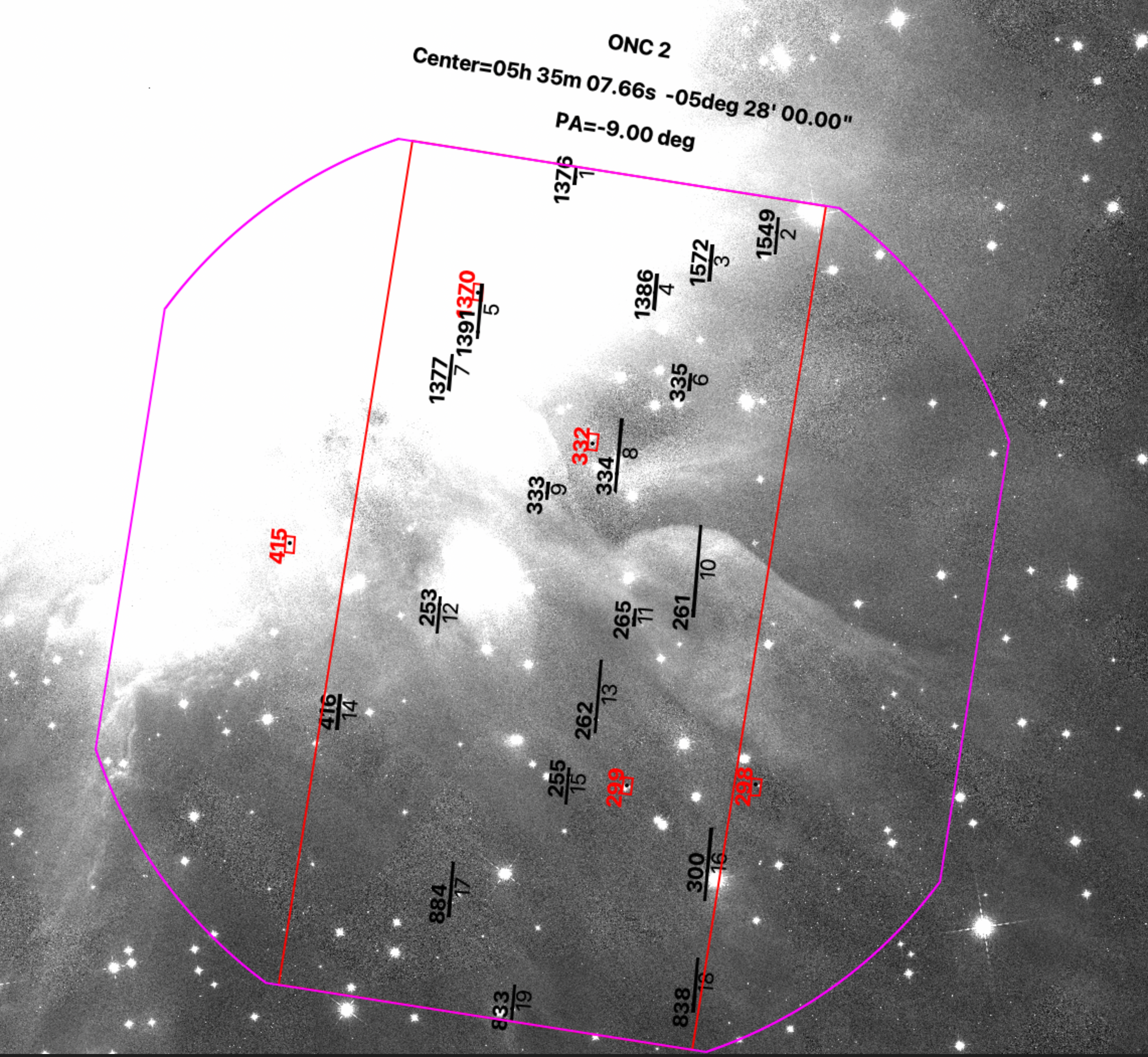}
    \caption{Region file for the second field observed on the 2021 October 28 in the ONC. The objects in black indicate our science targets, and the objects in red are stars used for aligning the MOSFIRE mask.}
    \label{fig:ONC_2}
\end{figure}

\begin{figure}
    \centering
    \includegraphics[width=0.99\linewidth]{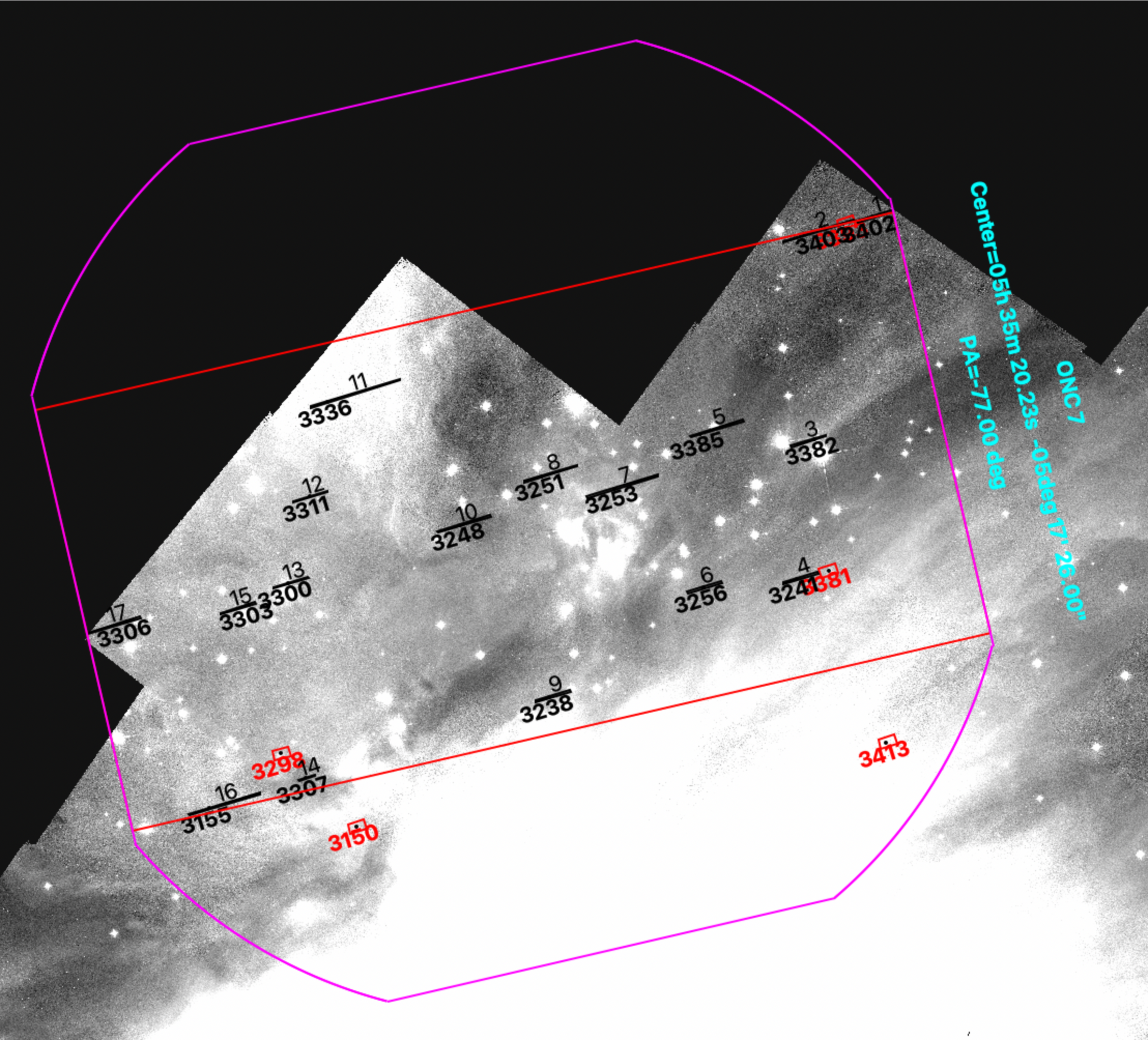}
    \caption{Region file for the thrid field observed on the 2021 November 12 in the ONC. The objects in black indicate our science targets, and the objects in red are stars used for aligning the MOSFIRE mask.}
    \label{fig:ONC_3}
\end{figure}

We carried out our observations using the MOSFIRE instrument on the Keck\,I telescope (program ID N154, P.I. Gennaro). MOSFIRE is a multi-object near-infrared spectrograph that can perform simultaneous spectroscopy of up to 46 objects in a {6$\farcm$1 $\times$ 6$\farcm$1} field of view, using the Configurable Slit Unit (CSU), a cryogenic robotic slit mask system that is reconfigurable electronically in less than 5 minutes without any thermal cycling of the instrument. A single photometric band is covered in each instrument setting ($Y$, $J$, $H$, or $K$). {We used three separate MOSFIRE mask configurations on three different nights}, {on 2021 November, 27 and 28, and December 12}, during a half night on each date.  In Fig. \ref{fig:all_ONC_masks} we show the layout of all the masks on the ONC. The three masks used are shown in Fig. \ref{fig:ONC_1}, \ref{fig:ONC_2}, \ref{fig:ONC_3} for the three different nights. We obtained spectra of the objects in the three masks in the $J$-, $H$- and $K$-bands using 1" slits that provide an approximate resolution of R$\sim$3000. We used an A B A' B' dithering pattern to remove the sky contamination. We obtained individual integrations of 120~s in the $J$-band, 180~s in the $H$-band, and 180~s in the $K$-band in each nodding point, with the MCDS sampling mode, and 16 reads, following the recommended times in the MOSFIRE webpage\footnote{\url{https://www2.keck.hawaii.edu/inst/mosfire/home.html}} to optimize the signal-to-noise of the spectra per dither position.

In addition, we observed one telluric standard per night {to perform the telluric} calibration, and to perform a relative flux calibration between the three filters (see Section~\ref{sec:data_reduction}). We observed the telluric stars using the \textit{long2pos} observation strategy recommended on the MOSFIRE webpage to obtain the full $J$, $H$ and $K$ wavelength coverage of the calibration stars' spectra. We integrated for $\sim$10~s in each \textit{long2pos} position. We observed the telluric star HIP~30761 ($\mathrm{V_{mag}}$ = 8.09, spectral type  B7V) in the night of {the 2021 October 27}, and the star HIP~22435 ($\mathrm{V_{mag}}$ = 8.25, spectral type  A0V) on the nights of {the 2021 October 28, and 2021 November 12}. 

For data reduction, we obtained 13 dome flats in each filter, plus 13 off-lamp flats for the $K$-band. 

\section{Data Reduction} \label{sec:data_reduction}

We reduced all data, including the telluric calibration star spectra, using version 1.17.1 of the \textit{Pypeit}\footnote{https://pypeit.readthedocs.io/en/stable/} pipeline {\citep{Prochaska2019, Prochaska2020}}. The pipeline corrects for dark current, creates a bad-pixel mask, and traces the edges of the slits using the dome flat fields obtained. To obtain a wavelength calibration \textit{Pypeit}  uses the telluric skylines present in each observation. After removing the skylines, the pipeline extracts the 1D spectra from the 2D sky-corrected, wavelength-calibrated frames. Finally, we combined all the individual 1D extracted spectra using the \texttt{coadd\_1dspec} script within \textit{Pypeit}. After the data reduction and combining all our spectra, our spectra have signal-to-noise (SNR) {per resolution element} between {8.5  for our faintest object, and 145.0 for our brightest (see Table \ref{table:parameters})} . For our telluric stars, we obtained {a SNR of $\sim$290 for HIP~30761, and a SNR of $\sim$250 for HIP~22435}. 

\textit{Pypeit} does not perform an automatic flux calibration, and in addition, our targets only have photometry from \cite{Robberto2020} in the F130M and F139N HST/WFC3 filters, which covers only the $J$-band, but not the $H$- and $K$-bands. However, our telluric calibration stars do have 2MASS $JHK$ photometry. To perform an initial flux and telluric calibration of our targets we followed this procedure:

\begin{enumerate}
    \item { Anchor MOSFIRE flux into 2MASS photometric system:} we calibrated the $J$, $H$ and $K$-band MOSFIRE spectra of the telluric standard star spectra using the 2MASS photometry in the respective filters. For this, {we individually integrated the the $J$, $H$, and $K$ MOSFIRE spectra of the standard stars under the 2MASS filter profile throughput}. We integrated the resulting flux, and divided it by the integrated profile of the filter. With the result of this integration,  we find a scaling factor between the MOSFIRE and the flux provided in 2MASS for each filter to bring the MOSFIRE to the 2MASS fluxes in each of the filters. {We repeated this procedure for all three nights with their individual telluric calibration stars}

    \item  { Calibrate colors of ONC targets into 2MASS photometric system:} we applied this same scaling factor to the targets observed in the corresponding nights to obtain a preliminary flux calibration for our targets in $J$, $H$, and $K$ bands, and more importantly, to provide a relative flux calibration between the three bands. Since the Spectral Energy Distributions (SED) of our targets are potentially affected by the presence of disks or reddening due to the extinction of the ONC cloud, and since we do not have any other photometric measurements in the $H$- and $K$-bands for these targets, this is the best option to obtain a relative flux calibration between the $J$, $H$ and $K$ bands for our targets.

    \item { Remove telluric lines:} using the flux-calibrated telluric stars, we performed the telluric calibration on our targets using the corresponding telluric star observed each night, in each band independently. We first removed the outliers present in the spectra of the telluric calibration star, due probably to hot pixels or cosmic rays. We then removed the intrinsic H and He lines present in the spectra in each band. Finally, we divided the spectra of the telluric star by a black body SED with the same effective temperature as the telluric star at the distance of the star, which is available in the Gaia DR3 Catalog, to flatten the spectra of the target, and end up only with the telluric contamination. Finally, we divided the spectra of our targets, in each band independently, by the flattened, line-free spectra of the telluric star to remove the telluric contamination from our targets.

    \item { Anchor photometry to HST F130M data:} finally, we refined the flux calibration of our targets using the HST/WFC3 photometry in the F130M filter from \cite{Robberto2020}, which overlaps with the $J$-band. We followed the same procedure as in step 1, but now we perform the same exercise using the HST/WFC3 F130M filter and the photometry of the targets in that filter from \cite{Robberto2020}. Since in step 2 we calculated the relative flux between the $J$, $H$ and $K$ bands, we can now flux calibrate our spectra using the same scaling factor between the MOSFIRE and the HST/WFC3 F130M filter for all three MOSFIRE filters.

    \item {Six objects still showed spurious emission features related to skylines that were not completely eliminated by the data reduction. These objects are 1376, 372, 3256, 1386, 3311, and 379. To avoid confusion during the analysis with features, we interpolated those features. In the Appendix, in Figure \ref{fig:spectra_with_skylinesJ}, \ref{fig:spectra_with_skylinesH} and \ref{fig:spectra_with_skylinesK} we show those six spectra with the sky lines overlapped, showing that some of the spurious emission features in fact are found at wavelengths where we expect sky lines.}

\end{enumerate}

\section{Spectral typing of  ONC library }\label{sec:data_analysis}

In this Section we present the main spectral characteristics shown in the 7--76~$\mathrm{M_{Jup}}$ brown dwarf spectra. For visualization purposes, we divide the spectra into those belonging to high-mass brown dwarfs (41--76~$\mathrm{M_{Jup}}$, Fig. \ref{fig:J-high_mass}, \ref{fig:H-high_mass}, \ref{fig:K-high_mass}), intermediate-mass brown dwarfs (14--40~$\mathrm{M_{Jup}}$, \ref{fig:J-intermediate_mass}, \ref{fig:H-intermediate_mass}, \ref{fig:K-intermediate_mass}), and planetary-mass objects (7--13~$\mathrm{M_{Jup}}$, Fig. \ref{fig:J-planetary_mass}, \ref{fig:H-planetary_mass}, \ref{fig:K-planetary_mass}). The {overall $J$-, $H$- and $K$- SEDs, on which the colors of each object can be observed}, are shown in the Appendix \ref{app:all_spectra} Fig. \ref{fig:JHK-high_mass} through \ref{fig:JHK-planetary_mass}. {The reduced and flux-calibrated spectra in Fig. \ref{fig:JHK-high_mass} through \ref{fig:JHK-planetary_mass} are available as "data behind the figures".}

\subsection{Identification of Spectral Lines and Molecular Bands}\label{sec:spectral_lines_bands}

\begin{figure}
    \centering
    \includegraphics[width=0.99\linewidth]{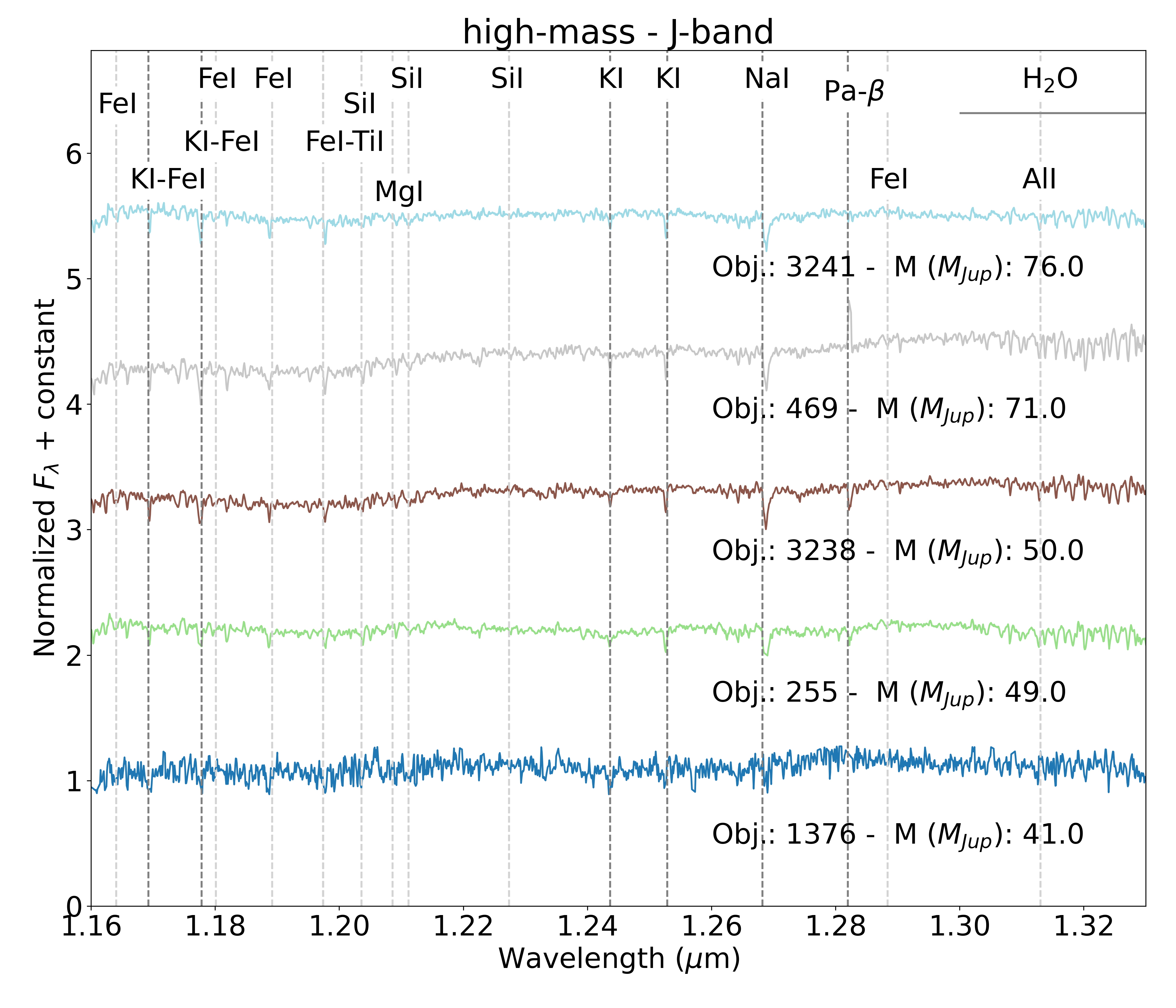}
    \caption{$J$-band spectra of the high-mass brown dwarfs with masses between 41 and 76~$\mathrm{M_{Jup}}$. {See main body discussion for significance of identified lines.}}
    \label{fig:J-high_mass}
\end{figure}

\begin{figure}
    \centering

    \includegraphics[width=0.99\linewidth]{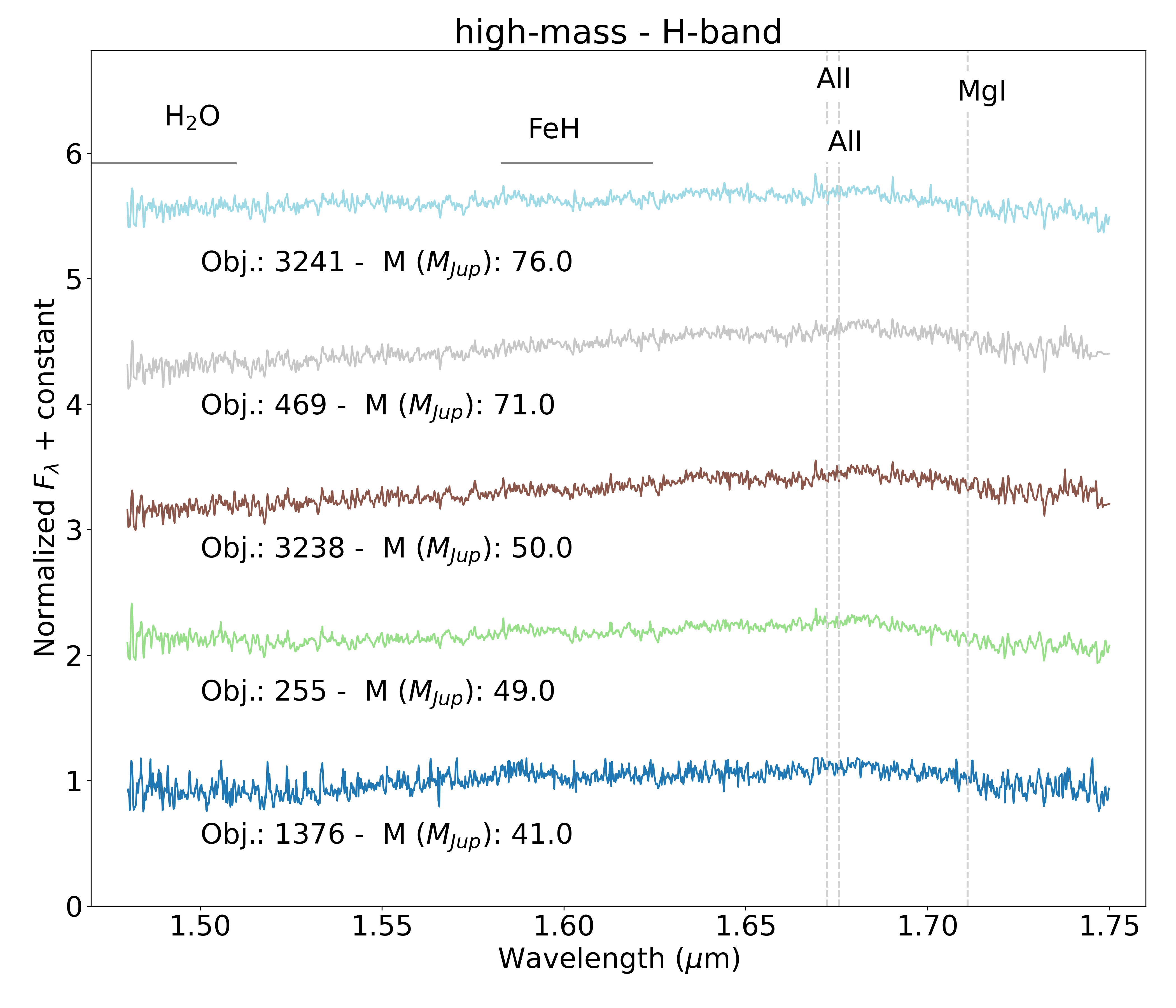}
    \caption{$H$-band spectra of the high-mass brown dwarfs with masses between 41 and 76~$\mathrm{M_{Jup}}$. {See main body discussion for significance of identified lines.}}
    \label{fig:H-high_mass}
\end{figure}

\begin{figure}
    \centering
 
    \includegraphics[width=0.99\linewidth]{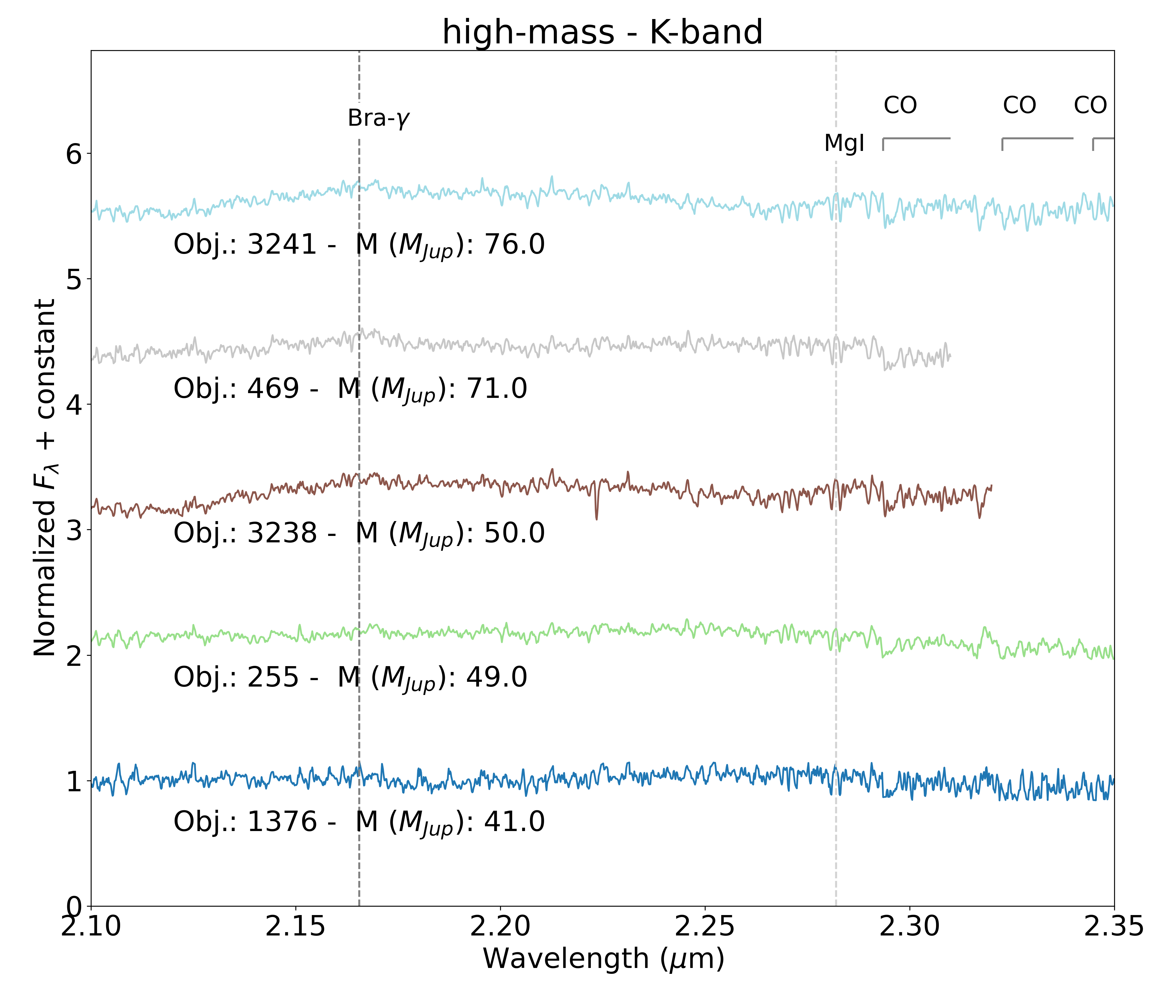}
    \caption{$K$-band spectra of the high-mass brown dwarfs with masses between 41 and 76~$\mathrm{M_{Jup}}$. {See main body discussion for significance of identified lines.}}
    \label{fig:K-high_mass}
\end{figure}

\begin{figure}
    \centering
    \includegraphics[width=0.99\linewidth]{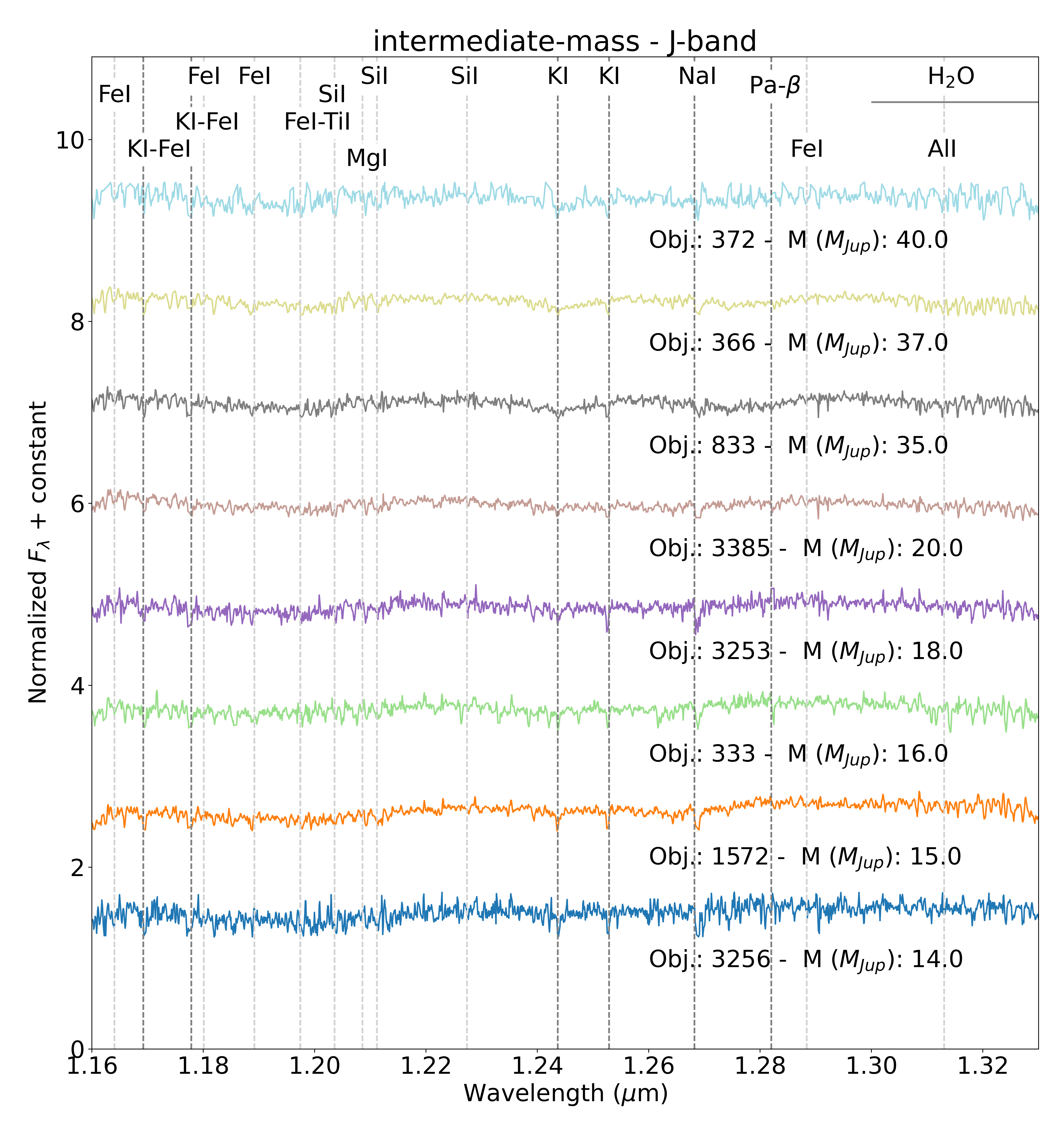}
    \caption{$J$-band spectra of the intermediate-mass brown dwarfs with masses between 14 and 40~$\mathrm{M_{Jup}}$. {See main body discussion for significance of identified lines.}}
    \label{fig:J-intermediate_mass}
\end{figure}

\begin{figure}
    \centering
    \includegraphics[width=0.99\linewidth]{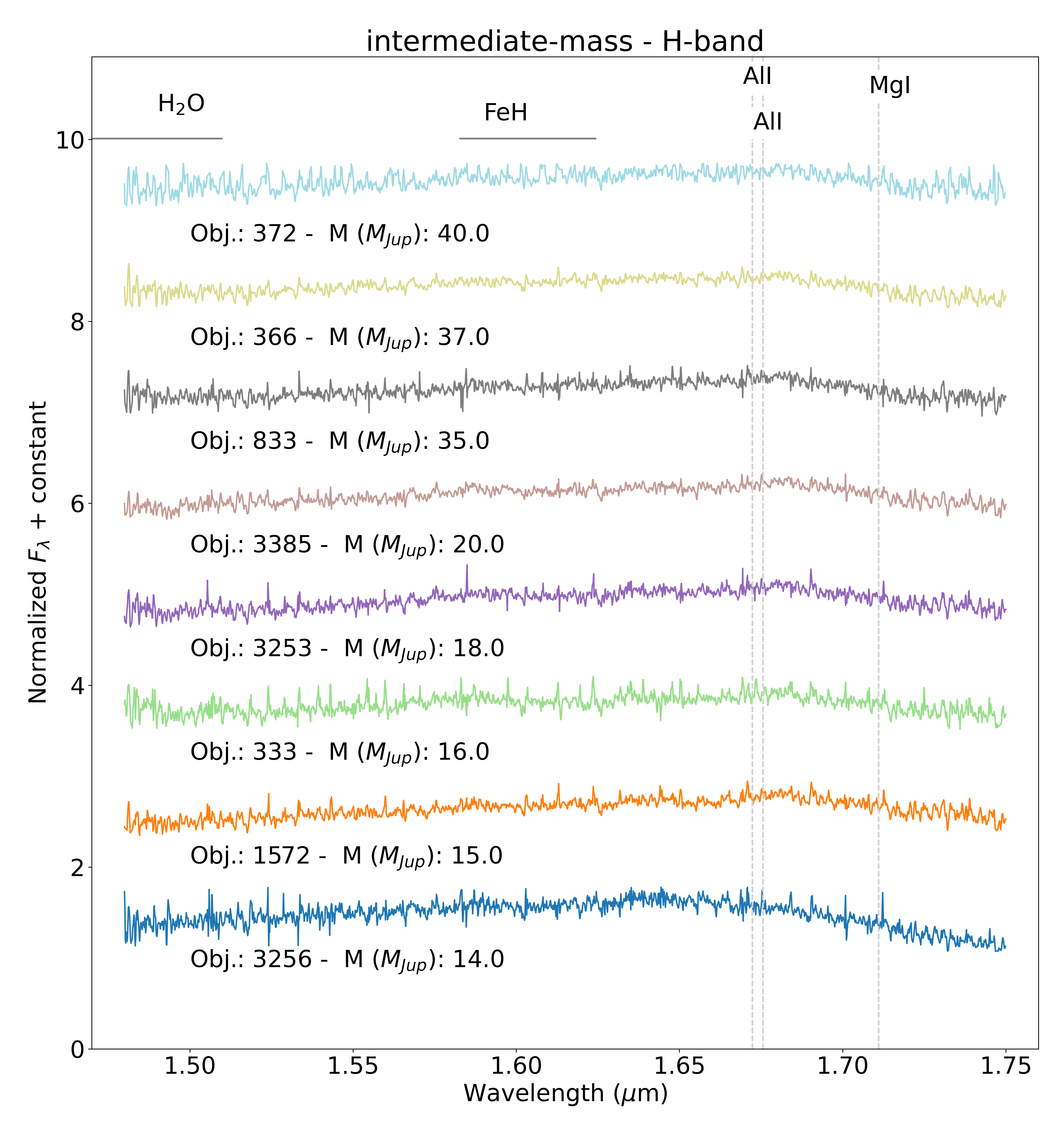}
    \caption{$H$-band spectra of the intermediate-mass brown dwarfs with masses between 14 and 40~$\mathrm{M_{Jup}}$.{See main body discussion for significance of identified lines.}}
    \label{fig:H-intermediate_mass}
\end{figure}

\begin{figure}
    \centering
    \includegraphics[width=0.99\linewidth]{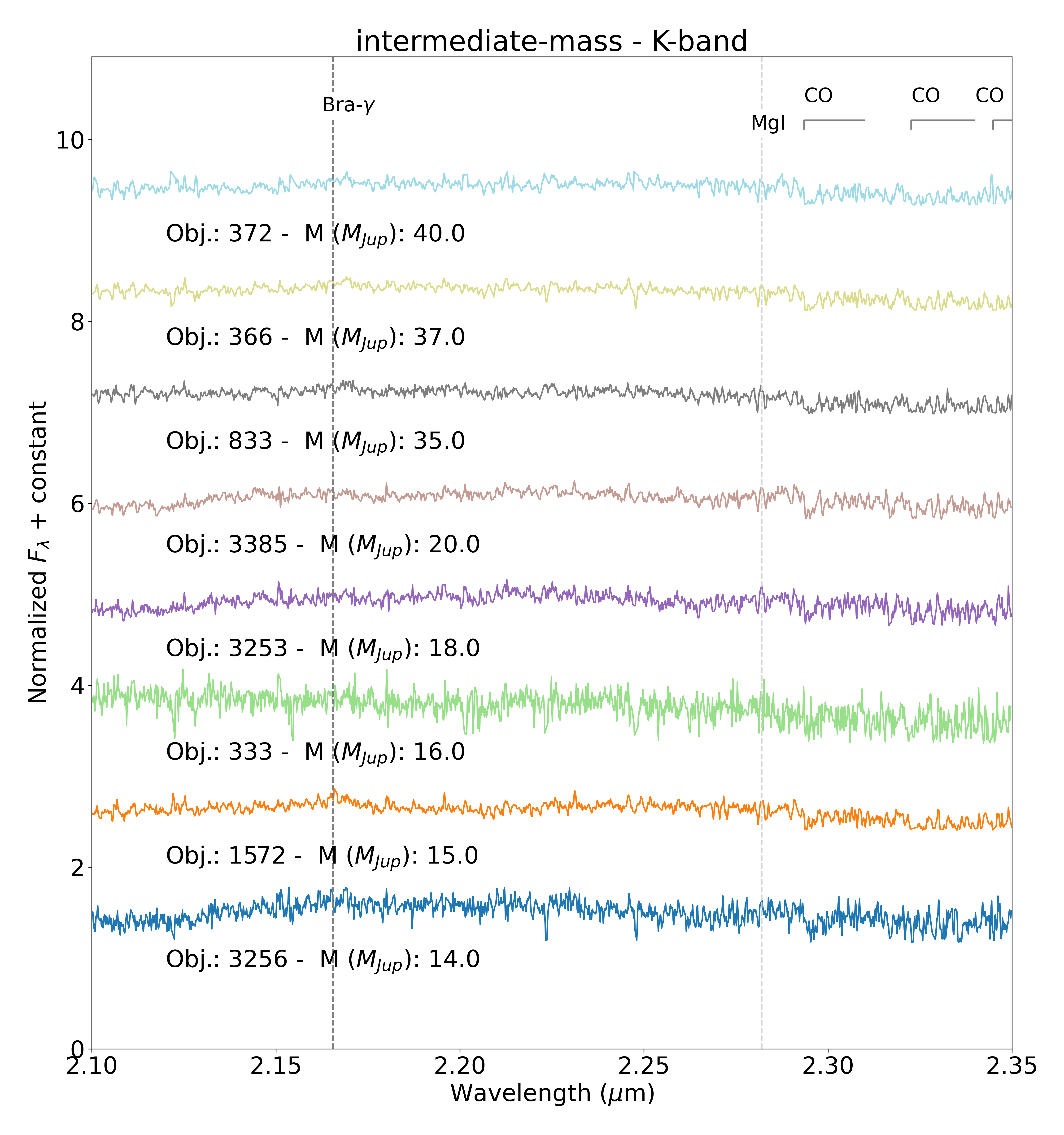}
    \caption{$K$-band spectra of the intermediate-mass brown dwarfs with masses between 14 and 40~$\mathrm{M_{Jup}}$. {See main body discussion for significance of identified lines.}}
    \label{fig:K-intermediate_mass}
\end{figure}

\begin{figure}
    \centering
    \includegraphics[width=0.99\linewidth]{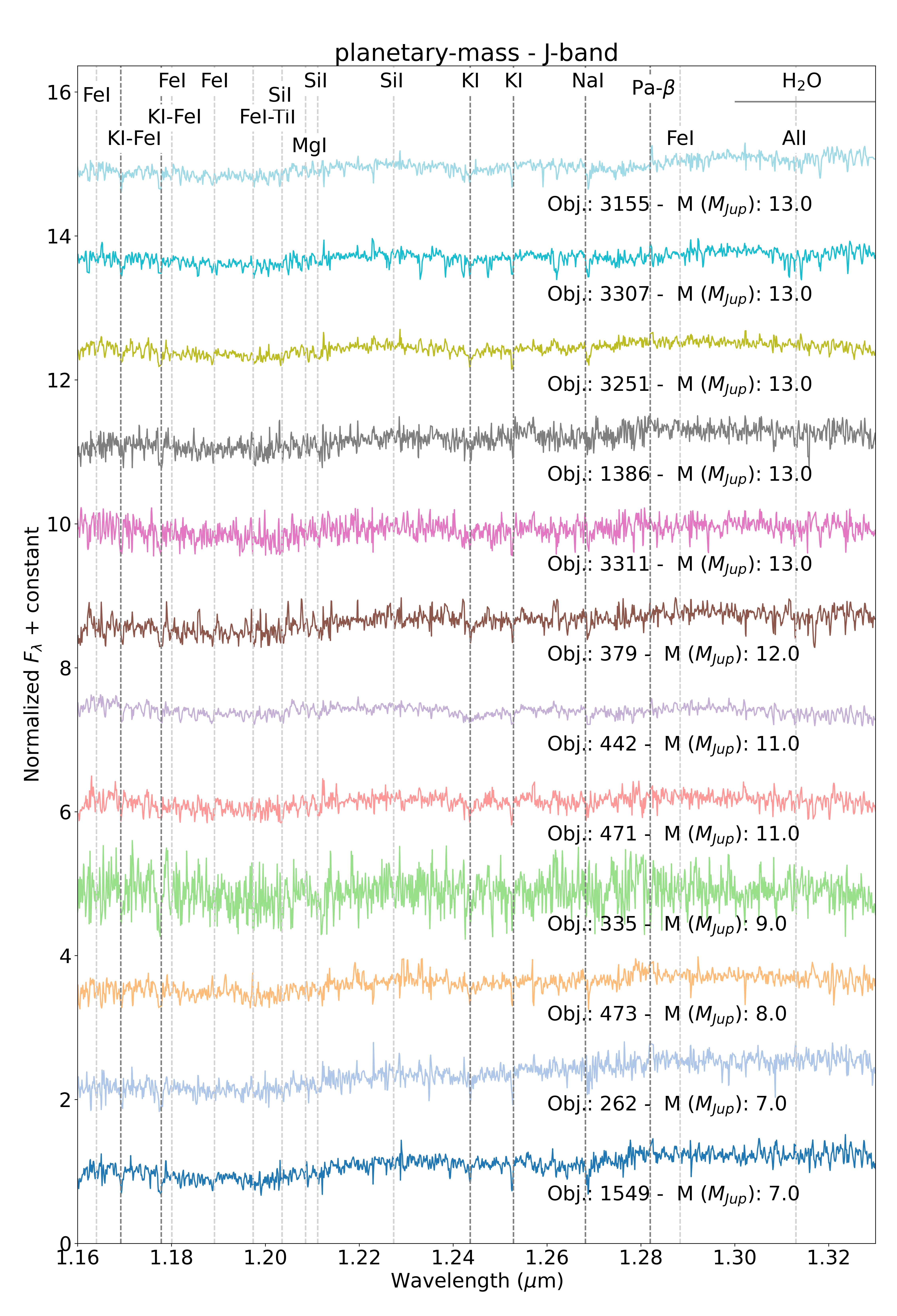}
    \caption{$J$-band spectra of the planetary-mass brown dwarfs with masses between 7 and 13~$\mathrm{M_{Jup}}$. {See main body discussion for significance of identified lines.}}
    \label{fig:J-planetary_mass}
\end{figure}

\begin{figure}
    \centering
    \includegraphics[width=0.99\linewidth]{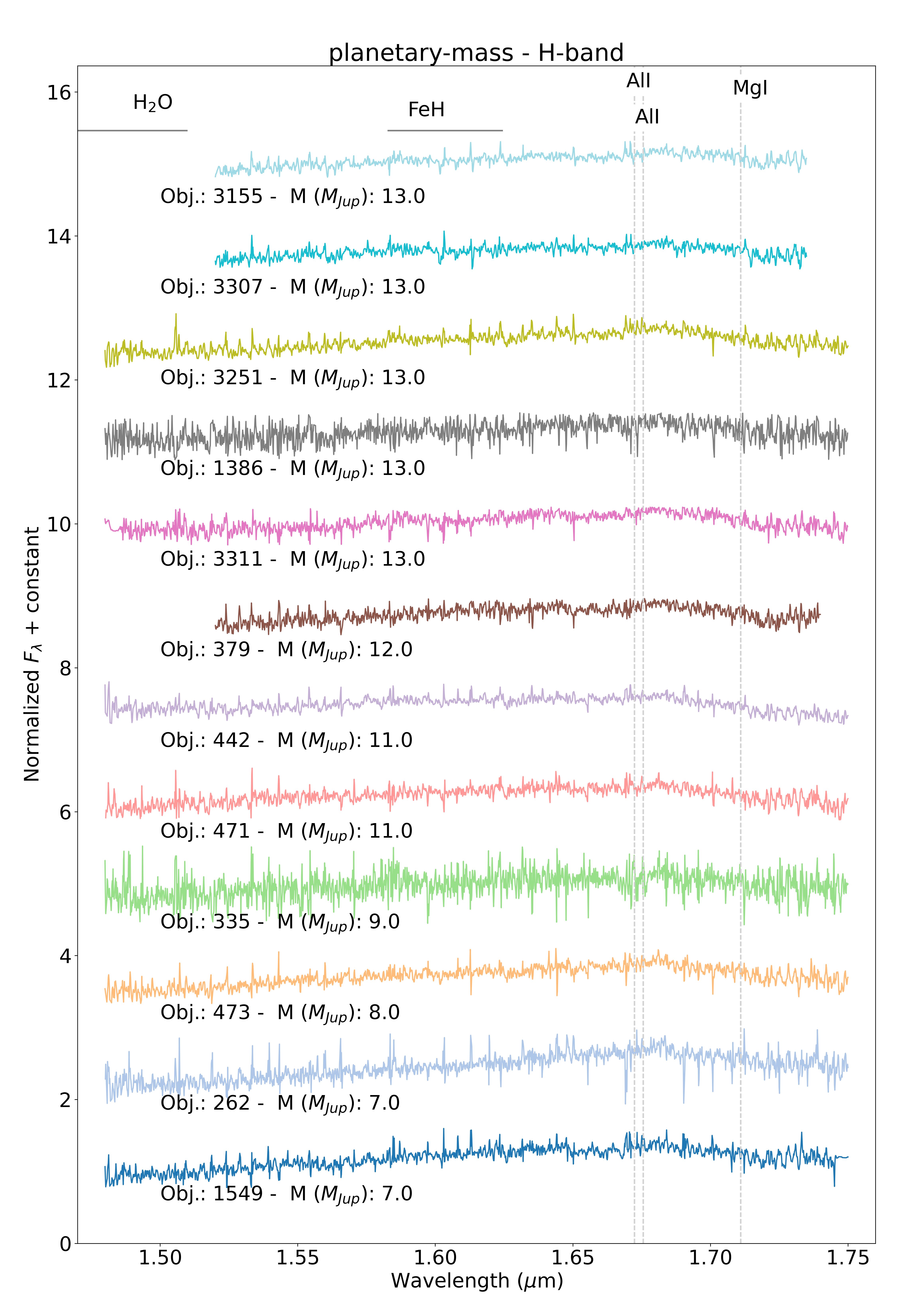}
    \caption{$H$-band spectra of the planetary-mass brown dwarfs with masses between 7 and 13~$\mathrm{M_{Jup}}$. {See main body discussion for significance of identified lines.}}
    \label{fig:H-planetary_mass}
\end{figure}

\begin{figure}
    \centering
    \includegraphics[width=0.99\linewidth]{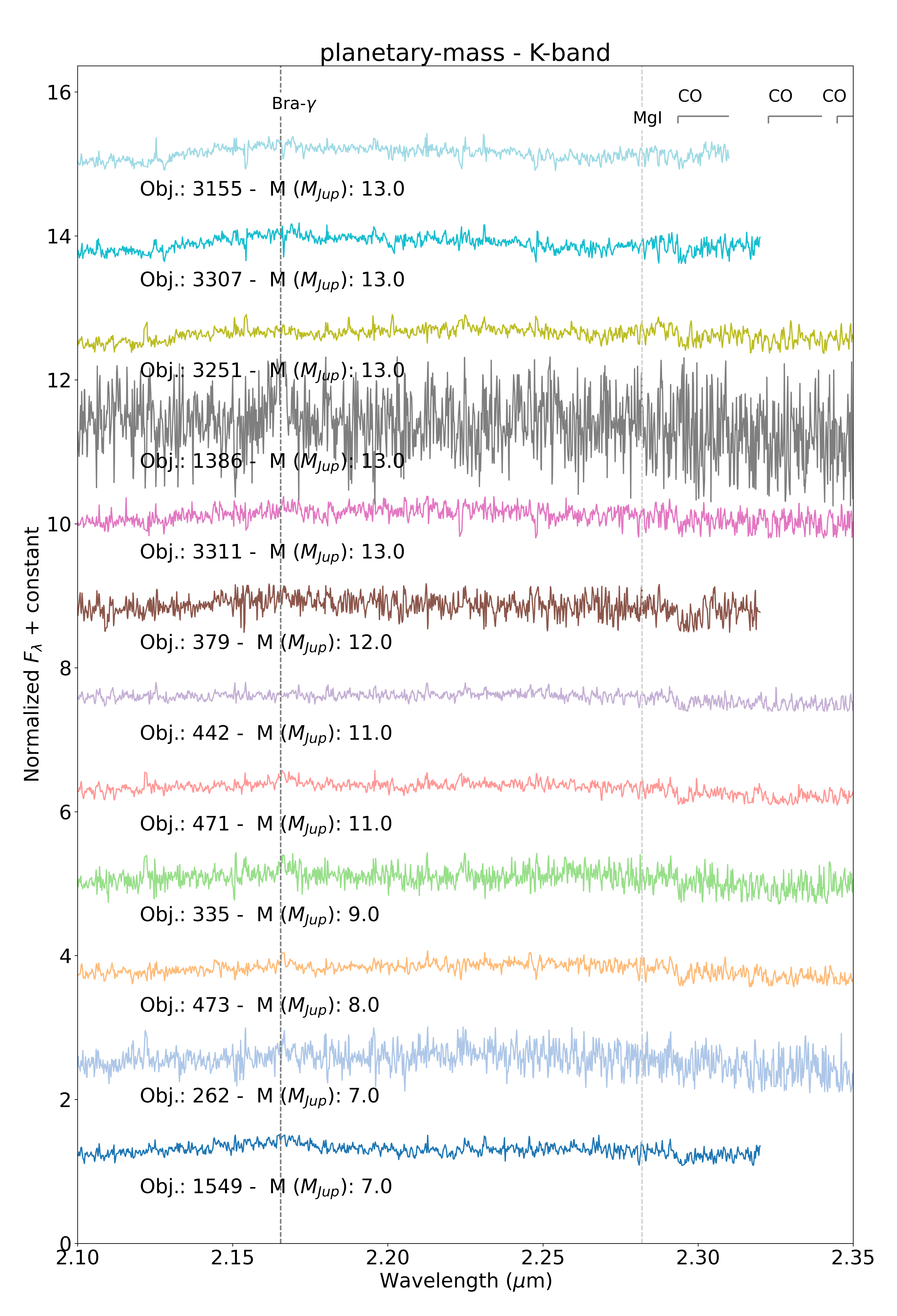}
    \caption{$K$-band spectra of the planetary-mass brown dwarfs with masses between 7 and 13~$\mathrm{M_{Jup}}$. {See main body discussion for significance of identified lines.}}
    \label{fig:K-planetary_mass}
\end{figure}

We observe the following spectral characteristics typical of brown dwarfs: 

\begin{itemize}
    \item In the $J$ band, we observe the K\,I doublet at around 1.169 and 1.177~$\mu$m, and at 1.2430 and 1.252~$\mu$m, and the Na\,I line at 1.268~$\mu$m. We also indicated other lines {(noted as light grey dashed lines)}, but those are not as prominent, with the exception of the Pa-$\beta$ emission line at 1.282~$\mu$m in few objects, in particular in objects 262, 3155, 3251, 3253, and 469 (Fig: \ref{fig:J-high_mass}, \ref{fig:J-intermediate_mass}, \ref{fig:J-planetary_mass}). A more thorough analysis {and interpretation} of these lines will be shown in Section \ref{sec:emission_lines}. 

    \item The $H$-band shows the characteristic triangular shape of young brown dwarfs \citep{Allers2013}. We indicate where the FeH molecule at around 1.6~$\mu$m and the Al\,I and Mag\,I lines should appear at 1.672, 1.675, and 1.711~$\mu$m respectively (see Fig. \ref{fig:H-high_mass}, \ref{fig:H-intermediate_mass}, \ref{fig:H-planetary_mass}), {but we do not clearly detect them, and we show them in light grey dashed lines}. This is expected for the FeH molecule that is usually not present is young brown dwarfs as indicated in previous studies (e.g. \citealt{Allers2013}, \citealt{Lodieu2018}).

    \item Finally, in the $K$-band we identified the CO bands between 2.3 and 2.35~$\mu$m, characteristic of M- and L-dwarfs, {the Mg\,I line at 2.282~$\mu$m (non-detected, light grey line)}, and the Bra-$\gamma$ line only in object 1549 (Fig. \ref{fig:K-high_mass}, \ref{fig:K-intermediate_mass}, \ref{fig:K-planetary_mass}).
    
\end{itemize}

\subsection{Spectral Typing}\label{sec:spectral_typing}

We estimated the spectral types of our targets and compared them with the estimated temperatures from \cite{Robberto2020} using the BT-Settl models {\citep{Allard2012a}}. To obtain a correct spectral type, we first removed the reddening produced by the extinction of the ONC cloud estimated by \cite{Robberto2020}. 


To remove the extinction, we used the Python function "extinction"\footnote{https://extinction.readthedocs.io/en/latest/} {\citep{Barbary2016}} with $R_V$=3.1, which is related to the grain size distribution, and 3.1 is the typical number for Milky Way-like dust. We used the extinction law from \cite{Cardelli1989}.
After {removing the extinction of the ONC cloud}, we aimed at spectral typing our targets by comparing them to field and young brown dwarf near-infrared template spectra published in the literature. Namely, we compared the field L  brown dwarf spectra published in \cite{Cushing2005}, and the young brown dwarf spectra published in \cite{Lodieu2008}, \cite{Bonnefoy2010}, \cite{Bonnefoy2013}, \cite{Allers&Liu}, \cite{Muench2007}, \cite{Manjavacas2014}, \cite{Manjavacas2016}, \cite{Manjavacas2020}, and \cite{Luhman2024}. We used equations 1 and 2 from \cite{Cushing2008} to find the best overall matches to our spectra. {Note that those template spectra cover a more mature age span than our sample. In this Section we focus on the broad similarities for spectral typing/effective temperate determination, and in the next section we will instead discuss the specific unique characteristics of our $\sim 1-2$ Myrs sample.}

We found that 11 out of our 25 targets showed a best match with at least one of the templates mentioned above in all $J$-, $H$-, and $K$-band simultaneously. Fourteen of the targets had overluminous $K$-bands,  indicating either a near-infrared excess due to a disk or underestimated extinctions. Six of the objects showed either Pa$\beta$ or Bra$\gamma$ emission lines, but only three of them showed also near-infrared excess. {In Appendix \ref{app:best_matches} (see Fig. \ref{fig:best_matches_templates1}, \ref{fig:best_matches_templates2} and \ref{fig:best_matches_templates3}) we show the best matches to all our targets with their respective templates}. In Table \ref{table:spectral_types} we specify the best matching template spectra to our targets, {whether or not} they show an overluminous $K$-band, and if the addition of a disk to the best matching template in the $J$- and $H$-band {explains this over-luminosity}. Finally, we also {indicate} if the target shows a Paschen $\beta$ or Brackett $\gamma$ emission line. For comparison, in the first two columns we show the derived masses and effective temperatures by \cite{Robberto2020}. 

\begin{table*}[!ht]
    \small
    \caption{Best matches to our ONC targets after comparing with templates of young and field brown dwarfs from the literature. {The best matches that have a * next to their name are templates with evidence of near-infrared excess or a protoplanetary disk.} }
	\label{table:spectral_types}
	\centering
    \begin{tabular}{llllllllll}
    \hline
    \hline

Target  & Mass ($\mathrm{M_{Jup}}$) & $\mathrm{T_{eff}}$ (K) & SpT & Best Match & Ref. & $K$-overlum.? & Disk? & Pa-$\beta$ & Bra-$\gamma$ \\ \hline 
262 & 7 & 2102 & M9.5 & OTS44* &  1 & No & ~ & Yes & ~ \\ 
{1549} & 7 & 2081 & L3.0 & 2MASS J03552337+1133437 &  2 & Yes &  & ~ & ~ \\ 
473 & 8 & 2141 & L3.0 & {G196-3B} &  {4} & No & ~ & ~ & Yes \\ 
{335} & 9 & 2179 & L3.0 & 2MASS J212650.40-814029.3 &  8 & Yes & ~ & ~ & ~ \\ 
442 & 11 & 2276 & M9.0 & 2MASS J15474719--242349 &  4 & No & ~ & ~ & ~ \\ 
471 & 11 & 2265 & M9-L2 & ONC95 &  5 & No &  & ~ & ~ \\ 
379 & 12 & 2308 & L0.0 & 2MASS J19355595--2846349 & 6 & No &  & ~ & ~ \\ 
1386 & 13 & 2328 & M8.5 & KPNO-Tau 6 & 7 & No &  & ~ & ~ \\ 
{3155} & 13 & 2343 & M9-L2 & ONC95 &  5 & Yes &  & Yes & ~ \\ 
{3251} & 13 & 2333 & M9.5 & OTS44* &  1 & Yes & Yes ($\mathrm{T_{eff}}$=250K) & Yes & ~ \\ 
{3307} & 13 & 2336 & M6.0 & 2MASS J04221413+15305* &  4 & Yes &  & ~ & ~ \\ 
{3311} & 13 & 2322 & M8.0 & 2MASS J11085497--763240* &  3 & Yes & Yes ($\mathrm{T_{eff}}$=1800K) & ~ & ~ \\ 
3256 & 14 & 2345 & M8.0 & 2MASS J11074665--761517* &  3 & No & ~ & ~ & ~ \\ 
{1572} & 15 & 2386 & M8.0 & 2MASS J11074665--761517* &  3 & Yes &  & ~ & ~ \\
333 & 16 & 2402 & M7.0 & 2MASS J11123099--765334 &  3 & No & ~ & ~ & ~ \\ 
{3253} & 18 & 2434 & M8.5 & ONC85 &  5 & Yes &  & Yes & ~ \\ 
{3385} & 20 & 2466 & M8.0 & 2MASS 11085497--763240* &  3 & Yes &  & ~ & ~ \\ 
{833} & 35 & 2649 & M9.0 & 2MASS J1935559-284634 & 4 & Yes &  & ~ & ~ \\ 
366 & 37 & 2675 & M8.0 & 2MASS J11085497--763240* &  3 & No & ~ & ~ & ~\\ 
372 & 40 & 2720 & M8.0 & 2MASS J11085497--763240* & 3 & No & ~ & ~ & ~ \\ 
{1376} & 41 & 2732 & M9.0 & 2MASS J19355595--284634 &  4 & Yes &  & ~ & ~ \\ 
{255} & 49 & 2761 & M9.0 & 2MASS J19355595--284634 &  4 & Yes &  & ~ & ~ \\ 
{3238} & 50 & 2763 &  M6.0 & 2MASS J04221413+15305* &  4 & Yes & ~ & ~ & ~ \\
469 & 71 & 2846 & M7.0 & 2MASS J04351455--141446 &  4 & No & ~ & Yes & ~ \\ 
{3241} & 76 & 2852 & M8.0 & 2MASS J11085497--763240* &  3 & Yes & ~ & ~ & ~ \\ 
\hline

    \end{tabular}
    
    \begin{tablenotes}
\item Best matches references: [1] - \cite{Bonnefoy2013}; [2] - \cite{Manjavacas2016}; [3] - \cite{Manjavacas2020}; [4] - \cite{Allers2013}; [5] - \cite{Luhman2024}; [6] - \cite{Cushing2005}; [7] - \cite{Muench2007}; [8] - \cite{Manjavacas2014}.
   \end{tablenotes}
   
\end{table*}

{Next, we briefly discuss the best fit templates for each object.}

\subsubsection{Targets with no {$K$-band over-luminosity}}\label{sec:no_ir_excess_targets}

In Table \ref{table:spectral_types} we show the 11 targets that have best matches to some of the templates found in the references listed above. Namely, these targets are 262, 473, 442, 471, 379, 1386, 3256, 333, 366, 372, and 469. Of the 11 targets, 10 show best matches to young brown dwarf templates. 

 In the following, we describe the best matches we found for each target. The plots comparing each object to their respective best matches are shown in the Appendix \ref{app:best_matches} in Fig. \ref{fig:best_matches_templates1},  \ref{fig:best_matches_templates2} and \ref{fig:best_matches_templates3}.

\begin{itemize}

\item Object 262 (M$\sim$7~$\mathrm{M_{Jup}}$, \citealt{Robberto2020}) shows a best match to OTS~44 {\citep{Bonnefoy2013}}, a M9.5 planetary-mass free-floating brown dwarf which shows strong H-$\alpha$ and Pa-$\beta$ emission lines, and a disk with an estimated mass of 0.1~$\mathrm{M_{Jup}}$. Object 262 also has Pa-$\beta$ emission, indicating that these two objects might share similar characteristics.

\item {Object 473 (M$\sim$8~$\mathrm{M_{Jup}}$) has a best match to G196-3B, a very-low surface L3 low-mass companion to a 20-300~Myr star \citep{Allers2013} with similar characteristics to the 2MASS~1207b planet companion. }

\item Object 442 (M$\sim$11~$\mathrm{M_{Jup}}$) has a best match with 2MASS~J15474719-2423493, an intermediate gravity M9.0 brown dwarf \citep{Allers2013}.

\item Object 471 (M$\sim$11~$\mathrm{M_{Jup}}$) has a best match with ONC~95 \citep{Luhman2024}, a M9.0-L2.0 planetary-mass brown dwarf, also a member of the ONC.

\item Object 379 (M$\sim$12~$\mathrm{M_{Jup}}$) has a best match with 2MASS J19355595–2846349 \citep{Allers2013}, a young L0 brown dwarf. 2MASS J19355595–2846349 {\citep{Allers2013}} shows some more flux in the $K$-band than 379, although nearly consistent with our target.

\item Object 1386 (M$\sim$13~$\mathrm{M_{Jup}}$) has a best match with KPNO-Tau 6, a $\sim$20~$\mathrm{M_{Jup}}$, M8.5 brown dwarf member of the Taurus-Auriga star-forming region Age 1-2 Myr, with a similar age to the ONC {\citep{Muench2007}}. 

\item Object 3256 (M$\sim$14~$\mathrm{M_{Jup}}$) has a best match with 2MASS~J11074665–761517 \citep{Manjavacas2020}, a M8.0 brown dwarf with a disk \citep{Manara2017} that is a member of the Chamaeleon I region \citep{Luhman2007}, which has a similar age to the ONC. {While this target is over-luminous in $K$-band, adding an extra blackbody component was not necessary  since it most likely shares the similar disk properties as 2MASS~J11074665–761517 .}

\item Object 333 (M$\sim$14~$\mathrm{M_{Jup}}$) has a best match with 2MASS~J11123099–765334 \citep{Manjavacas2020}, a M7.0 brown dwarf that is also a member of the Chamaeleon I region \citep{Luhman2007} with an estimated mass of $\sim$27~$\mathrm{M_{Jup}}$.

\item Objects 366 (M$\sim$37~$\mathrm{M_{Jup}}$) and 372 (M$\sim$40~$\mathrm{M_{Jup}}$) have best matches to 2MASS~J11085497--763240 \citep{Manjavacas2020}, a M8.0 brown dwarf member of the Chamaeleon I region with a protoplanetary disk \citep{Long2017}. {While this target is over-luminous in K band, adding an extra blackbody component was not necessary  since it most likely shares the similar disk properties as 2MASS~J11085497--763240.}

\item Finally, object 469 (M$\sim$71~$\mathrm{M_{Jup}}$) has a best match with 2MASS J04351455--141446, a very red M7.0 brown dwarf with very-low surface gravity \citep{Allers2013}.

\end{itemize}


\subsection{Targets with {$K$-band over-luminosity}}\label{sec:ir_excess_targets}

From the 25 targets in our sample, 14 {are over-luminous} in $K$-band {(e.g there is no empirical template that fits  $J$, $H$ and $K$-bands simultaneously}. To try to reproduce their SEDs, we fit their $J$- and $H$-bands with the brown dwarf spectral templates mentioned above, and then we try to reproduce their near-infrared excess using two different methods: modifying the extinction applied to the spectra, and adding the flux contribution of a protoplanetary disk.

Since the extinctions we applied to our spectra were measured by \cite{Robberto2020} using the BT-Settl models that could not fit the area of the color-magnitude diagram where the objects with $<$20~$\mathrm{M_{Jup}}$ lie, there is the possibility that some of those extinctions are not accurate. Thus, we add additional extinction to the spectra to try to reproduce their SEDs in these cases. However, none of the best matches of our spectra were improved in the $K$-band by adding additional extinction.

We also added to the $J$- and $H$-band best matching templates the emission of a black body with varying effective temperature ($\mathrm{T_{eff}}$) between 50 and 3000~K to {test if the additional $K$-band flux can be explained} the contribution of a protoplanetary disk, and {its} contribution to the $K$-band flux that we were not able to fit with any of the templates. Of the 14 targets that show near-infrared excess in their spectra, two were successfully reproduced in their entirety by adding the contribution of a protoplanetary disk {represented as a pure black body}. In the following,  we will specify the best matches for each target. The individual best matches are shown in the Appendix \ref{app:best_matches} in Fig. \ref{fig:best_matches_templates1}, \ref{fig:best_matches_templates2} and \ref{fig:best_matches_templates3}:

\begin{itemize}

    \item Object 3251 has a best match with the planetary-mass object OTS~44 (M9.5) described in Section \ref{sec:no_ir_excess_targets}. In addition, to fit the $K$-band, we need to add the SED of a protoplanetary disk of 250~K {\citep{Bonnefoy2013}}.

    \item Object {3311} is best fitted by  2MASS~J11085497–763240 \citep{Manjavacas2020}, a M8.0 dwarf member of the Chameleon I region. 2MASS~J11085497–763240 has a protoplanetary disk \citep{Long2017}. To fit the $K$-band of object 3311 we need to add a black body simulating the presence of a protoplanetary disk of an approximate temperature of 1800~K.

\end{itemize}

Finally, objects {3155}, 3385, 3307, 1572, 3253, 833, 1376, 255, 1549, 335,  3238, and 3241 show a near-infrared excess in the $K$-band that we were not able to reproduce using a black body disk emission or modifying their extinction. {Given the relative youth of our sample compared to existing templates, we hypothesize that these sources feature atmospheric properties that are ``unique'' to ONC's age. Clouds are known to produce broad features that impact near-infrared colors and it might be possible that the very low gravity of these sources induces cloud properties that are not captured by existing spectral templates. Follow-up observations of these sources at longer wavelengths would test this hypothesis. Regardless,} we could reproduce their $J$- and $H$-bands using one of the templates available in the spectral libraries mentioned above.

In the following, we specify the best matches for each of these objects. The best matches are shown in the Appendix \ref{app:best_matches} in Fig. \ref{fig:best_matches_templates1}, \ref{fig:best_matches_templates2} and \ref{fig:best_matches_templates3}:

\begin{itemize}
    \item The $J$- and $H$-bands of the spectrum of object 1549 have a best match to object 2MASS J03552337+1133437. As mentioned above, 2MASS J03552337+1133437 is an extremely red L3 young brown dwarf \citep{Faherty2013, Manjavacas2016} with similar characteristics to the 2MASS~1207b planet companion.

    \item The $J$- and $H$-bands of the spectrum of object 335 have a best match to 2MASS J212650.40-814029.3, and L3.0 young object \citep{Manjavacas2014}, with a similar spectrum to CD-352722B, a brown dwarf companion with member to the AB Doradus association (75--150 Myr old, \citealt{Delorme2012}).


    \item The $J$- and $H$-bands of 3238 and 3307 have a best match to 2MASS~J04221413+15305 \citep{Allers2013}, a very red and very young M6.0, $\sim$10$^{\circ}$ to the South of the Taurus star-forming region on the sky. 

    \item The $J$- and $H$-bands of objects 3241, 3311, and 3385 have a best match to 2MASS~J11085497–763240 \citep{Manjavacas2020}. As explained above, it is a M8.0 dwarf member of the Chameleon I region with a protoplanetary disk \citep{Long2017}. Even though object 3241 (76~$\mathrm{M_{Jup}}$) has a higher mass in comparison to 3238 (50~$\mathrm{M_{Jup}}$), 3241 has a best match to a later spectral type object, at least in the $J$- and $H$-bands. However, we need to take these estimated masses with {caution}, since they were estimated using evolutionary models by \cite{Robberto2020}.

    \item Object 1572 is best reproduced by the M7.0 Chamaelon\,I member 2MASS~J11074665–761517 \citep{Manjavacas2020}, but adding the contribution of a protoplanetary disk did not significantly improve the fit in the $K$-band. 2MASS~J11074665–761517 has a protoplanetary disk \citep{Manara2017}.

    \item Object 3253 is best reproduced by ONC~85,  M8.5 brown dwarf member of the ONC \citep{Luhman2024} that shows no near-infrared excess, but adding the contribution of a protoplanetary disk did not significantly improve the fit in the $K$-band.

    \item {Object 3155 is best reproduced by ONC~95,  an M9-L2 brown dwarf member of the ONC \citep{Luhman2024} that shows no near-infrared excess. Adding the contribution of a protoplanetary disk did not significantly improve the fit in the $K$-band.}

    \item Finally, objects 833, 1376, and 255 have a best match to 2MASS~J19355595--284634, a M9.0 with very-low surface gravity \citep{Allers2013}, but adding the contribution of a protoplanetary disk did not significantly improve the fit the flux excess in the $K$-band.
    
\end{itemize}

\section{Unique characteristics of our ONC sample}\label{sec:characteristics_ONC_sample}

\subsection{Gravity Spectral Indices}\label{sec:spectral_indices}

{As discussed above, all our sources qualitatively feature low surface gravity the gravity spectral indices from \cite{Allers2013}. This confirms the youth of our targets and their membership to the ONC and thus spectroscopically validates the photometric criteria based on water absorption in \cite{Robberto2020}. We now quantitatively assess the strength of these features. To do so,} we measured the pseudo-equivalent widths of the K\,I alkali lines at 1.169, 1.177, 1.243 and 1.253~$\mu$m for all our targets.
Similarly, we measured the $H$-cont, the $\mathrm{K\,I_{J}}$, and the $\mathrm{FeH_{J}}$ indices following the same procedure as \cite{Allers2013}. These three spectral indexes measure, respectively, the shape of the $H$-band, which is more triangular for low surface gravity brown dwarfs,  the depth of the FeH molecule in the $J$-band, and the depth of the $\mathrm{K\,I_{J}}$ lines. The specific wavelength ranges that each of these indices span, and how the final index values are obtained are shown in Table~4 and Equation~1 in \cite{Allers2013}.

Using brown dwarfs with ages from 2~Myr to $\sim$1~Gyr, \cite{Allers2013} provided a range of values for each of their indices and spectral types that would determine the surface gravity category of brown dwarfs. Those three categories are field (FLD-G, gravity score 0, {log $g$ = 5.0-5.5}), intermediate-gravity (INT-G, gravity score 1, {log $g$ = 4.0-4.5}), and very-low gravity (VL-G, gravity score 2, {log $g$ $<$ 4.0}). {After measuring the pseudo-equivalent widths of the alkali lines and the gravity indices, we classified the targets by gravity types following the procedure in \cite{Allers2013} using their Tables 10 and 9, respectively. As expected for very young brown dwarfs as the members of ONC, most of our targets were classified as very-low gravity objects in almost all the indices independently, according to the values of their pseudo-equivalent widths and gravity index values (see Tables~\ref{table:ew_all_lines_nir} and \ref{table:gscores}). The final gravity score is computed as the median of the individual gravity scores of the pseudo-equivalent width of the 1.169, 1.177 and 1.253~$\mu$m K\,I lines, and the $H$-cont, the $\mathrm{K\,I_{J}}$, and the $\mathrm{FeH_{J}}$ indexes. If this score is $\geq$1.5, then the object is classified as a very-low surface gravity object. All our targets had a final gravity score $\geq$1.5, confirming that our qualitative analysis targets are all very-low surface gravity brown dwarfs, as expected for members of the ONC and {further corroborating the finding on the Initial Mass Function by \cite{Gennaro2020}.}}

\begin{figure*}
    \centering
    \includegraphics[width=0.48\linewidth]{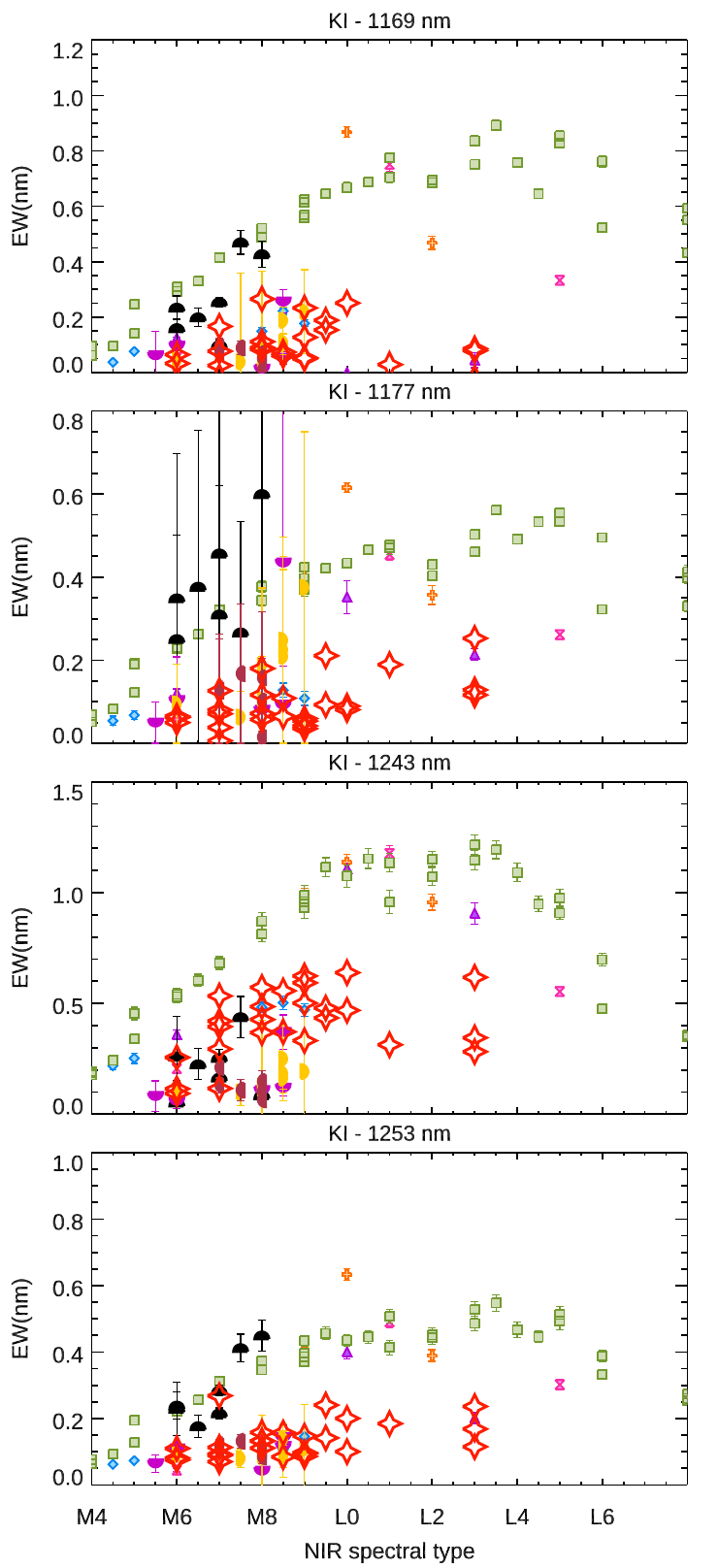}
    \includegraphics[width=0.48\linewidth]{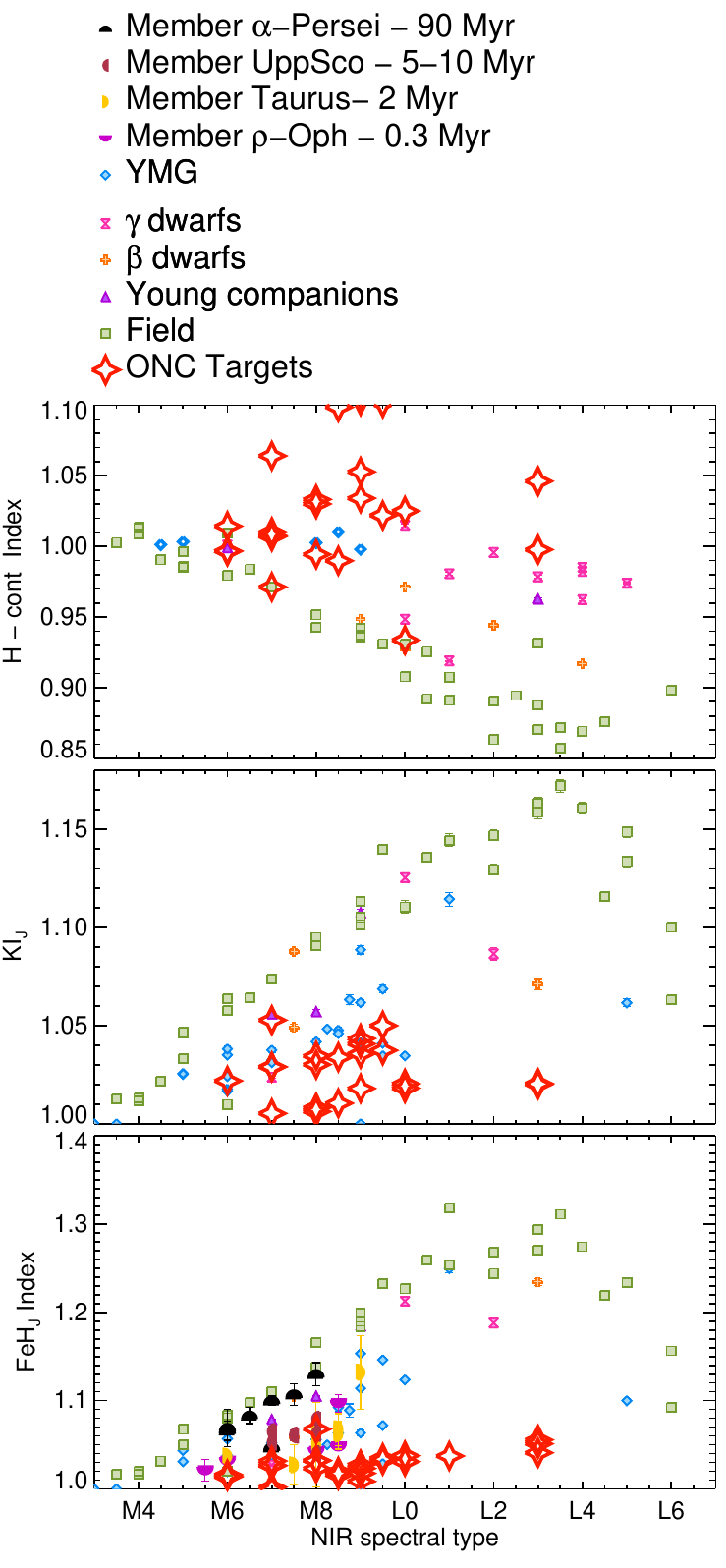}
    \caption{Pseudo equivalent widths {(left)} and spectral indices {(right)} calculated for our targets (red stars) following the methodology presented in \cite{Allers2013}. We include the same indices calculated for brown dwarfs and low-mass stars that are members of young moving groups (YMG), $\gamma$-dwarfs and $\beta$-dwarfs \citep{Allers2007, Bonnefoy2014a}, young companions \citep{Allers2007}, $\alpha$ Persei (90~Myr), Upper Scorpious (5--10~Myr), Taurus (2~Myr), and $\rho$ Ophiucus (0.3~Myr) from \cite{Martin2017} and field brown dwarfs \citep{McLean, Cushing2005}.}
    \label{fig:indices}
\end{figure*}

In Fig. \ref{fig:indices}, we show the values of the pseudo-equivalent widths and the gravity spectral indices measured for our targets using \cite{Allers2013} compared to brown dwarfs and low-mass stars of different ages. {For all diagnostics, our sample spans a very distinct track than the one of field brown dwarfs \citep{McLean, Cushing2005}. This is not surprising given the extreme age difference between {the ONC (1-3~Myr, \citealt{Jeffries2011}) and the field {($>$500~Myr, \citealt{Manjavacas2020}}}). Moreover our sample of 25 object appears to provide a robust ``young ages'' boundary on the progression of these gravity diagnostics with spectral type. Up to now,  establishing this empirical boundary was plagued by small number statistics with only a handful of sources at ages comparable to the ONC: Taurus ($\sim$2~Myr), and $\rho$ Ophiucus ($\sim$0.3~Myr), also indicated on  Fig. \ref{fig:indices}. Our sample more than triples the number of sources below $\sim$2~Myr and extends to much later spectral types, and confirms the tentative trends already identified in the literature. Finally there appears to be an age evolution. Indeed, the values of these diagnostics for intermediate ages -e.g members of young moving groups (YMG), $\gamma$-dwarfs / $\beta$-dwarfs \citep{Allers2007, Bonnefoy2014a}, young companions \citep{Allers2007}, and  members of open clusters and associations, $\alpha$ Persei ($\sim$90~Myr), Upper Scorpious (5--10~Myr)- lies between the field and young tracks. Extending this track to later spectral types is the next observational challenge, best suited for JWST observations \citep{Luhman2024}.}




	\begin{table*}
		\small
		\caption{Equivalent widths in nm for alkali lines measured  in the near-infrared.}  
		\label{table:ew_all_lines_nir}
		\centering
		\begin{center}
			\begin{tabular}{lllllll}
				\hline
				\hline 
				
				Target & NIR SpT & K\,I~(1169~nm)  & K\,I~(1177~nm) & K\,I~(1243~nm) & K\,I~(1253~nm) & GS $\mathrm{^{a, b}}$\\		
				\hline              
				262  & M9.5 & 0.188$\pm$0.012 & 0.211$\pm$0.021   & 0.434$\pm$0.021   & 0.241$\pm$0.020  &  22--2   / VL-G  \\
                1549 & L3.0 & 0.086$\pm$0.010   & 0.129$\pm$0.029   & 0.282$\pm$0.026    & 0.169$\pm$0.013   &    22--2   / VL-G  \\
                473  & L3.0 & 0.079$\pm$0.025  & 0.116$\pm$0.022  & 0.618$\pm$0.046   & 0.115$\pm$0.008  &    22--2    / VL-G  \\
                335  & L3.0 & -- & 0.253$\pm$0.032 & 0.345$\pm$0.054 &	0.236$\pm$0.025    &  --2--2 / VL-G   \\
                442  & M9.0 & 0.054$\pm$0.026   & 0.0355$\pm$0.021   & 0.624$\pm$0.043   & 0.147$\pm$0.0158   &   22--2     / VL-G    \\
                471  & M9-L2 & --    &	0.089$\pm$0.022   & 0.639$\pm$0.040   & 0.200$\pm$0.016   &  --2--2    / VL-G     \\
                
                379  & L1.0	& 0.029$\pm$0.019   & 0.189$\pm$0.025   & 0.313$\pm$0.058   & 0.185$\pm$0.015    &   22--2  / VL-G   \\
                1386 & M8.5	    & 0.061$\pm$0.009   & 0.108$\pm$0.021   & 0.556$\pm$0.048   & 0.085$\pm$0.027    &  22--2   / VL-G    \\
                3155 & M9-L2    & 0.251$\pm$0.016   & 0.081$\pm$0.019   & 0.468$\pm$0.027   & 0.100$\pm$0.016    &   22--2  / VL-G    \\
                3251 & M9.5     & 0.153$\pm$0.025   & 0.093$\pm$0.020   & 0.479$\pm$0.031   & 0.142$\pm$0.006    &  22--2   / VL-G   \\
                3307 & M8.0   & 0.265$\pm$0.017 & 0.072$\pm$0.021   & 0.486$\pm$0.051 & 0.111$\pm$0.016  &   22--2  / VL-G    \\
                3311 & M8.0    & 0.111$\pm$0.047  & 0.180$\pm$0.026   & 0.427$\pm$0.062   & 0.132$\pm$0.025    &   22--2  / VL-G    \\
                3256 & M8.0    & 0.083$\pm$0.016   & 0.117$\pm$0.026   & 0.57$\pm$0.025   & 0.158$\pm$0.010  &    22--2  / VL-G    \\
                
                1572 & M8.0     & 0.089$\pm$0.003  & 0.057$\pm$0.019   & 0.36$\pm$0.020    & 0.133$\pm$0.002    &   22--2  / VL-G     \\
				333  & M7.0	    & 0.167$\pm$0.015	  & 0.071$\pm$0.017   & 0.419$\pm$0.022 & 0.093$\pm$0.008    &   22--2/ VL-G     \\  
                3253 &	M8.5    & 0.074$\pm$0.014	  & 0.065$\pm$0.019   & 0.372$\pm$0.019   & 0.157$\pm$0.002    &  22--2   / VL-G     \\

				3385 & M8.0     &	0.025$\pm$0.017 & 0.038$\pm$0.018	& 0.294$\pm$0.031	 & 0.113$\pm$0.004  &     22--2   / VL-G   	\\
                833  &	M9.0    &	0.127$\pm$0.011 & 0.060$\pm$0.018	&	0.593$\pm$0.033   &	0.097$\pm$0.016	 & 22--2 / VL-G        \\
                366  &	M8.0    & 0.077$\pm$0.014 & --     & 0.533$\pm$0.034  & 0.070$\pm$0.013  & 2-- --2   / VL-G      \\

                372  &	M8.0    & --	  & 0.126$\pm$0.028	& 0.395$\pm$0.074    & 0.269$\pm$0.024  &   --2--1   / VL-G      \\
                1376 &	M9.0    & 0.233$\pm$0.015 & 0.054$\pm$0.015	& 0.501$\pm$0.036    & 0.086$\pm$0.003	         &   22--2   / VL-G      \\
                255  &	M9.0    & 0.049$\pm$0.014 & 0.042$\pm$0.010	& 0.332$\pm$0.021    & 0.094$\pm$0.008         &   22--2   / VL-G      \\
                3238 &	M6.0    & 0.064$\pm$0.011 & 0.062$\pm$0.014	&0.114$\pm$0.014    & 0.075$\pm$0.003  &   222 --   / VL-G      \\
                469  &	M7.0    & --	  & 0.080$\pm$0.012	& 0.116$\pm$0.013    & 0.089$\pm$0.005   &   --2--2  / VL-G      \\
                3241 &	M8.0    & --	  & 0.065$\pm$0.016	& 0.093$\pm$0.010   & 0.082$\pm$0.001         &   --2--2   / VL-G      \\
				
				\hline
				
			\end{tabular}
		\end{center}
				\begin{tablenotes}
		\small
		\item a: GS: Gravity scores calculated as in \cite{Allers2013}. b: Gravity scores are ordered according to the alkali line that they correspond to. The dash symbol indicates that none gravity score can be determined with that particular line. \\
		
	\end{tablenotes}		
	\end{table*}

	\begin{table*}
		\small
		\caption{Gravity scores for our sample derived from spectral indices defined in the literature \citep{Allers2013}.}  
		\label{table:gscores}
		\centering
		\begin{center}
			\begin{tabular}{llllll}
				\hline
				\hline 
				
				Name &  SpT     &   $\mathrm{FeH_{J}}$ & $\mathrm{KI_{J}}$ & $H$-cont         & G$\mathrm{S^{a}}$   \\		
				\hline              
				262 & M9.5 &     1.032$\pm$0.001  &  1.037$\pm$0.001 &   1.101$\pm$0.003     &  222 / VL-G  \\
                1549 & L3.0    & 1.051$\pm$0.001   & 0.990$\pm$0.002   & 1.046$\pm$0.002      &    222    / VL-G  \\
                473  & L3.0  &   1.055$\pm$0.001  &  1.020$\pm$0.002  &  0.998$\pm$0.001    &    222    / VL-G  \\
                335 & L3.0     &  1.041$\pm$0.001 &  1.020$\pm$0.002 &   1.132$\pm$0.002 &  222 / VL-G    \\
                442 & M9.0     &  1.023$\pm$0.001 &	 1.041$\pm$0.001 &	 1.103$\pm$0.001  &   222     / VL-G    \\
                471 & M9-L2  &    1.038$\pm$0.001 &	 1.018$\pm$0.001 &	 0.934$\pm$0.001  &   221    / VL-G     \\
                
                379 & L1.0	&    1.037$\pm$0.001 &   0.991$\pm$0.001 &  1.122$\pm$0.002  &   222  / VL-G   \\
                1386 & M8.5 &    1.019$\pm$0.001 &	 1.010$\pm$0.002 &	1.099$\pm$0.002     &  222   / VL-G    \\
                3155 & M9-L2   &  1.031$\pm$0.001&   1.020$\pm$0.001 & 1.025$\pm$0.001   &   222  / VL-G    \\
                3251 & M9.5    &  1.035$\pm$0.001 &  1.050$\pm$0.002 & 1.022$\pm$0.002     &  222   / VL-G   \\
                3307 & M8.0   &   1.032$\pm$0.001 &  1.030$\pm$0.001 & 1.030$\pm$0.001 &   222  / VL-G    \\
                3311 & M8.0    &  1.068$\pm$0.001 &	 1.034$\pm$0.001 & 1.144$\pm$0.002 &   222  / VL-G    \\
                3256 & M8.0    &  1.032$\pm$0.001 &  1.006$\pm$0.001 & 1.033$\pm$0.001 &    222  / VL-G    \\
                
                1572 & M8.0     & 1.023$\pm$0.001 &	1.008$\pm$0.002 & 0.994$\pm$0.002 &   222  / VL-G     \\
				333 & M7.0	    & 0.999$\pm$0.001 & 1.053$\pm$0.002 &  1.064$\pm$0.002  &   212 / VL-G     \\  
                3253 &	M8.5    & 1.016$\pm$0.001 & 1.034$\pm$0.001 &  0.989$\pm$0.001  &  221   / VL-G     \\

				3385 & M8.0     &  1.027$\pm$0.001 & 1.029$\pm$0.001 & 1.010$\pm$0.001 &     222   / VL-G   	\\
                833  &	M9.0    &  1.027$\pm$0.001 & 1.036$\pm$0.001 & 1.053$\pm$0.001 & 222 / VL-G        \\
                366 &	M8.0    &  1.032$\pm$0.001 & 1.029$\pm$0.001 &  1.008$\pm$0.001 &   222  / VL-G      \\

                372 &	M8.0    & 1.002$\pm$0.001 & 1.005$\pm$0.001 & 0.971$\pm$0.001 &   221   / VL-G      \\
                1376 &	M9.0    & 1.008$\pm$0.001 & 1.043$\pm$0.001 & 1.108$\pm$0.001 &   222   / VL-G      \\
                255 &	M9.0    & 1.018$\pm$0.001 & 1.018$\pm$0.001 & 1.034$\pm$0.001  &  222   / VL-G      \\
                3238 &	M6.0    & 1.016$\pm$0.001 & 0.987$\pm$0.001 & 0.997$\pm$0.001 &   222   / VL-G      \\
                365 &	M6.0    & 1.014$\pm$0.001 & 1.022$\pm$0.001 & 1.014$\pm$0.001 &   222   / VL-G      \\
                469 &	M7.0    & 1.026$\pm$0.001 & 0.983$\pm$0.001 & 1.007$\pm$0.001 &   222   / VL-G      \\
                3241 &	M8.0    & 1.015$\pm$0.001 & 0.983$\pm$0.001 & 0.997$\pm$0.001 &   222   / VL-G      \\
                \hline
			\end{tabular}
		\end{center}		
		\begin{tablenotes}
		\small
		\item a: GS: Gravity scores calculated as in \cite{Allers2013}\\		
	\end{tablenotes}		
	\end{table*}

\subsection{Emission Lines \& Circumstellar Disks}\label{sec:emission_lines}

\begin{figure}
    \centering
    \includegraphics[width=0.99\linewidth]{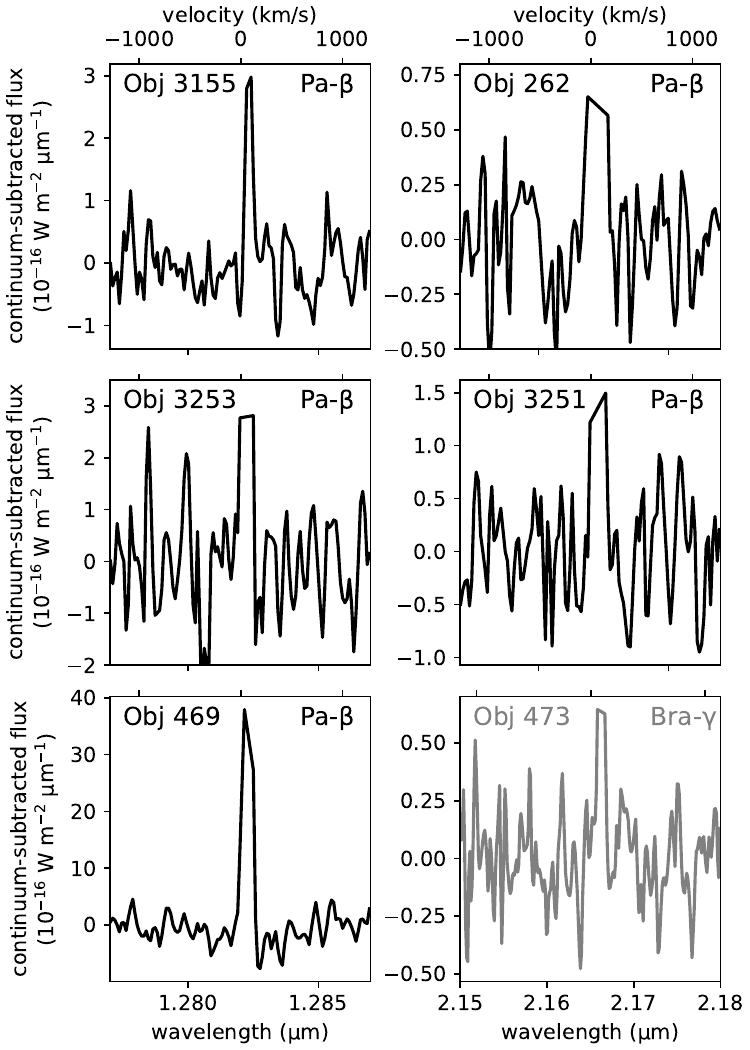}
    \caption{NIR emission lines for the six accreting objects.  Five show accretion from Pa-$\beta$, while one (Obj 473) shows accretion in Bra-$\gamma$ (gray spectrum)}
    \label{fig:accretion_emission_lines}
\end{figure}

As shown in Table~\ref{table:spectral_types} and in Fig.~\ref{fig:accretion_emission_lines}, six of the targets show 3$\sigma$ detections in emission from Pa-$\beta$ (5 targets) and Bra-$\gamma$ (1 target).  These targets are 262, 473, 3155, 3251, 3253, and 469, described in Sections~\ref{sec:no_ir_excess_targets}-\ref{sec:ir_excess_targets}. In the infrared, the Paschen and Brackett series emission are considered unambiguous signs of accretion \citep{Betti2022}.  In order to compute their accretion rates, we use \texttt{species} {\citep{Stolker2020}}{package\footnote{https://species.readthedocs.io/en/latest/tutorials/emission\_line.html}} to fit each emission line and determine the line flux ($F_{line}$) and line luminosity ($L_{line}$). The line fluxes are computed by directly integrating the continuum-subtracted spectra, which is found by first modeling and fitting a 3rd order polynomial to the local continuum. The uncertainty is found by sampling 1000 random spectra from the flux errors (see the \texttt{species} package for more details).  The line luminosity is then found as $L_{line} = 4\pi d^2 F_{line}$, where the distance ($d$) is assumed to be the distance to the ONC \citep[$d=414\pm7$ pc;][]{Menten2007}.  To compute the mass accretion rate ($\dot{M}$), we convert the $L_{line}$ to accretion luminosities ($L_{acc}$). In stars, the accretion luminosity linearly correlates with the emission line luminosities \citep{Rigliaco2012, Alcala2014, Alcala2017} and as
\begin{equation}
    \log(L_{acc}/L_\odot) = a \times \log(L_{line}/L_\odot) + b,
\end{equation}
where $a$ and $b$ are fit coefficients for each emission line.  These relationships have been assumed to extend in the substellar brown dwarf and planet regime, although \citet{Betti2023} has shown empirically that these might not hold for brown dwarfs.  For objects in the planetary mass regime, \citet{Aoyama2020, Aoyama2021} has theorized that the physical conditions of the accretion likely differ compared to stars due to slower shock velocities and determined theoretical $L_{acc}-L_{line}$ relationships to account for these differences. For all {ONC} objects, we determine their $L_{acc}$ using the stellar scaling relationship \citep[Pa-$\beta$: $(a,b)=(1.06, 2.76)$, Br-$\gamma$: $(a,b)=(1.19, 4.02)$;][]{Alcala2017}; for the objects below the deuterium burning limit, we additionally compute $L_{acc}$ using the theoretical planetary relationship \citep[Pa-$\beta$: $(a,b)=(0.86, 2.21)$, Br-$\gamma$: $(a,b)=(0.85, 2.84)$;][]{Aoyama2021}.       

The mass accretion rate is then found as:
\begin{equation}
    \dot{M} = \left(1- \frac{R_*}{R_{in}}\right)^{-1} \frac{L_{acc}R_*}{GM_*},
\end{equation}
where $R_{in}$ is the inner disk radius assumed to be 5 $R_*$ \citep[e.g.][]{Gullbring1998, Herczeg2008, Rigliaco2012, Alcala2014, Alcala2017}, $R_*$ is the stellar radius, and $M_*$ is the stellar mass. We give the line and accretion measurements in Table~\ref{table:accretion}. 

\begin{figure*}
    \centering
    \includegraphics[width=0.7\linewidth]{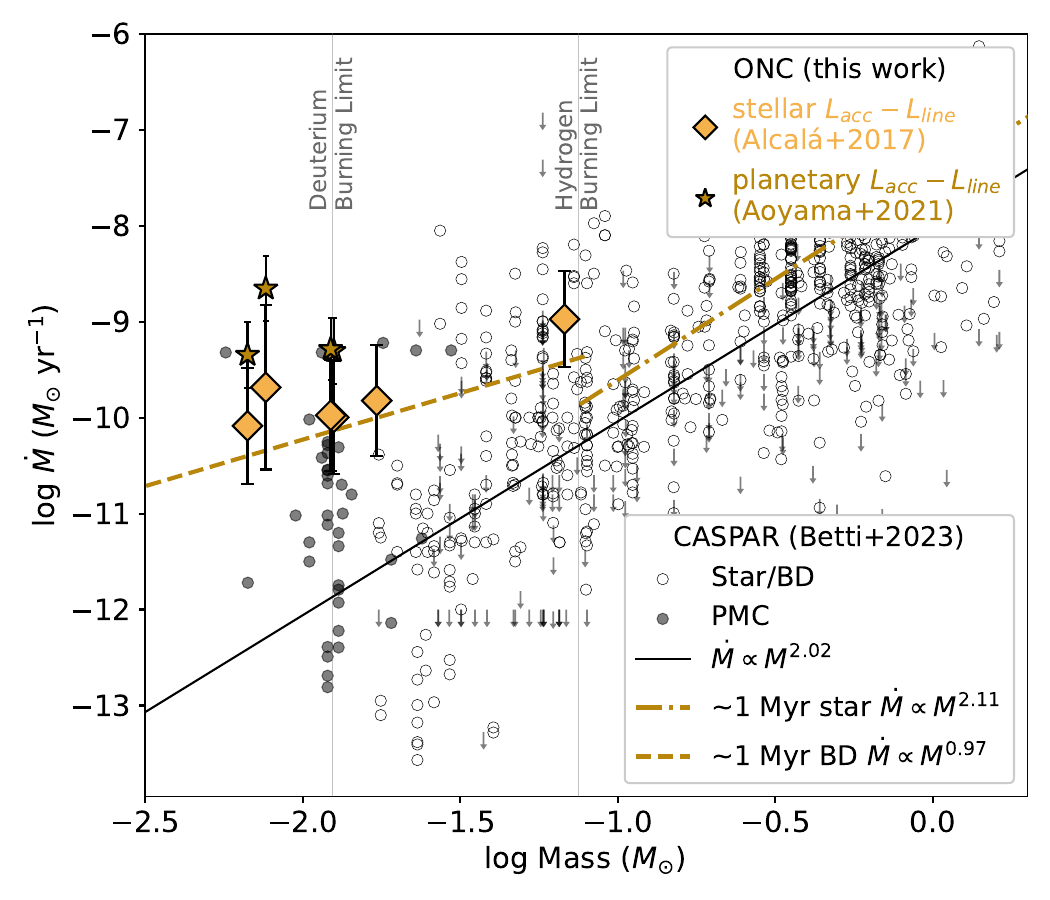}
    \caption{Mass accretion rate vs mass for the six ONC accretors (gold diamonds).  The four objects with masses below the deuterium burning limit have accretion rates calculated using planetary $L_{acc}-L_{line}$ scaling relations (gold stars).  The black circles indicate the stars, brown dwarfs, and planetary mass companions (PMC; filled circles) from CASPAR \citep{Betti2023}.  The black line is the $\dot{M} \propto M^{2.02}$ relation for all objects in CASPAR, while the gold lines are the $\dot{M}-M$ relationships for stars (dash-dot line) and brown dwarfs (dash line) at 1 Myr from CASPAR \citep[eqns 10-11 from ][]{Betti2023}.}
    \label{fig:MMdot}
\end{figure*}

In Fig.~\ref{fig:MMdot}, we show the mass accretion rate as a function of mass for the six accretors compared to accretion rates from the Comprehensive Archive of Substellar and Planetary Accretion Rates \cite[CASPAR;][]{Betti2023}. We find that the ONC objects are strongly accreting compared to similar mass objects. However, we find that their accretion rates are consistent with predicted accretion rates at 1 Myr taking into account by mass and age \citep[gold dashed line in Fig.~\ref{fig:MMdot};][]{Betti2023}, indicating these systems are depleting their disks quickly at young ages.  


\begin{deluxetable*}{lccccc}
\tablecaption{Line and accretion characteristics for ONC targets.\label{table:accretion}}
\tablewidth{0pt}
\tablehead{\colhead{Target} & \colhead{Line} & \colhead{$F_{line}$} & \colhead{$L_{line}$} & \colhead{$\log(L_{acc})$} & \colhead{$\log(\dot M)$} \\
\colhead{} & \colhead{} & \colhead{($10^{-20}$ W m$^{-2}$)} & \colhead{($10^{-8}\ L_\odot$)} & \colhead{($L_\odot$)} & \colhead{($M_\odot$ yr$^{-1}$)}
}
\startdata
262	&	Pa-$\beta$	&	5.84	$\pm$	1.92	&	3.16	$\pm$	1.02	&	-4.17	$\pm$	0.60	&	-10.08	$\pm$	0.61	\\
	&		        &				            &				            &	-3.43	$\pm$	0.33	&	-9.35	$\pm$	0.34	\\
473	&Bra-$\gamma$	&	5.84	$\pm$	1.79	&	3.08	$\pm$	0.95	&	-3.74	$\pm$	0.85	&	-9.69	$\pm$	0.86	\\
	&		        &				            &		             		&	-2.71	$\pm$	0.32	&	-8.66	$\pm$	0.33	\\
3155&	Pa-$\beta$	&	9.38	$\pm$	2.87	&	5.00	$\pm$	1.53	&	-3.94	$\pm$	0.58	&	-9.99	$\pm$	0.59	\\
	&		        &           				&			            	&	-3.25	$\pm$	0.33	&	-9.31	$\pm$	0.34	\\
3251&	Pa-$\beta$	&	9.59	$\pm$	2.15	&	5.07	$\pm$	1.15	&	-3.92	$\pm$	0.57	&	-9.98	$\pm$	0.58	\\
	&	        	&		             		&		            		&	-3.23	$\pm$	0.31	&	-9.29	$\pm$	0.32	\\
3253&	Pa-$\beta$	&	15.70	$\pm$	4.96	&	8.49	$\pm$	2.65	&	-3.70	$\pm$	0.57	&	-9.82	$\pm$	0.58	\\
	&		        &				            &			            	&		--	            	&	--	            		\\
469	&	Pa-$\beta$	&	167.00	$\pm$	20.50	&	89.30	$\pm$	10.9	&	-2.60	$\pm$	0.49	&	-8.97	$\pm$	0.49	\\
	&		        &           				&			            	&		--	            	& --		        		\\
\hline
\enddata
\tablecomments{$\log(L_{acc})$ and $\log(\dot M)$: first line computed using $L_{acc}-L_{line}$ scaling relations from \citet{Alcala2017}; second line computed using $L_{acc}-L_{line}$ scaling relations from \citet{Aoyama2021}. }
\end{deluxetable*}

\section{The Spectral Evolution of a 7--10~$\mathrm{M_{Jup}}$ Planetary-Mass Object across time}\label{sec:evol_planetary_mass_objects}

{Immediately following their formation, planetary-mass objects and brown dwarfs start cooling down \citep{Burrows}}. Thus, {as they age}, these objects change spectral types and surface gravities. The spectra of very young planetary-mass objects members of the ONC presented in this paper allow us to illustrate, for the first time, the spectral evolution with time of a planetary-mass object with masses between 7--10~$\mathrm{M_{Jup}}$, since their very young ages ($1-3$~Myr), until they reach maturity ($\sim$200~Myr). In this Section, we aim to analyze how the decrease in effective temperature, and increase in surface gravity with time affects the spectra of planetary-mass objects. 

\begin{figure}[h]
    \centering
    \includegraphics[width=0.49\textwidth]{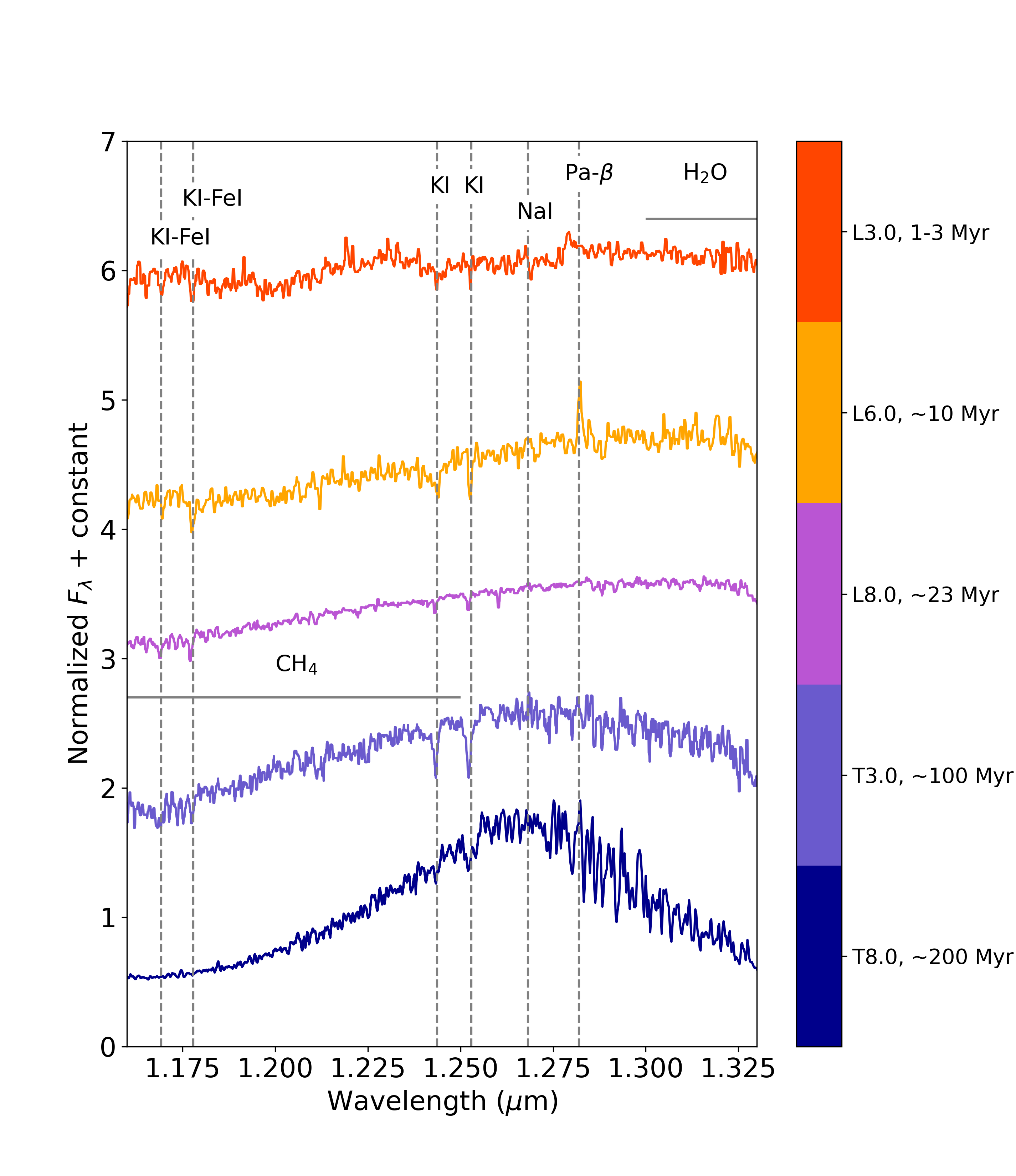}
   \caption{$J$-band spectra of the L3-T8 planetary-mass objects with masses between 7 and 10~$\mathrm{M_{Jup}}$. In red is the spectrum of target 473 of our sample (1--3~Myr old). In orange is the JWST/NIRSpec spectrum of the L6 spectral type planet, 2M1207b ($\sim$10~Myr old). In purple we show the VLT/Xshooter spectrum of the L8 free-floating brown dwarf PSO-318--22 ($\sim$23~Myr). In blue we show the VLT/Xshooter spectrum of GU~Psc~b (T3.0, $\sim$100~Myr, and in dark blue we show the JWST/NIRSpec spectrum of Ross~458c (T8.0, $\sim$200~Myr).}
    \label{fig:mass_sequence_J-band}
\end{figure}

\begin{figure}[h]
    \centering
    \includegraphics[width=0.49\textwidth]{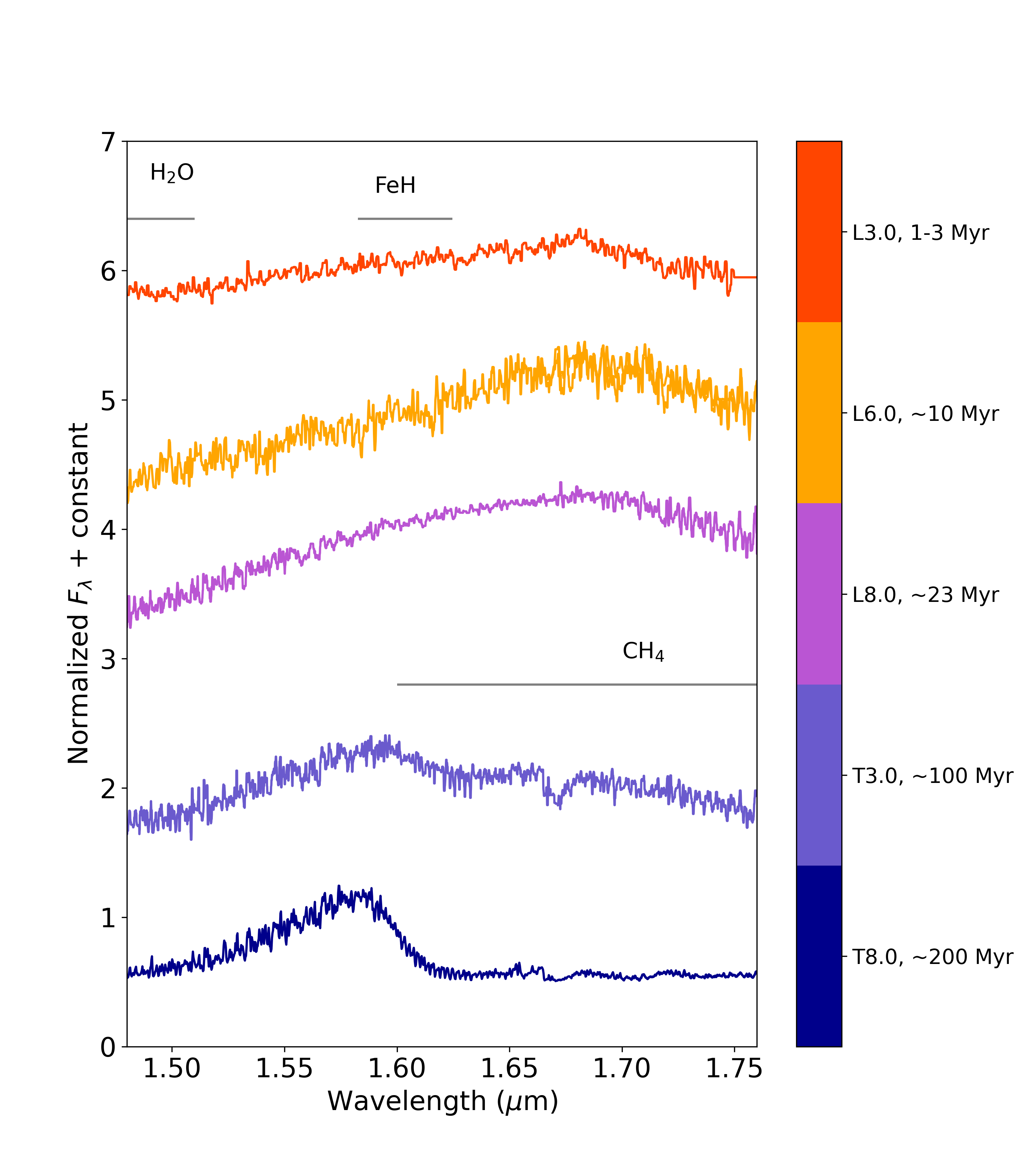}
   \caption{$H$-band spectra of the L3-T8 planetary-mass objects with masses between 7 and 10~$\mathrm{M_{Jup}}$. In red is the spectrum of target 473 of our sample (1--3~Myr old). In orange is the JWST/NIRSpec spectrum of the L6 spectral type planet, 2M1207b ($\sim$10~Myr old). In purple we show the VLT/Xshooter spectrum of the L8 free-floating brown dwarf PSO-318--22 ($\sim$23~Myr). In blue we show the VLT/Xshooter spectrum of GU~Psc~b (T3.0, $\sim$100~Myr, and in dark blue we show the JWST/NIRSpec spectrum of Ross~458c (T8.0, $\sim$200~Myr).}
    \label{fig:mass_sequence_H-band}
\end{figure}

\begin{figure}[h]
    \centering
    \includegraphics[width=0.49\textwidth]{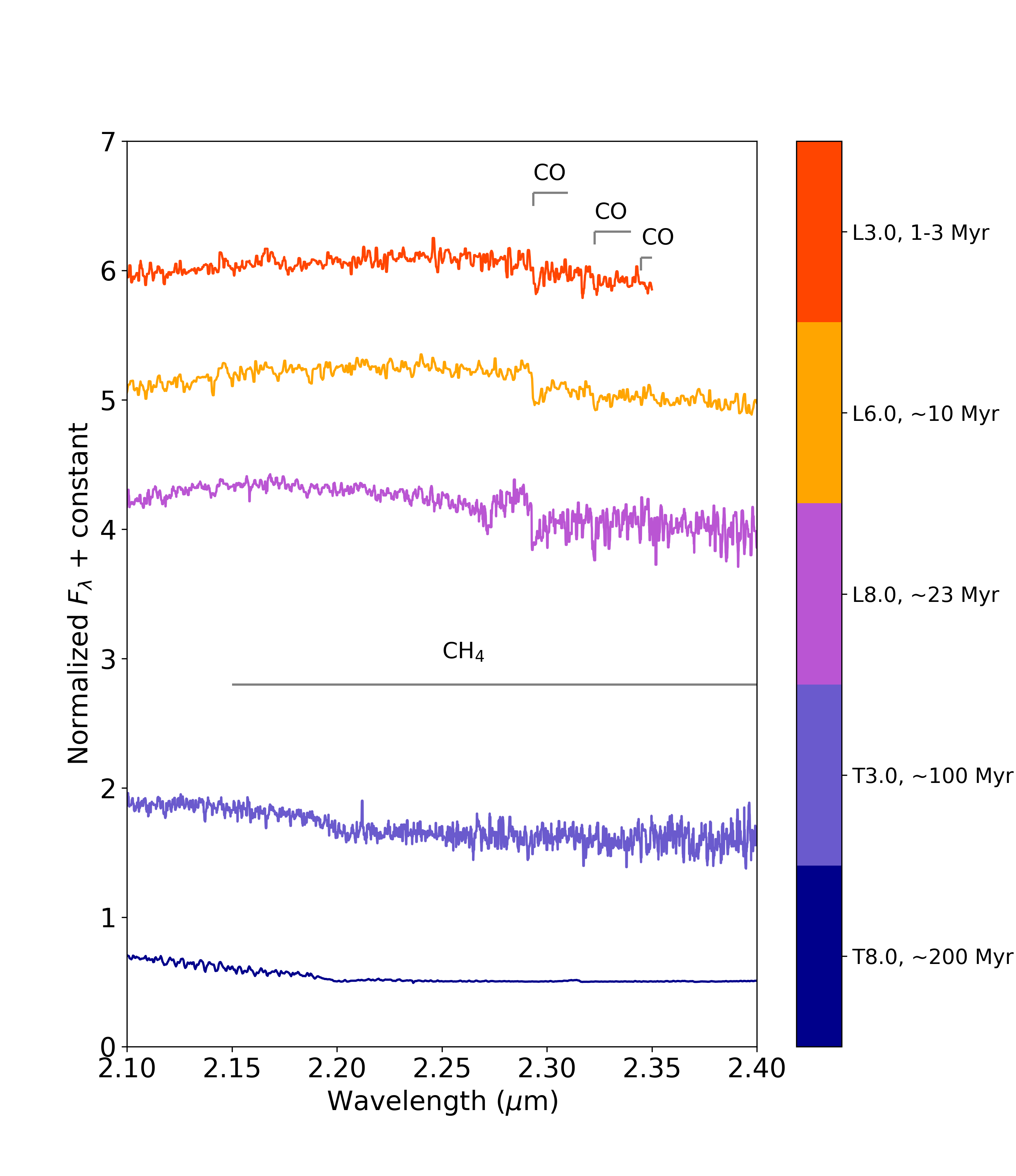}
   \caption{{$K$-band} spectra of the L3-T8 planetary-mass objects with masses between 7 and 10~$\mathrm{M_{Jup}}$. In red is the spectrum of target 473 of our sample (1--3~Myr old). In orange is the JWST/NIRSpec spectrum of the L6 spectral type planet, 2M1207b ($\sim$10~Myr old). In purple we show the VLT/Xshooter spectrum of the L8 free-floating brown dwarf PSO-318--22 ($\sim$23~Myr). In blue we show the VLT/Xshooter spectrum of GU~Psc~b (T3.0, $\sim$100~Myr, and in dark blue we show the JWST/NIRSpec spectrum of Ross~458c (T8.0, $\sim$200~Myr).}
    \label{fig:mass_sequence_K-band}
\end{figure}

We collected the $J$-, $H$-, and $K$-band medium-resolution spectra of five 7--10~$\mathrm{M_{Jup}}$ planetary-mass objects obtained with different instruments and telescopes (see Figs. \ref{fig:mass_sequence_J-band}, \ref{fig:mass_sequence_H-band} and \ref{fig:mass_sequence_K-band}). The youngest object is our target 473 (red spectra in Figs. \ref{fig:mass_sequence_J-band}, \ref{fig:mass_sequence_H-band} and \ref{fig:mass_sequence_K-band}), with an L3 spectral type and a member of the ONC (1--3~Myr). The second youngest object is TWA~27B or 2MASS~J1207334-393254~b (2M1207b). 2M1207b (orange spectra in Fig. \ref{fig:mass_sequence_J-band}, \ref{fig:mass_sequence_H-band} and \ref{fig:mass_sequence_K-band}) is the first directly imaged exoplanet with a dynamical mass of 5$\pm$2~$\mathrm{M_{Jup}}$ \citep{Chauvin2005}, a L6 spectral type, and is a member of the TWA association ($\sim$10~Myr, \citealt{Mamajek2005}). Here we collected the R$\sim$2700 spectrum published in \cite{Manjavacas2024} and \cite{luhman2023} obtained with the NIRSpec instrument onboard the \textit{James Webb Space Telescope}. The third oldest object we present is PSO-318-22 \citep{Liu2013}. PSO-318-22  (purple spectra in Fig. \ref{fig:mass_sequence_J-band}, \ref{fig:mass_sequence_H-band} and \ref{fig:mass_sequence_K-band}) is a $\mathrm{6.5^{+1.3}_{-1.0}}$~$\mathrm{M_{Jup}}$ free-floating planetary-mass object with an L8 spectral type. PSO-318-22 is a member of the $\beta$-Pictoris young-moving group ($\sim$23~Myr, \citealt{Mamajek2014}). Here we show the near-infrared spectrum published by \cite{Piscarreta2024} from the X-shooter spectrograph on the Very Large Telescope (VLT) with a resolution of $\sim$4000. 
The next older object is GU~Psc~b (blue spectra in Fig. \ref{fig:mass_sequence_J-band}, \ref{fig:mass_sequence_H-band} and \ref{fig:mass_sequence_K-band}), a T3 companion to a M3 star with an estimated age of $\sim$100~Myr \citep{Naud2014}. Here we use the VLT/Xshooter spectrum from \cite{Piscarreta2024}. Finally, the oldest object is Ross~458~c \citep{Goldman2010}, a T8 companion to a M0.5/M7.0 binary system, a member of the Carina Near Moving Group with an estimated age of $\sim$200~Myr (dark blue spectra in Fig. \ref{fig:mass_sequence_J-band}, \ref{fig:mass_sequence_H-band} and \ref{fig:mass_sequence_K-band}). Here we use the JWST/NIRSpec medium-resolution spectrum obtained in GTO~1292 (P.I. Lunine) available in the MAST Archive {(DOI: 10.17909/jszb-bm84)}. 

In Fig. \ref{fig:mass_sequence_J-band} through  \ref{fig:mass_sequence_K-band} we show the evolution of spectral characteristics of 7--10~$\mathrm{M_{Jup}}$ non-irradiated planets and planetary-mass objects as they change spectral types due to the decrease of their effective temperature as their heat is dissipated with time. A 7--10~$\mathrm{M_{Jup}}$ planetary-mass object that is born with a L3 spectral type ($\mathrm{T_{eff}}$$\sim$2100~K, red spectrum), like our target 473 (1--3~Myr), shows the typical characteristics of a young L3 brown dwarf, as discussed in Section~\ref{sec:spectral_indices}: weak K\,I and Na\,I lines in the $J$-band, and they might show emission lines, such as Pa-$\beta$ and Bra-$\gamma$ if they have protoplanetary disks, which are expected until they are at least $\sim$10~Myr, like objects 473 and 262. The $H$-band shows a triangular shape, since the FeH molecule absorption is depleted, as is known for very-young brown dwarfs. 

As the planetary-mass object cools down, by $\sim$10~Myr it has reached a L6 spectral type (2M1207b, orange spectrum). We observe that the K\,I alkali lines are deeper than the 1--3~Myr L3 object, even though the equivalent width of these lines decreases with spectral type up to the start of the L/T transition (L8-L9). At this age, planetary-mass objects can still show emission lines due to the accretion of material from a protoplanetary disk, since substellar objects seem to retain their disks longer than stars \citep{Luhman_Mamajek2012, Rilinger2021}, as is the case of 2M1207b. The $H$-band is still triangular.

At 23~Myr old, a 7--10~$\mathrm{M_{Jup}}$ planetary-mass object has reached a L8 spectral type, like PSO 318-22 (purple spectrum), and is at the start of the L/T transition. The K\,I alkali lines in the $J$-band are depleted due to a combination of low surface gravity and the condensation of the silicate clouds \citep{McLean}. The $H$-band is not as triangular as for 2M1207b.


At $\sim$100~Myr old, a 7--10~$\mathrm{M_{Jup}}$ planetary-mass object has passed the L/T transition and reached a T3 spectral type, and would be similar to GU~Psc~b (blue spectrum). The most remarkable feature, as expected, is the appearance of the $\mathrm{CH_{4}}$ molecular bands in the $J$-, $H$- and $K$-bands and the disappearance of the CO-band in the $K$-band. At $\sim$100~Myr old, brown dwarfs have reached a surface gravity similar to that of field brown dwarfs \citep{Baraffe, Manjavacas2020, Piscarreta2024}, thus, their alkali lines are not depleted with respect to other field T3 dwarfs.

Finally, at $\sim$200~Myr, a 7--10~$\mathrm{M_{Jup}}$ planetary-mass object would have a T8 spectral type, similar to Ross~458c (dark blue spectrum), showing deep $\mathrm{CH_{4}}$ bands in the $H$- and $K$-bands. The K\,I line at 1.25~$\mu$m is still present, but the rest of the alkali lines will not be present any more, due to the deep water and $\mathrm{CH_{4}}$ bands masking these features.

\section{Discussion}\label{sec:discussion}

The near-infrared medium-resolution spectra of brown dwarfs and planetary-mass ONC members presented in this paper allow us to characterize the early stages (1--3~Myr) of their lives for the first time at such high resolution and wide wavelength range. We obtained spectra of objects with masses between 7--76~$\mathrm{M_{Jup}}$, with effective temperatures between 2100--2800~K \citep{Robberto2020} {estimated using photometry and isochrone fitting}. 

In Section \ref{sec:spectral_typing} we obtained spectral types spanning L3 and M6.0 for all targets using young brown dwarf and planetary-mass object spectral templates previously published in the literature \citep{Bonnefoy2013, Allers2013, Manjavacas2016, Manjavacas2020}. However, as observed in Table \ref{table:spectral_types}, columns 2 and 3, the spectral types of the targets do not necessarily correspond to the same effective temperature predicted by \cite{Robberto2020}. As observed in column 3, while the predicted effective temperature by \cite{Robberto2020} decreases monotonically with mass (column 2), the spectral types found in Section \ref{sec:spectral_typing} do not monotonically transition to later spectral types. This could be due to several factors or a combination of several of them: first, as mentioned above, the effective temperature and masses estimated by \cite{Robberto2020} are model-based. As they pointed out in their paper, these parameters were estimated using the BT-Settl atmospheric models that struggled to reproduce the part of the color-magnitude diagram where most of these objects are located, in particular brown dwarfs and planetary-mass objects with masses below 40~$\mathrm{M_{Jup}}$. To fix this discrepancy, they adjusted the models to fit that part of the color-magnitude diagram (see \citealt{Robberto2020}, their Section 4.9). Thus, the estimated masses and effective temperatures might be over- or under-estimated. {In addition, some of these objects might have accretion disks and differential reddening, which would change the position of the targets in the color-magnitude diagram and the unique use of evolutionary models to estimate effective temperatures and masses would provide inaccurate results, as explained by \cite{Gennaro2020}}. The targets that show the highest discrepancy between spectral type and effective temperatures are: targets 262, 442, 3307, 833, 1376, and 255.  Targets 262, 442, and 3307 have earlier spectral types than expected for their predicted effective temperatures, and targets 833, 1376, and 255 have later spectral types than expected. 

The other possibility is that the spectral types of the spectral templates we used to estimate the spectral types of our targets are inaccurate. Spectral typing young brown dwarfs and planetary-mass objects is usually challenging since {the results may depend} on the wavelength range used (young brown dwarfs tend to show earlier spectral types in the optical). {Additionally, since our targets are very young, some of them might still keep their protoplanetary disks. This could lead to flux excess in the redder wavelength ranges, which seems to influence, at least, targets 3251 and 3311, for which their matches to their best matching template significantly improved when adding a disk contribution.} Also, since the ONC cloud is non-homogeneous, differences in extinction could also modify the spectral type that we find for our targets.

Finally, the other possibility is that the flux calibration of our targets is somewhat inaccurate. Given that we can only use the HST/WFC3 F130M photometry from \cite{Robberto2020}, the relative flux calibration between the MOSFIRE $J$-, $H$- and $K$-bands might not be accurate, which {could} lead to an offset spectral classification. However, we believe that this might not explain in its totality these differences, since the targets that show these differences were observed on different nights, and other targets observed simultaneously do not show that difference in spectral type. Since the flux calibration was performed simultaneously for all the targets observed on the same night, probably the two other possibilities mentioned above are more strongly influencing the spectral type obtained for our targets.

In Section \ref{sec:emission_lines}, we looked for Pa-$\beta$ and Bra-$\gamma$ emission lines in our targets, which indicate the existence of accretion due to a protoplanetary disk \citep{Betti2022}. We found that 5 of our targets showed Pa-$\beta$, and another Bra-$\gamma$ with a 3$\sigma$ detection. We used those lines to measure the mass accretion rate as a function of mass, as explained in Section \ref{sec:emission_lines}. We concluded that the ONC members are strongly accreting compared to similar mass objects, but still consistent with the predicted accretion rate for 1~Myr (see Fig. \ref{fig:accretion_emission_lines}), implying that planetary-mass objects deplete their disks quickly at very young ages.

These 6--7 objects, $\sim$25\% of our sample, show some evidence of protoplanetary disks. This disk fraction is smaller than the fraction found by other surveys in other clusters of similar age, like $\sigma$-Orionis ($\sim$3~Myr) by \cite{Luhman2008b}, where they found a disk ratio of $\sim60$\% for brown dwarfs. However, the MOSFIRE spectra do not cover the wavelength range that would be more affected by the flux of a circumstellar disk, thus, it is likely that some other targets in our sample also have circusmstellar disks, but cold enough that they would not be detectable in the near-infrared.

In Section \ref{sec:spectral_indices}, we use the spectral indices from \cite{Allers2013}, and the pseudo equivalent widths of the K\,I alkali lines present in the $J$-band to confirm the extreme youth of all our targets, and therefore their high probability of being bona fide members of the ONC, and not foreground or background contaminants. { When looking at these indices as a function of spectral type, our sample spans a very distinct track than the one of field brown dwarfs \citep{McLean, Cushing2005} and appears to provide a robust ``young ages'' boundary.}

In Section \ref{sec:evol_planetary_mass_objects} we show the spectral evolution of a 7--10~$\mathrm{M_{Jup}}$ across time, from 1--3~Myr to 200~Myr old, using different planetary-mass objects of similar estimated mass, but different age, namely, target 473, 2M1207b, PSO~318--22, GU~Psc~b and Ross~458c (see Figs. \ref{fig:mass_sequence_J-band}, \ref{fig:mass_sequence_H-band}, and \ref{fig:mass_sequence_K-band}). In this plot, we illustrate how substellar objects of a given mass cool with time, changing spectral types, and how surface gravity increases as the cool down. One caveat is that these five objects did not necessarily form under the same conditions, or even through the same physical mechanism (core collapse or core accretion), thus, if we could actually trace the spectral evolution of a planetary-mass object similar to our target 473 during $\sim$200~Myr, the spectral characteristics would probably not be exactly the same as for 2M1207b at 10~Myr, for PSO~318--22 at 23~Myr, for  GU~Psc~b at 100~Myr old and for Ross~458c at 200~Myr old. In particular at the very young ages, below 10~Myr old, when brown dwarfs or planetary-mass objects are still within the star-forming region where they were born, and  the influence on the environment on their spectra might be substantial.


\section{Conclusions}\label{sec:conclusions}

We presented medium-resolution $J$-, $H$- and $K$-band Keck/MOSFIRE spectra of a sample of 25 brown dwarfs and planetary-mass objects (7--76~$\mathrm{M_{Jup}}$, $\mathrm{T_{eff}}$ = 2100--2800~K) that are members of the Orion Nebula Cluster. In the following, we describe our analysis and main results:

\begin{enumerate}
    \item We estimated the spectral types of our sample using template spectra from other brown dwarfs and planetary-mass objects in the literature. Our targets have spectral types between M6.0 and L3.0, and they do not always correlate with the effective temperature estimated by \cite{Robberto2020} using the BT-Settl  models. {This suggests the need to develop more sophisticated self-consistent models with atmospheric properties matching the bulk evolutionary ones at very young ages}. From our sample, 11 targets show best matches in all $J$-, $H$- and $K$-band with other spectral templates in the literature, and the other 14 show and excess in the $K$-band. This could be due to the contribution of a protoplanetary disk, extinction from the ONC cloud or an imperfect flux calibration in the $K$-band.

    \item We labeled the expected atomic lines and molecular bands expected for M6.0-L3.0 objects. We identified the K\,I doublets and Na\,I atomic lines in the $J$-band for all our targets. For five of them we also identified the Pa-$\beta$ emission line (targets 262, 473, 3155, 3251, and 3253). In the $H$-band, we identified the FeH band that was rather depleted due to the low surface gravity of our targets. In the $K$-band, we identified the CO bandheads in all targets, and also the Bra-$\gamma$ emission line in one of them (target 469).

    \item We measured the mass accretion rate using the Pa-$\beta$ and Bra-$\gamma$ emission lines from the 6 targets mentioned above. We concluded that the accretion masses measured for our targets are of the similar order of more massive stars and brown dwarfs, implying that these systems are depleting their disks quickly at young ages. The mass accretion rate measured for our targets follows the prediction by the CASPAR \citep{Betti2023} for 1~Myr substellar objects.

    \item We measured the gravity spectral indices by \cite{Allers2013} in all our targets, which measures the depth of the gravity-sensitive spectral characteristics (K\,I lines, the H-band shape and the depth of the FeH molecular band). We concluded that all of them show signs of very-low surface gravity, as expected for members of the ONC. Thus, we confirmed that all the targets in our sample are bona-fide members of the cluster.

    \item We compared the near-infrared medium-resolution spectra of five 7--10~$\mathrm{M_{Jup}}$ planetary-mass objects of ages between 1--3~Myr to 200~Myr, namely, our target 473 (L3.0, 1--3~Myr), 2MASS~1207b (L6.0, $\sim$10~Myr), PSO~318--22 (L8.0, $\sim$23~Myr), GU~Psc~b (T3.0, $\sim$100~Myr) and Ross~458~c (T8.0, $\sim$200~Myr). This comparison illustrates how the spectra of a planetary-mass object evolve with time since their early stages until they almost reach their final radius, cooling down and evolving in spectral types, and simultaneously increasing surface gravity.

    \item {The spectra presented in this work are publicly available for the community's use as data behind the figures.}
    
\end{enumerate}


\begin{acknowledgments}

The authors wish to recognize and acknowledge the very significant cultural role and reverence that the summit of Maunakea has always had within the Native Hawaiian community. We are most fortunate to have the opportunity to conduct observations from this mountain.

This research has made use of the Keck Observatory Archive (KOA), which is operated by the W. M. Keck Observatory and the NASA Exoplanet Science Institute (NExScI), under contract with the National Aeronautics and Space Administration.

\end{acknowledgments}

%

\vspace{5mm}
\facilities{W.M. Keck observatory/MOSFIRE}


\software{astropy \citep{2013A&A...558A..33A,2018AJ....156..123A, 2022ApJ...935..167A}}

{This research made use of \ttfamily{PypeIt},\footnote{\url{https://pypeit.readthedocs.io/en/latest/}}
a Python package for semi-automated reduction of astronomical slit-based spectroscopy.}

\bibliography{ONC_BDs}{}
\bibliographystyle{aasjournal}



\appendix

\section{All MOSFIRE spectra of our sample}\label{app:all_spectra}

Here we show all the MOSFIRE spectra that we obtained in our sample. Some of the spectra were found to be contaminants, and we removed from them the final sample as explained in Section \ref{sec:targets}.

\begin{figure*}
    \centering
    \includegraphics[width=0.8\linewidth]{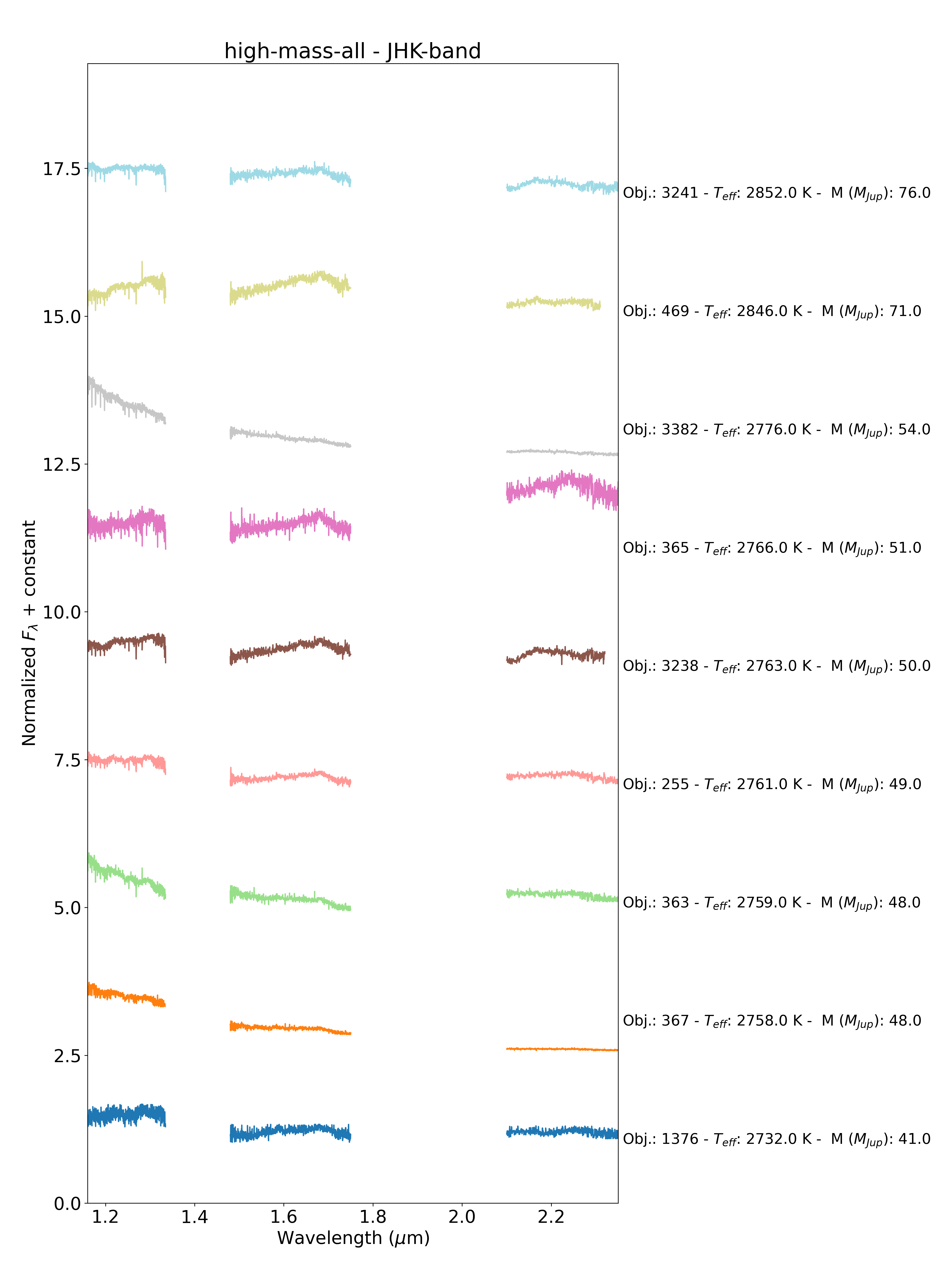}
    \caption{JHK MOSFIRE spectra of all the high-mass brown dwarfs in our sample with masses between 41 and 76~$\mathrm{M_{Jup}}$.}
    \label{fig:JHK-high_mass}
\end{figure*}

\begin{figure*}
    \centering
    \includegraphics[width=0.8\linewidth]{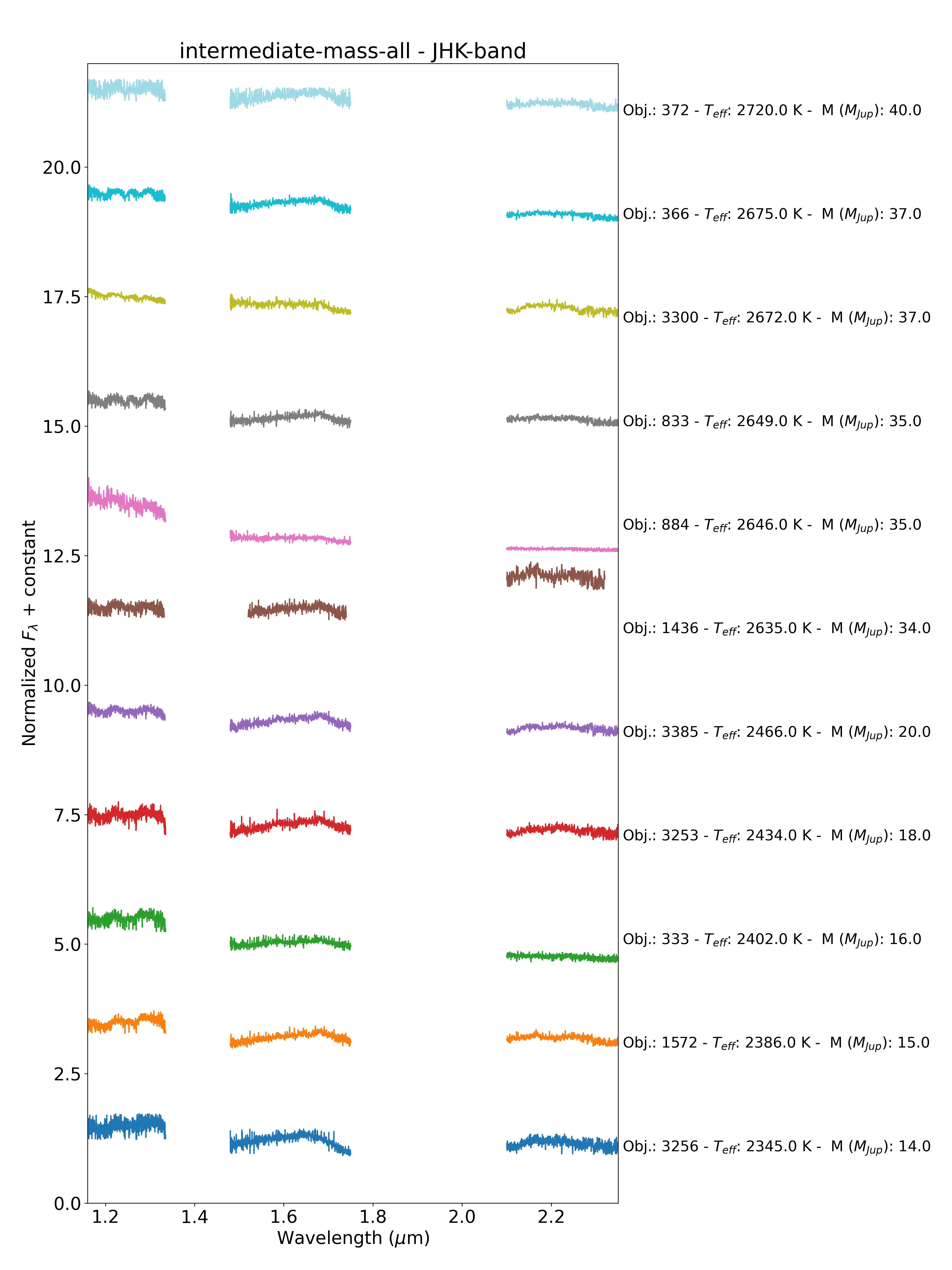}
    \caption{JHK MOSFIRE spectra of all the high-mass brown dwarfs in our sample with masses between 14 and 40~$\mathrm{M_{Jup}}$.}
    \label{fig:JHK-intermediate_mass}
\end{figure*}

\begin{figure*}
    \centering
    \includegraphics[width=0.8\linewidth]{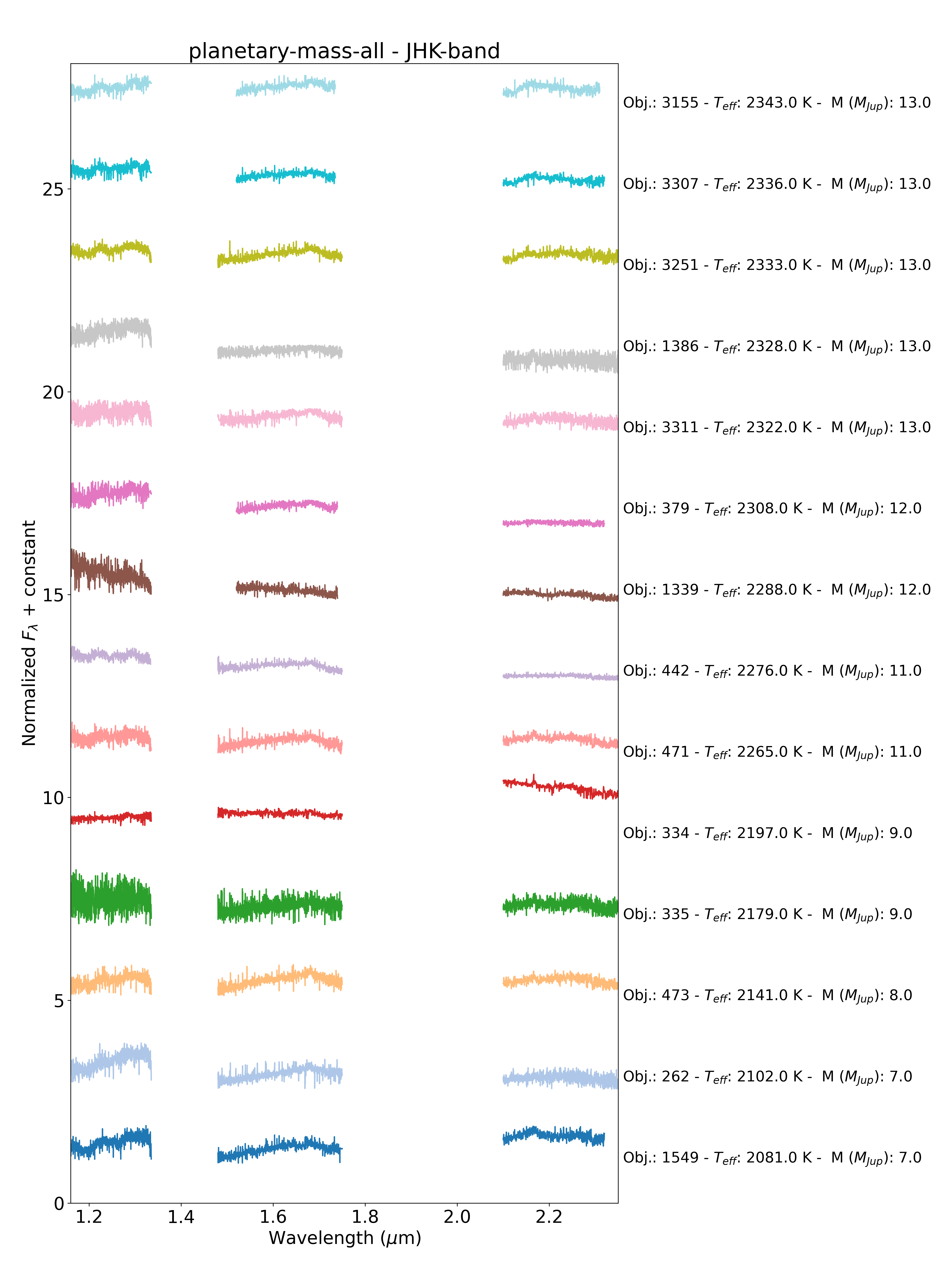}
    \caption{JHK MOSFIRE spectra of all the high-mass brown dwarfs in our sample with masses between 7 and 13~$\mathrm{M_{Jup}}$.}
    \label{fig:JHK-planetary_mass}
\end{figure*}

{In the following plots we show objects 1376, 372, 3256, 1386, 3311, and 379 before we remove the spurious emission features and we show that they are found at wavelengths where skylines are expected.}

\begin{figure*}
    \centering
    \includegraphics[width=0.48\linewidth]{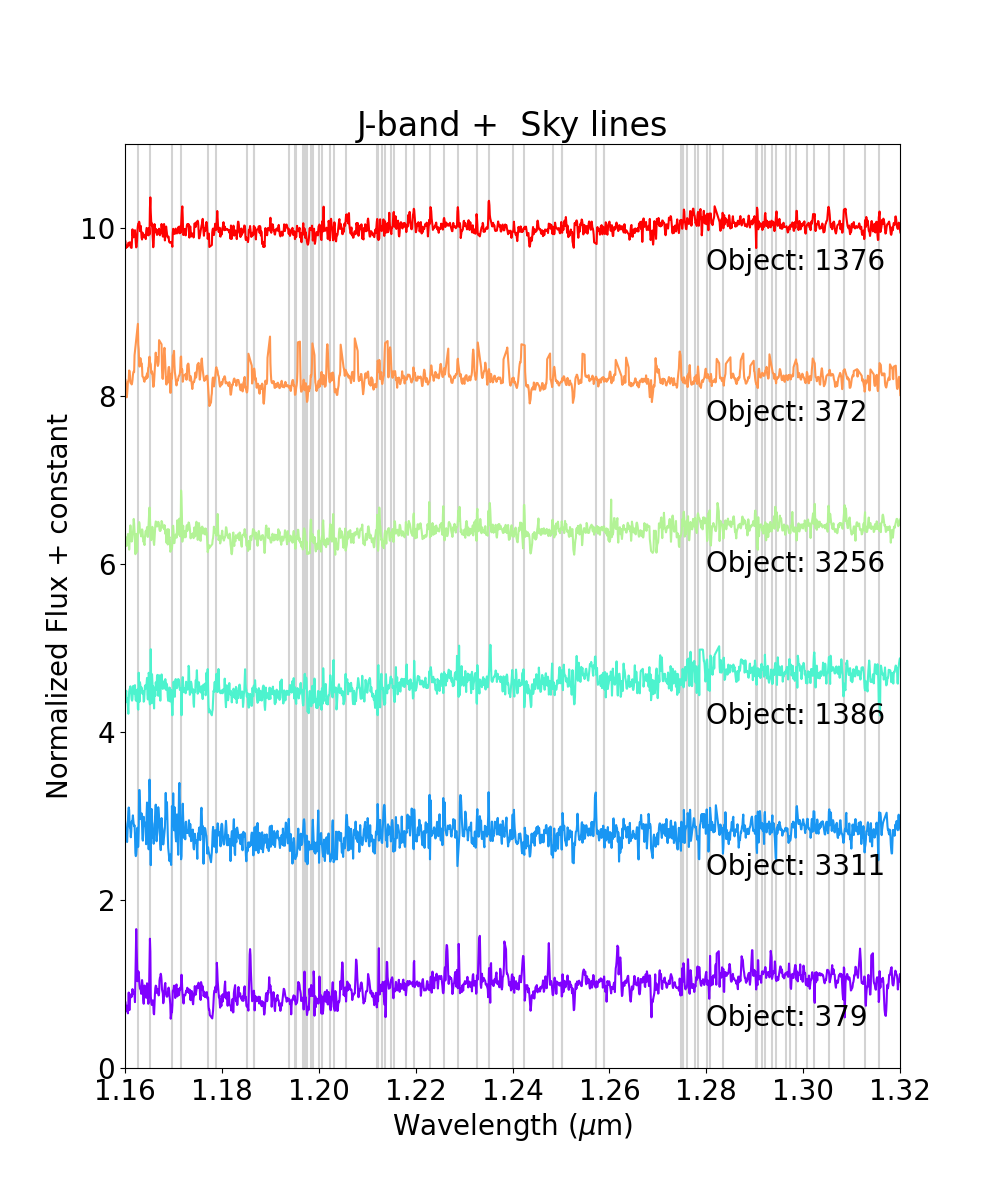}
    \includegraphics[width=0.48\linewidth]{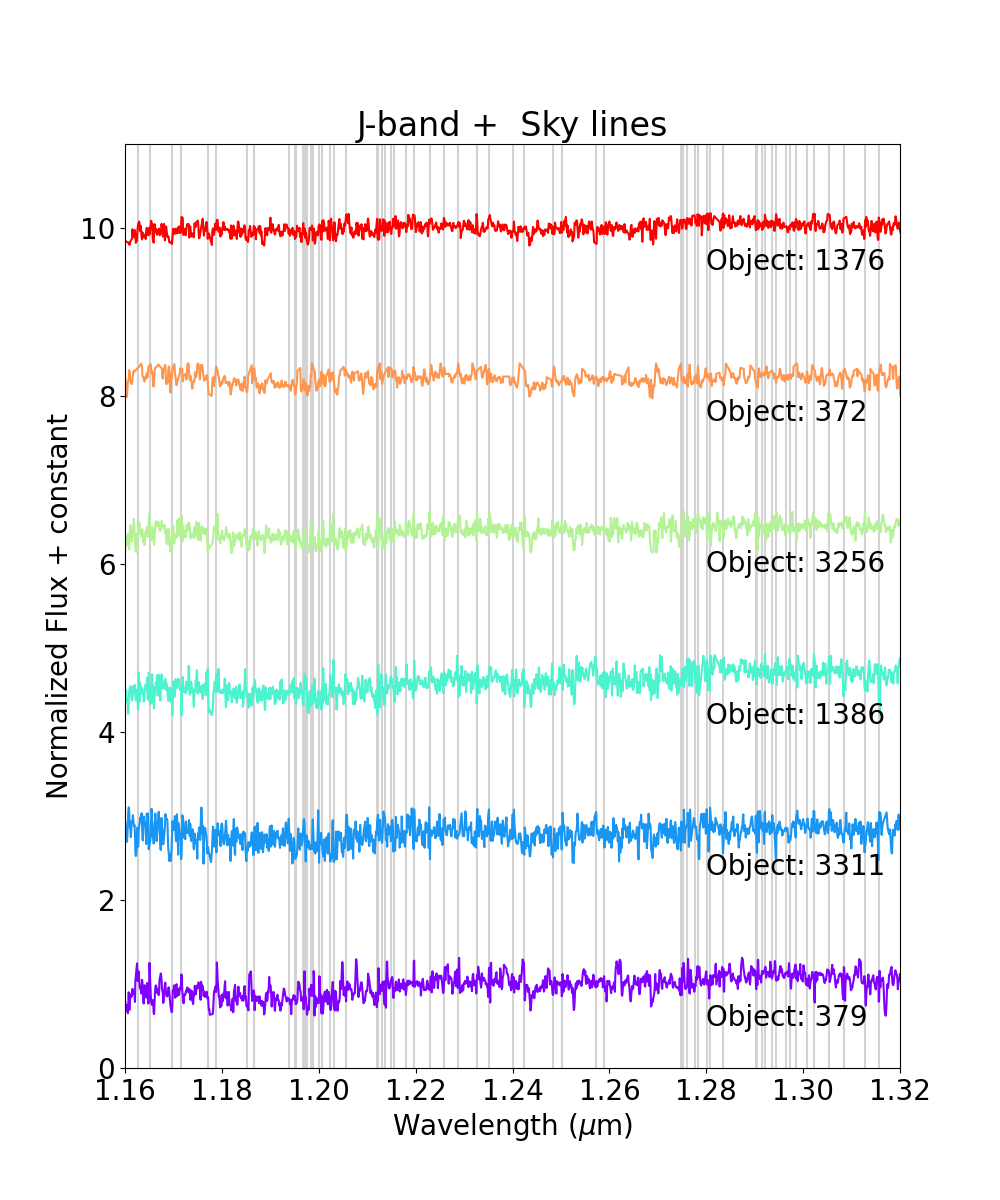}
     
    \caption{$J$-band MOSFIRE spectra of the objects showing spurious emission features before being removed (left-hand side plot), and after (right-hand side plot), and the sky lines expected overlapped to show that some of those might be introduced by an incomplete sky line removal in the data reduction process.}
    \label{fig:spectra_with_skylinesJ}
\end{figure*}

\begin{figure*}
    \centering
    \includegraphics[width=0.48\linewidth]{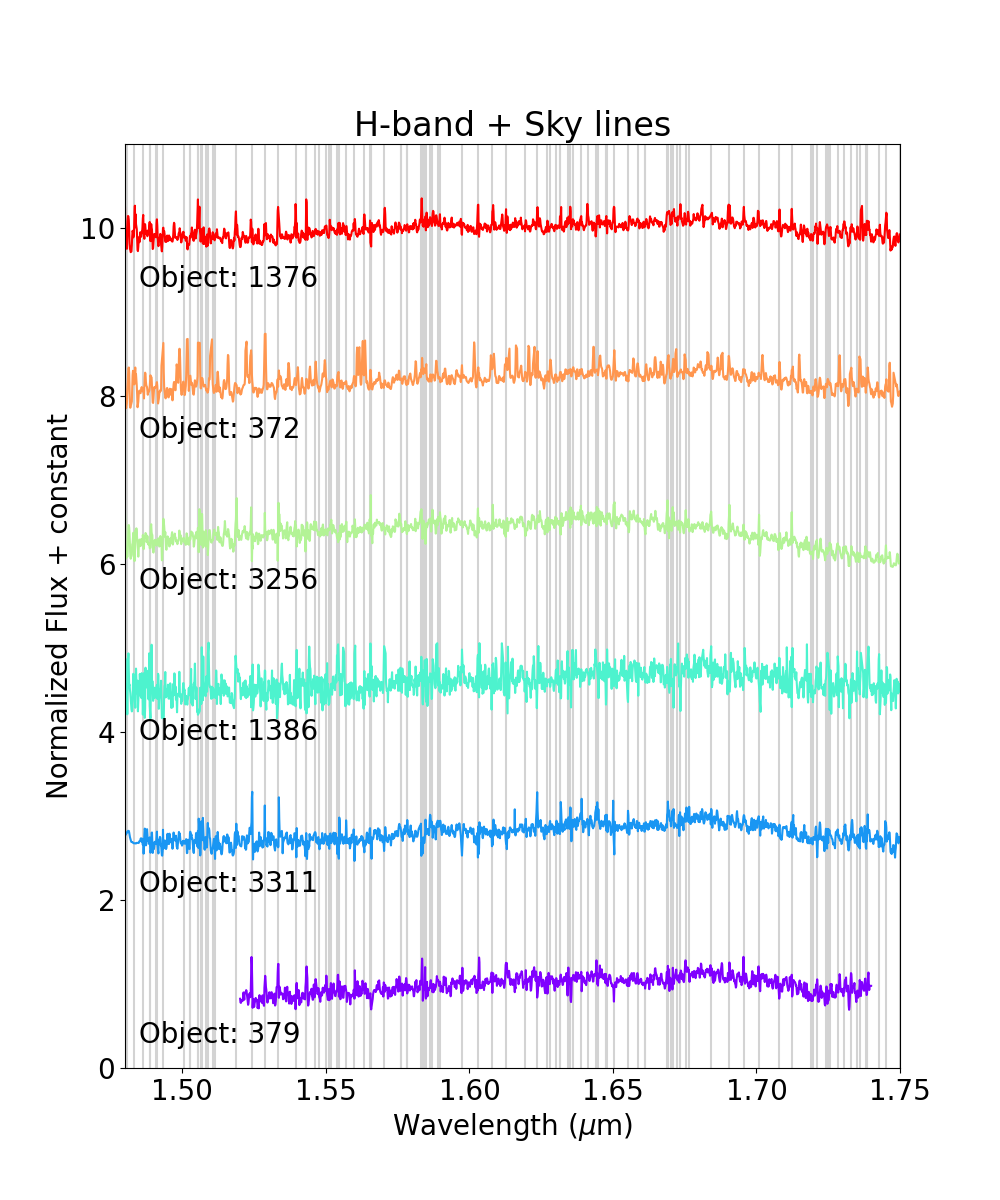}
    \includegraphics[width=0.48\linewidth]{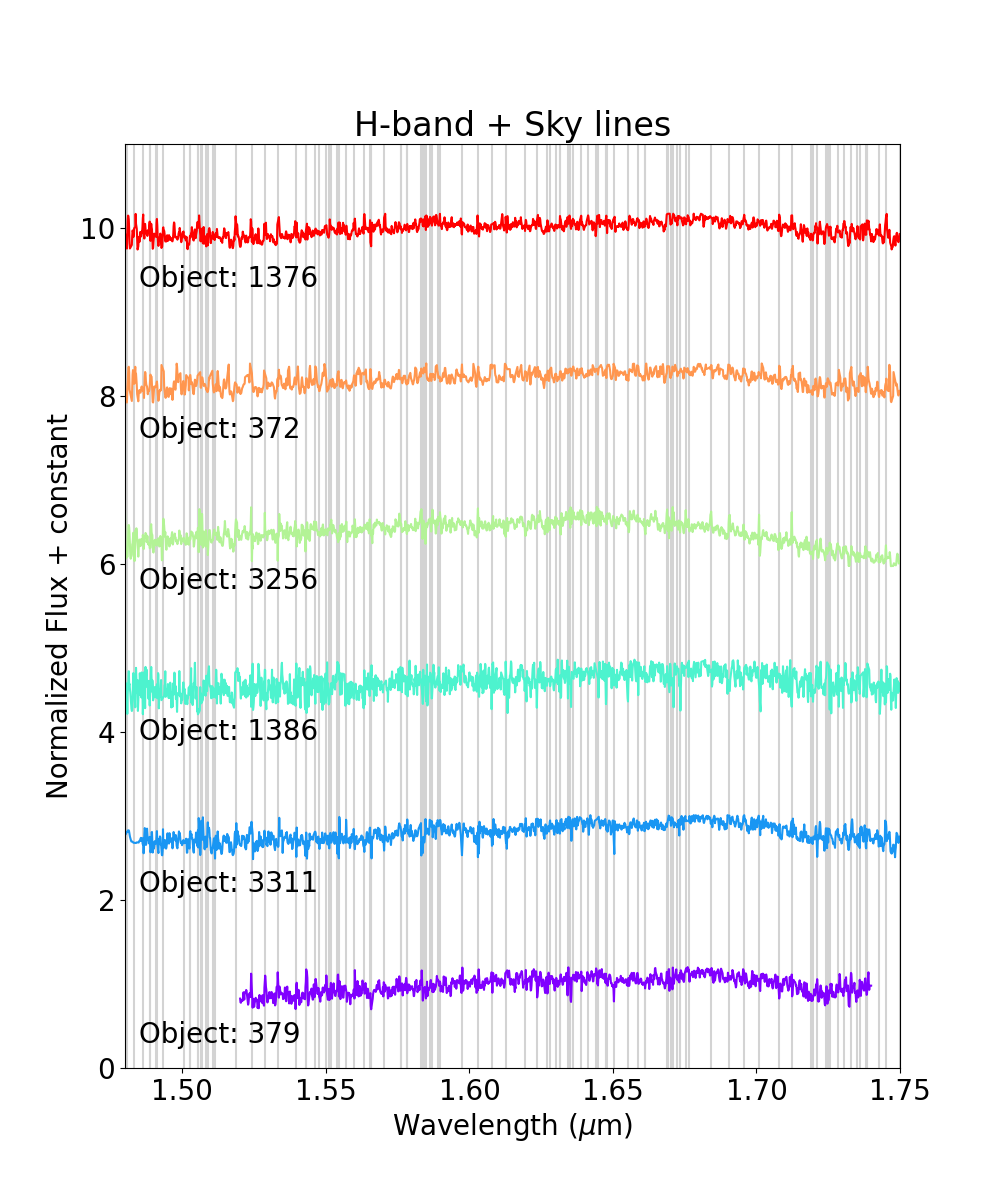}
    \caption{$H$-band MOSFIRE spectra of the objects showing spurious emission features before being removed (left-hand side plot), and after (right-hand side plot), and the sky lines expected overlapped to show that some of those might be introduced by an incomplete sky line removal in the data reduction process.}
    \label{fig:spectra_with_skylinesH}
\end{figure*}

\begin{figure*}
    \centering
    \includegraphics[width=0.48\linewidth]{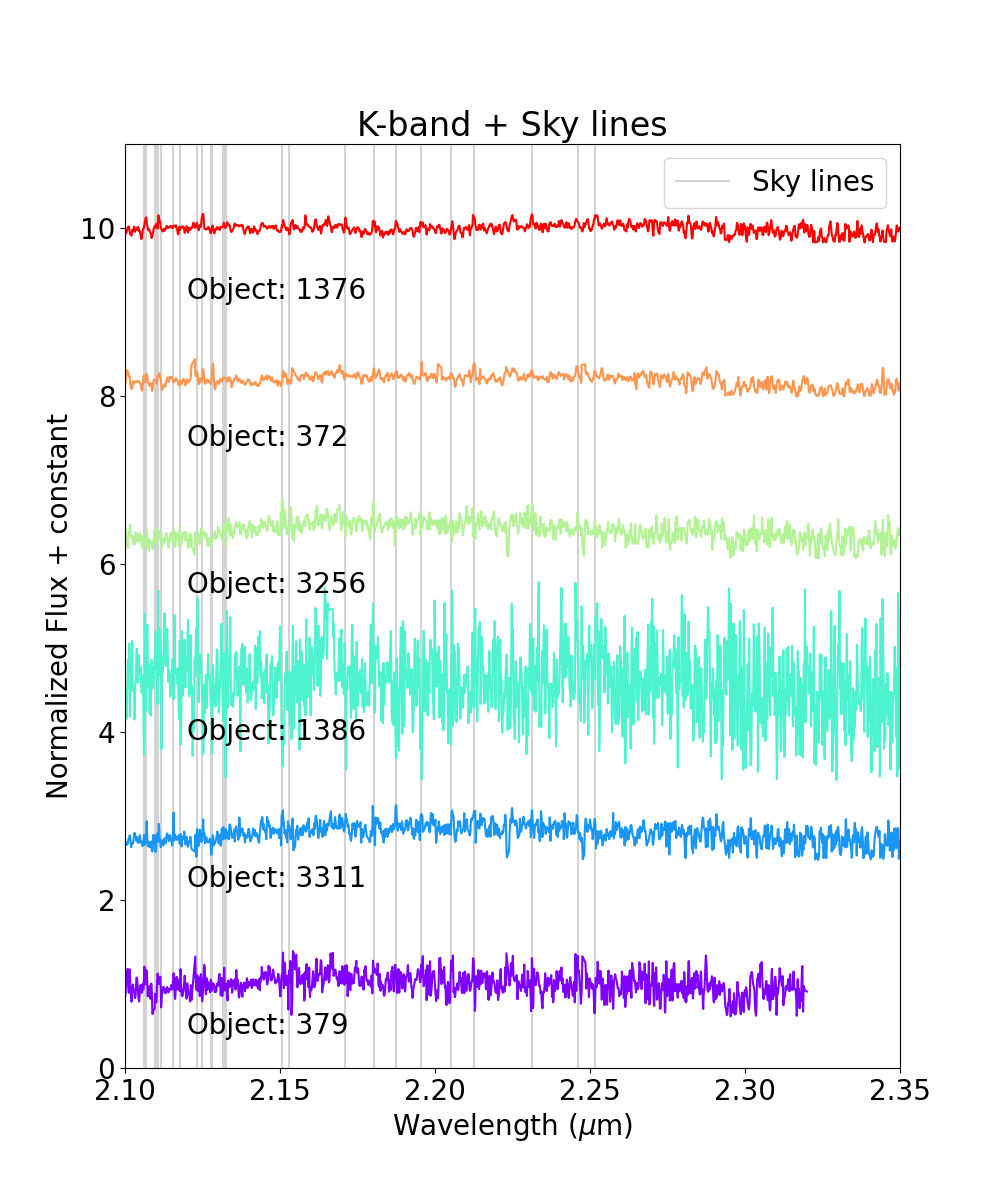}
    \includegraphics[width=0.48\linewidth]{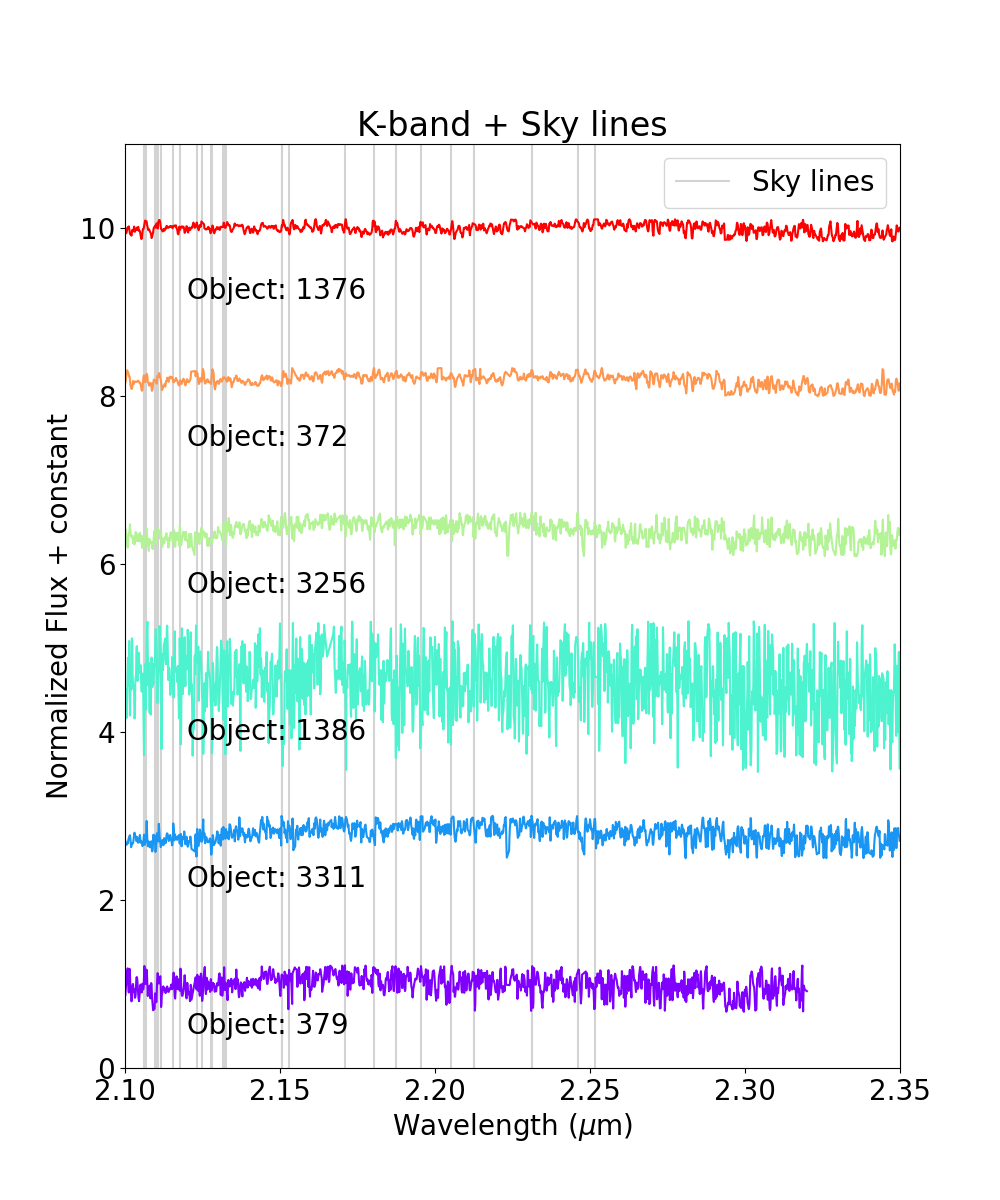}
    \caption{$K$-band MOSFIRE spectra of the objects showing spurious emission features before being removed (left-hand side plot), and after (right-hand side plot), and the sky lines expected overlapped to show that some of those might be introduced by an incomplete sky line removal in the data reduction process.}
    \label{fig:spectra_with_skylinesK}
\end{figure*}

\section{Best matches to other template spectra}\label{app:best_matches}

In this Section we show all the best matches of our MOSFIRE brown dwarf spectra with other template spectra from the literature, as described in Section \ref{sec:spectral_typing}.

\begin{figure*}
    \centering
    \includegraphics[width=0.49\linewidth]{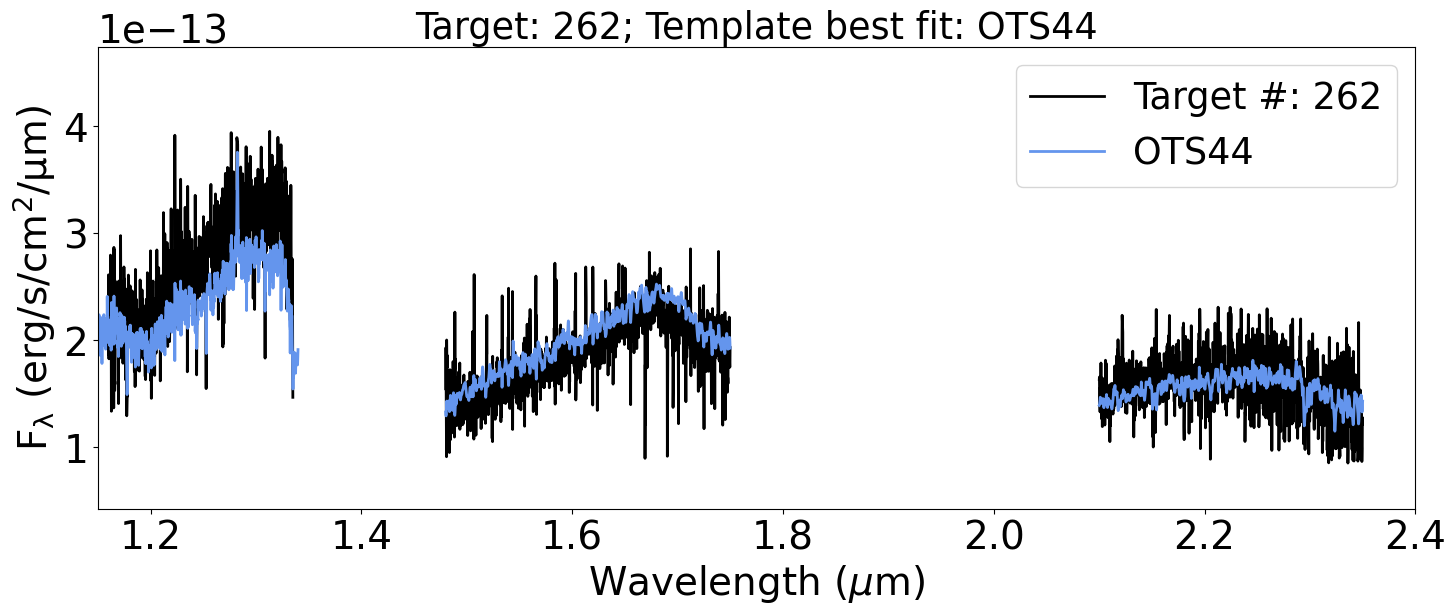}
    \includegraphics[width=0.49\linewidth]{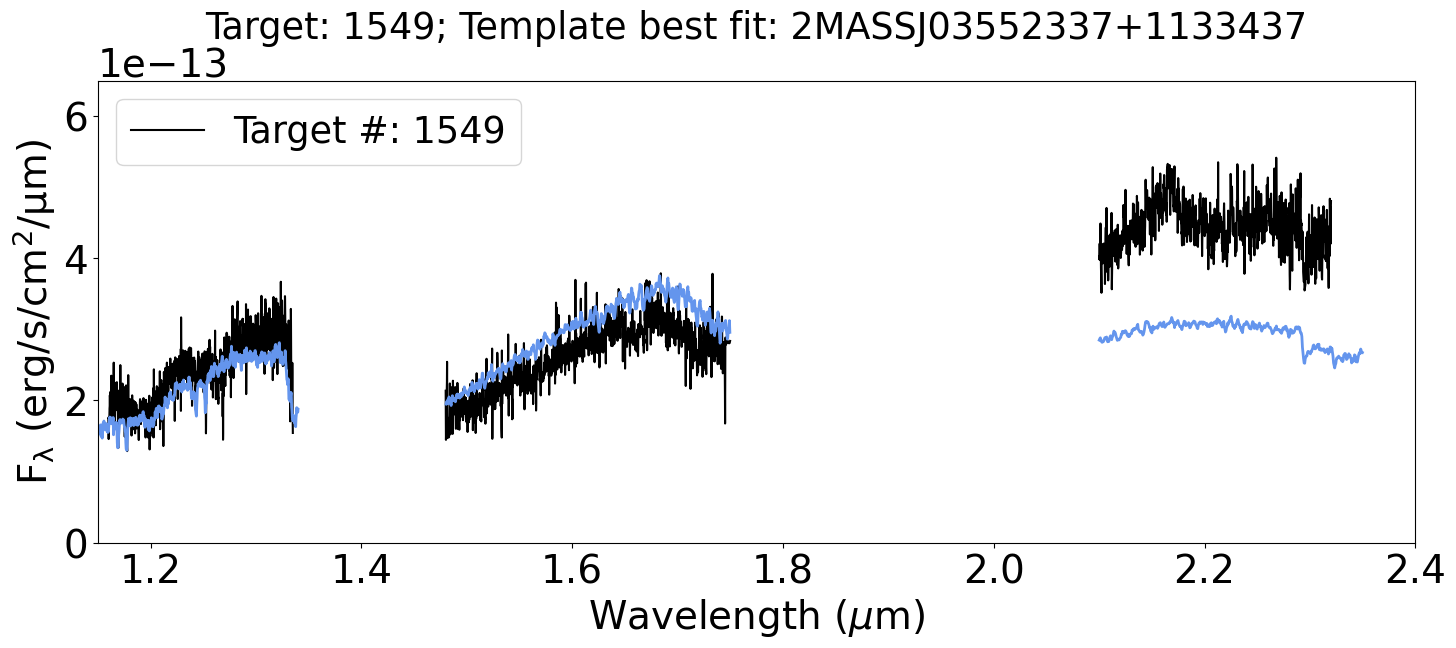}
    \includegraphics[width=0.49\linewidth]{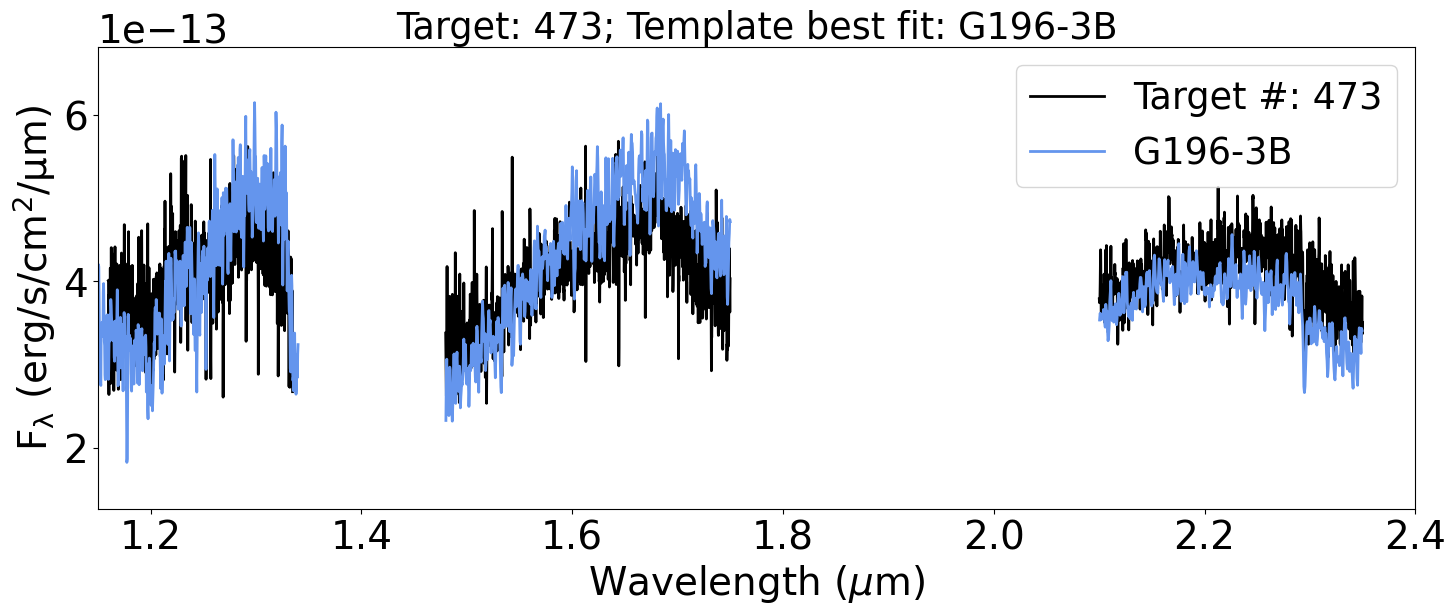}
    \includegraphics[width=0.49\linewidth]{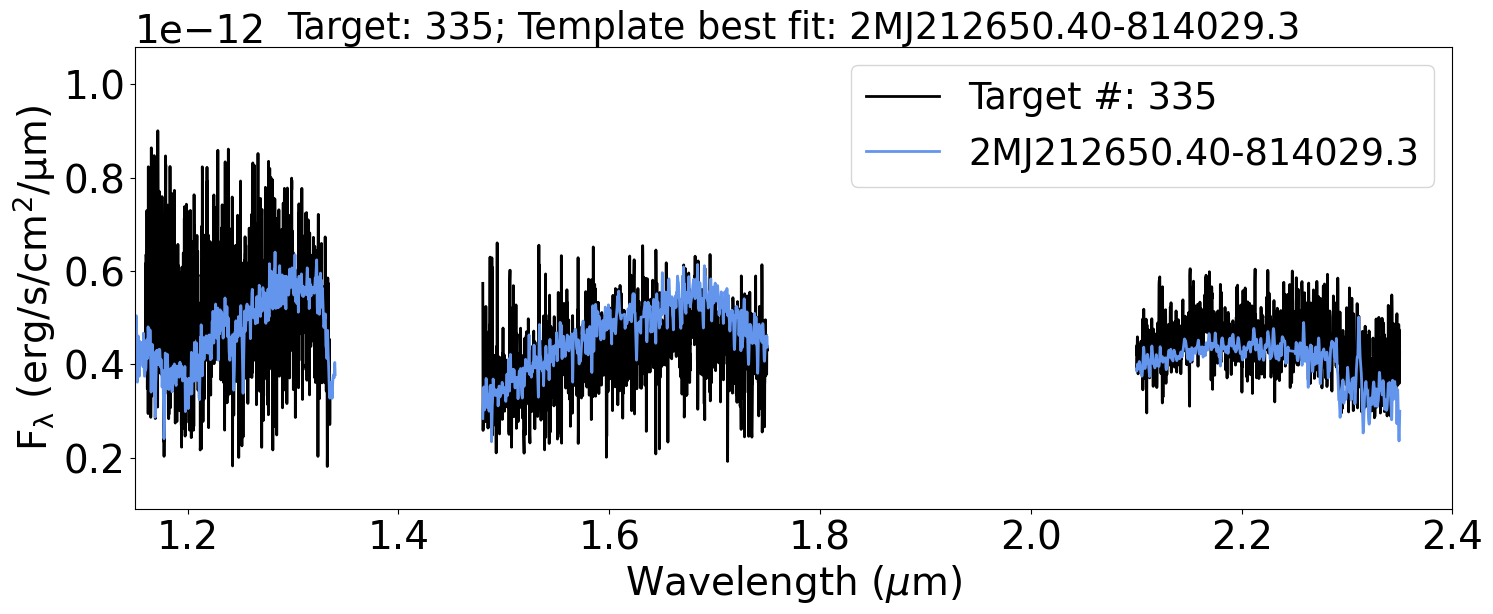}
    \includegraphics[width=0.49\linewidth]{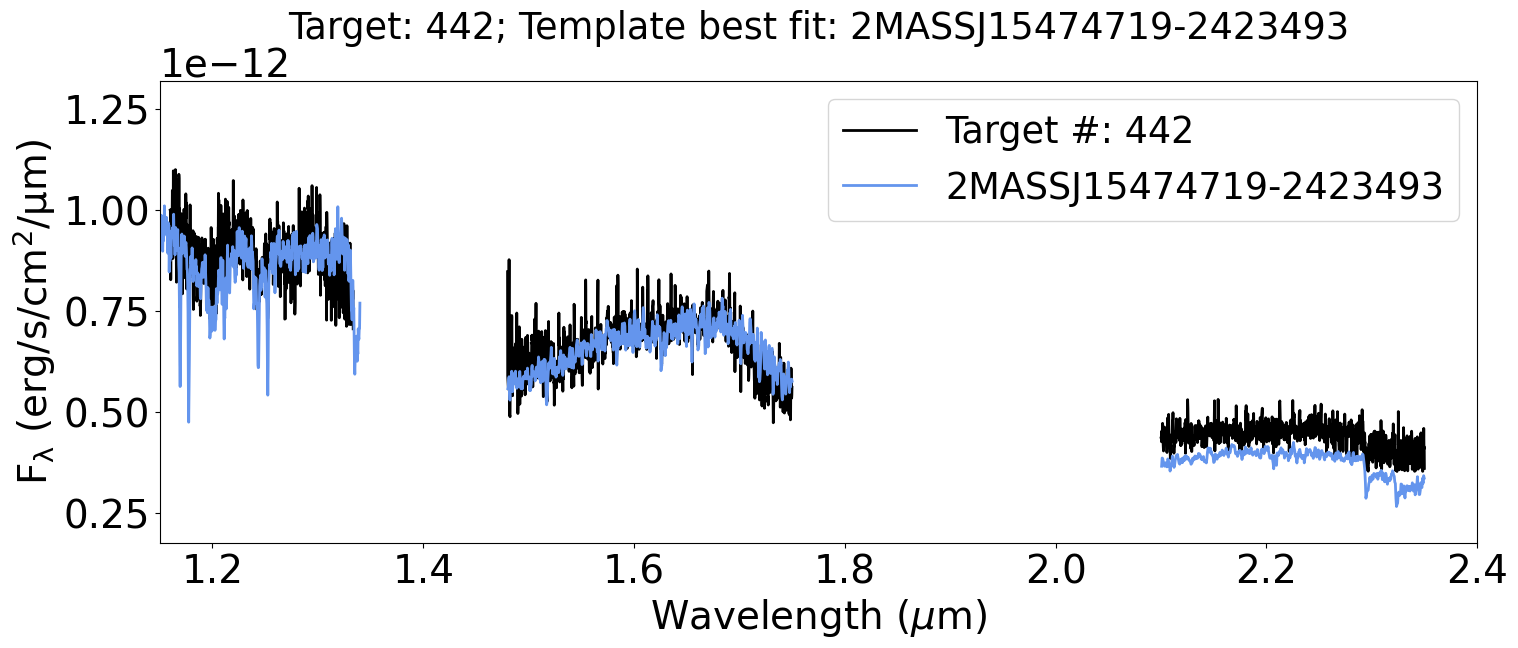}
    \includegraphics[width=0.49\linewidth]{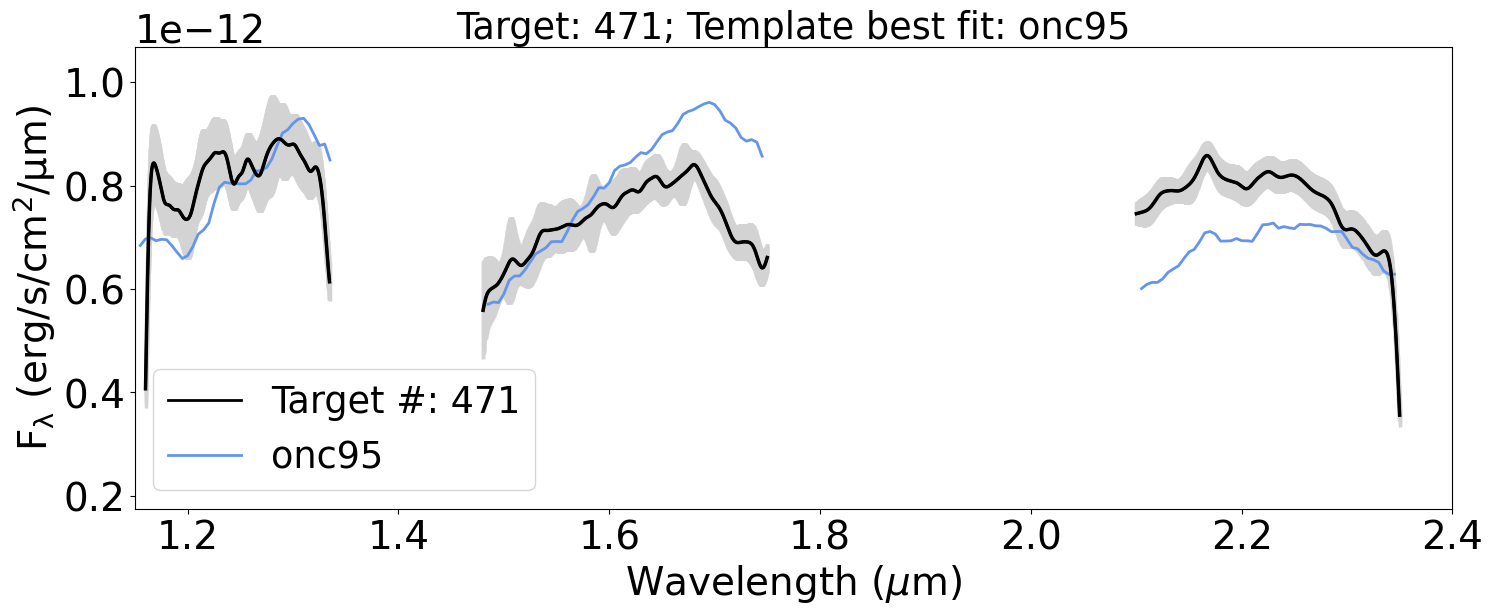}
    \includegraphics[width=0.49\linewidth]{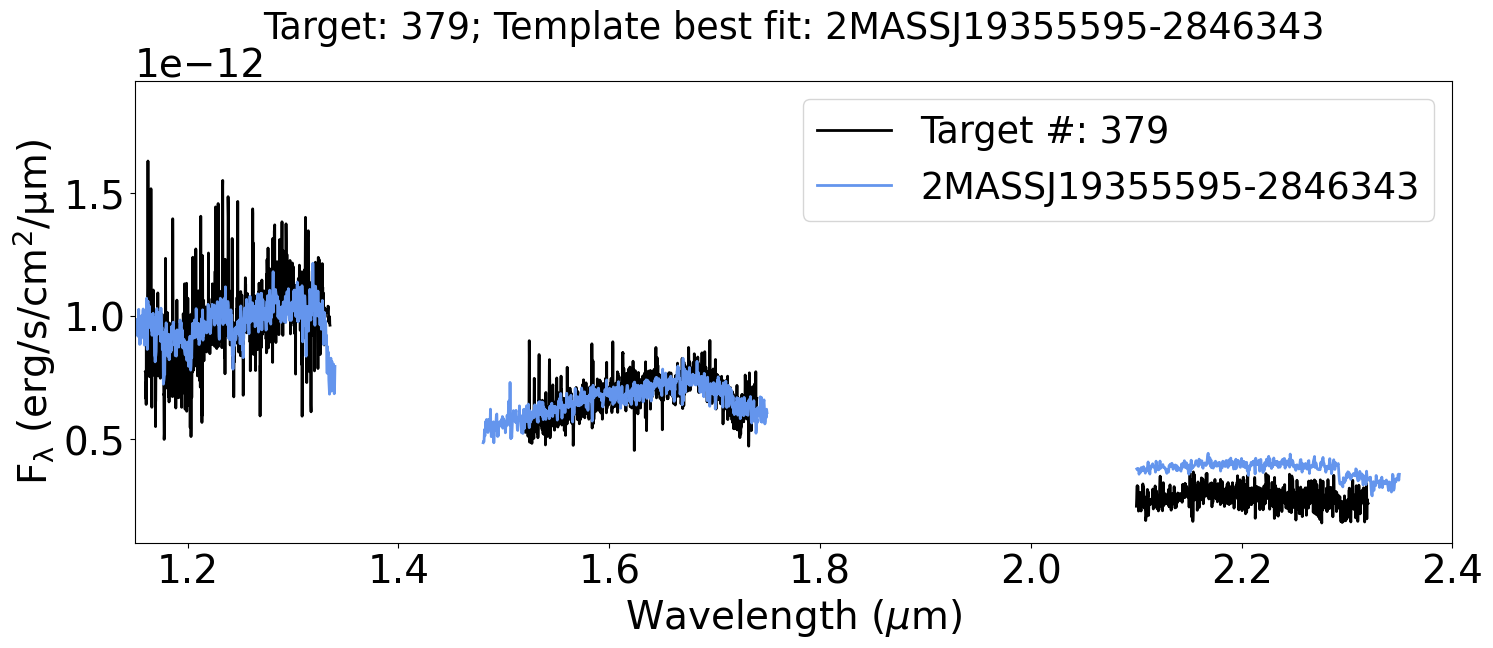}
    \includegraphics[width=0.49\linewidth]{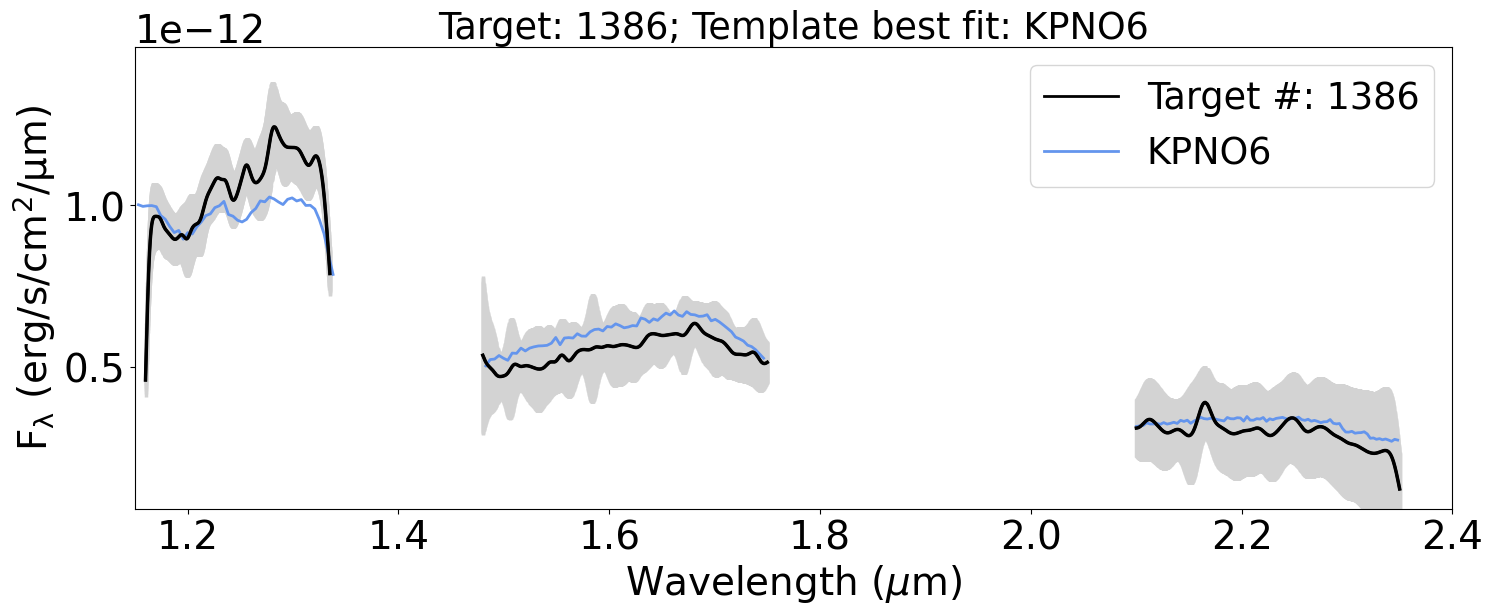}
    \includegraphics[width=0.49\linewidth]{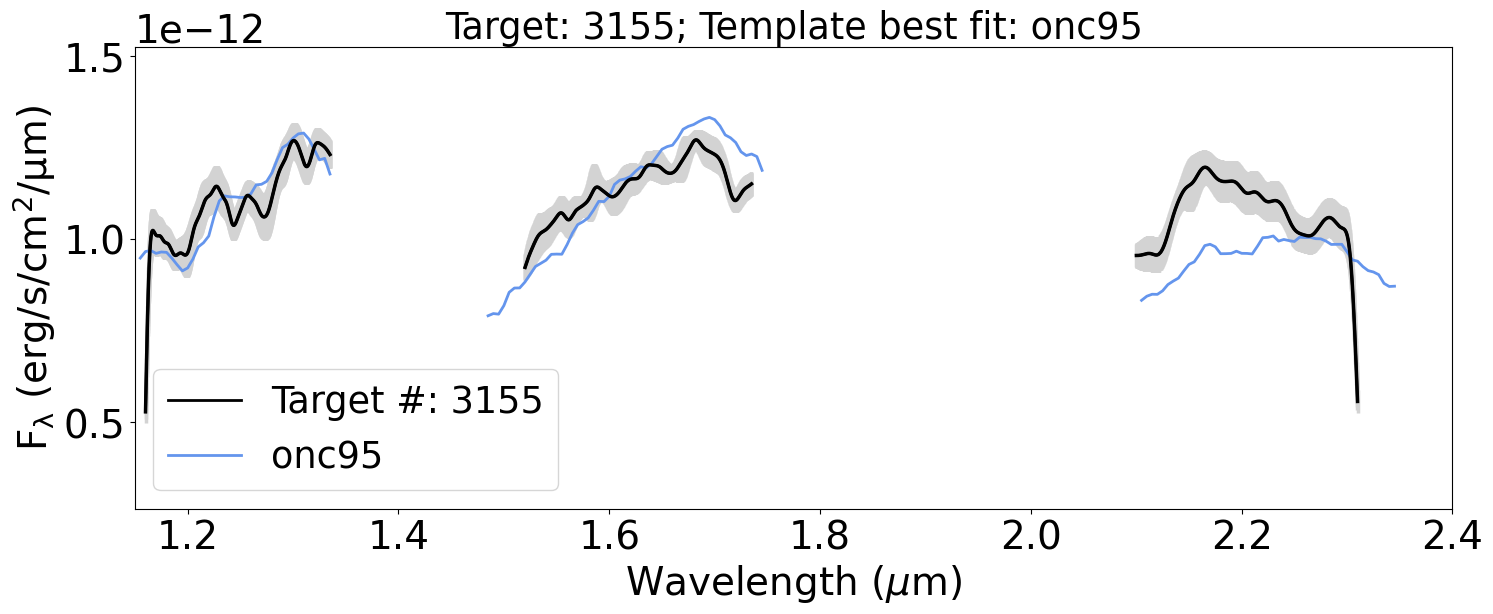}
    \includegraphics[width=0.49\linewidth]{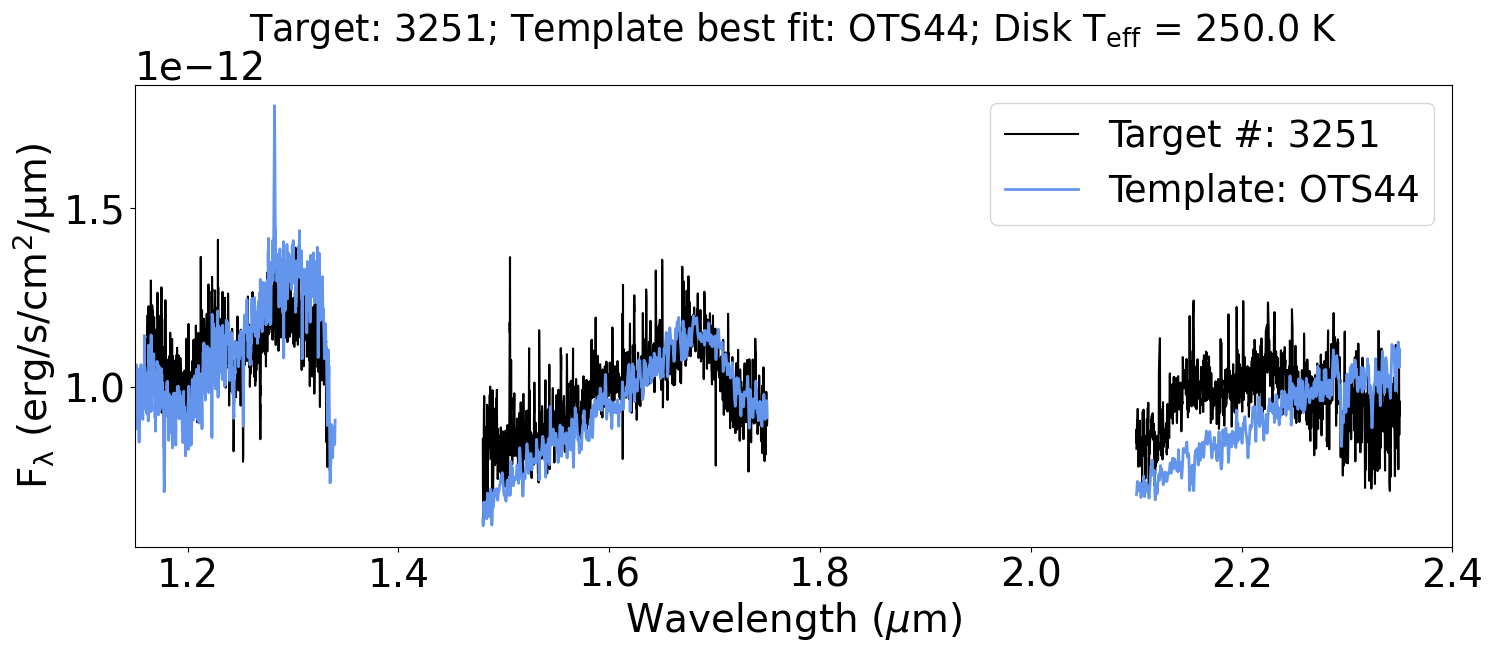}
    \includegraphics[width=0.49\linewidth]{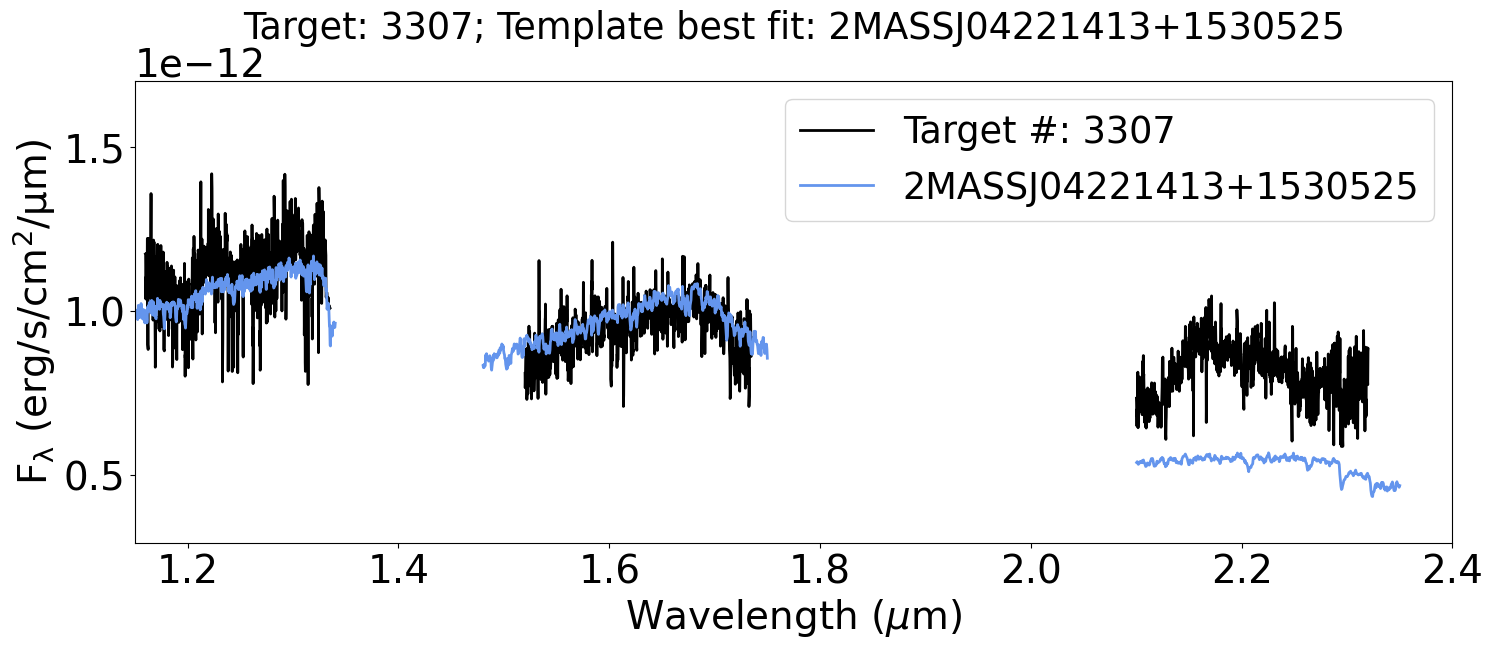}
    \includegraphics[width=0.49\linewidth]{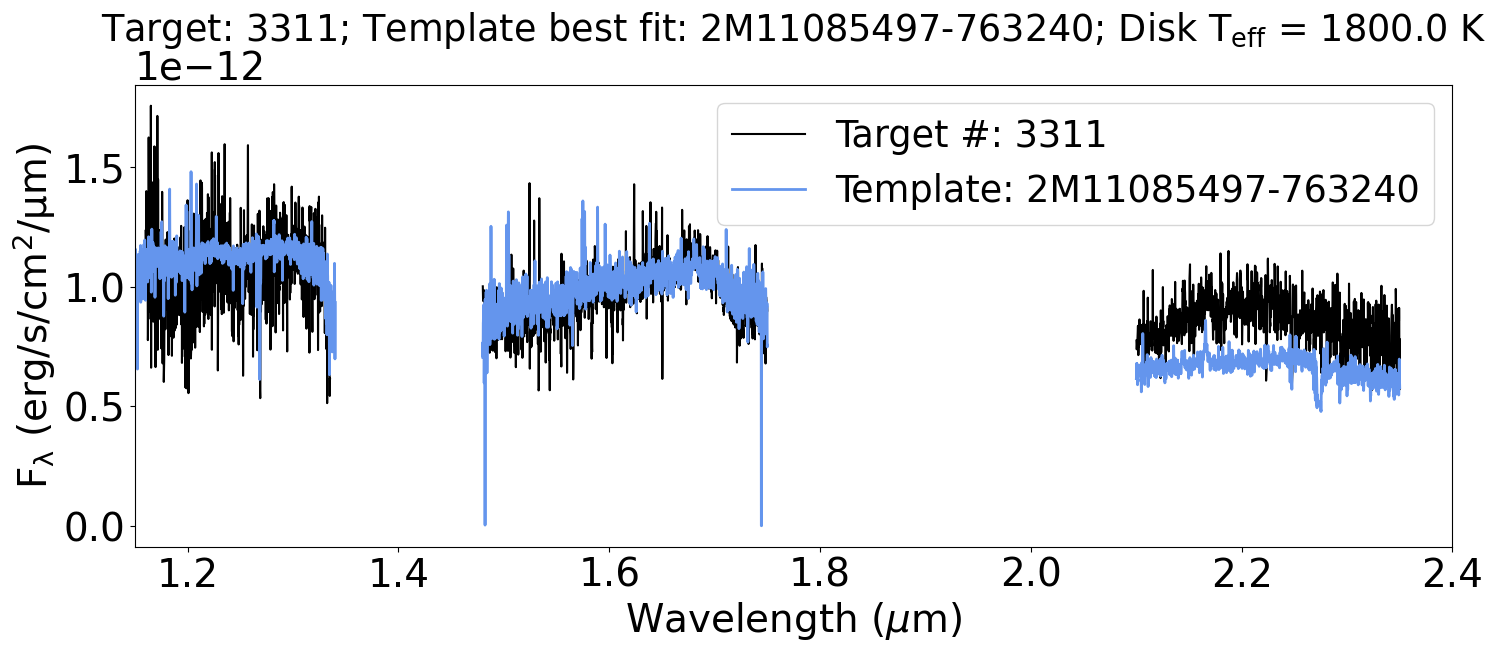}
    \caption{Best matches of our targets with templates from the literature.}
    \label{fig:best_matches_templates1}
\end{figure*}

\begin{figure*}
    \centering
    \includegraphics[width=0.49\linewidth]{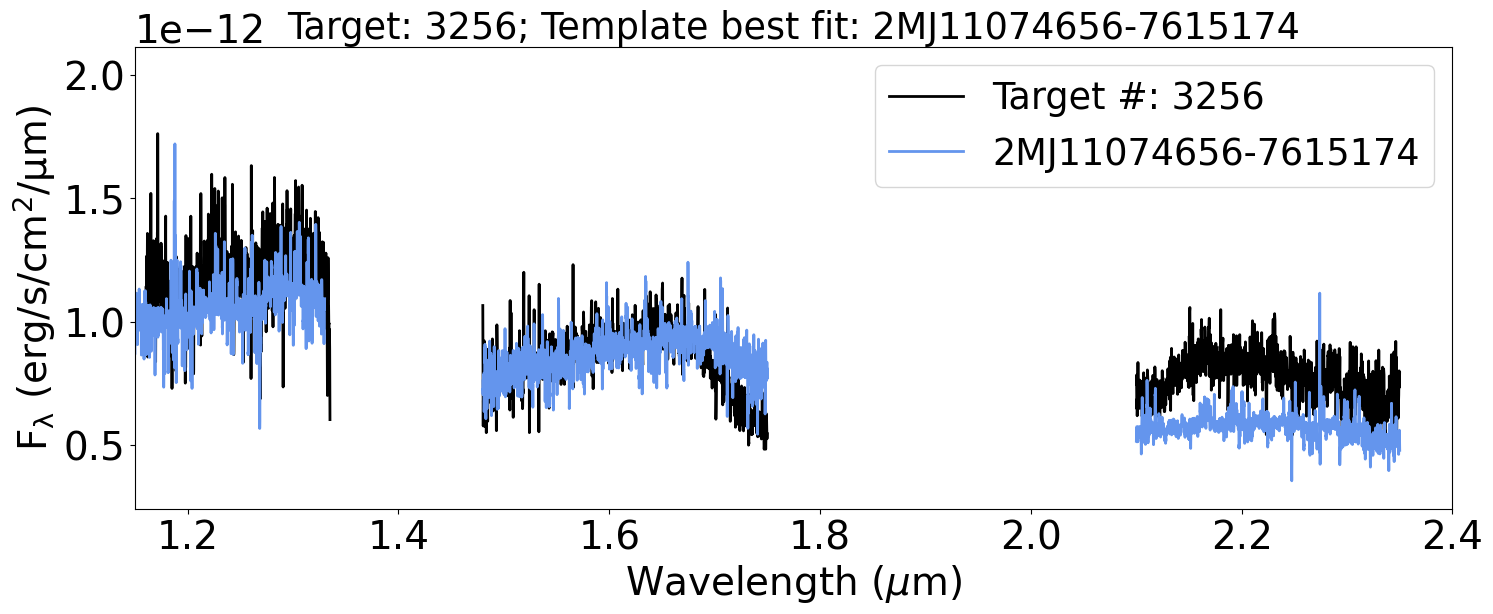}
    \includegraphics[width=0.49\linewidth]{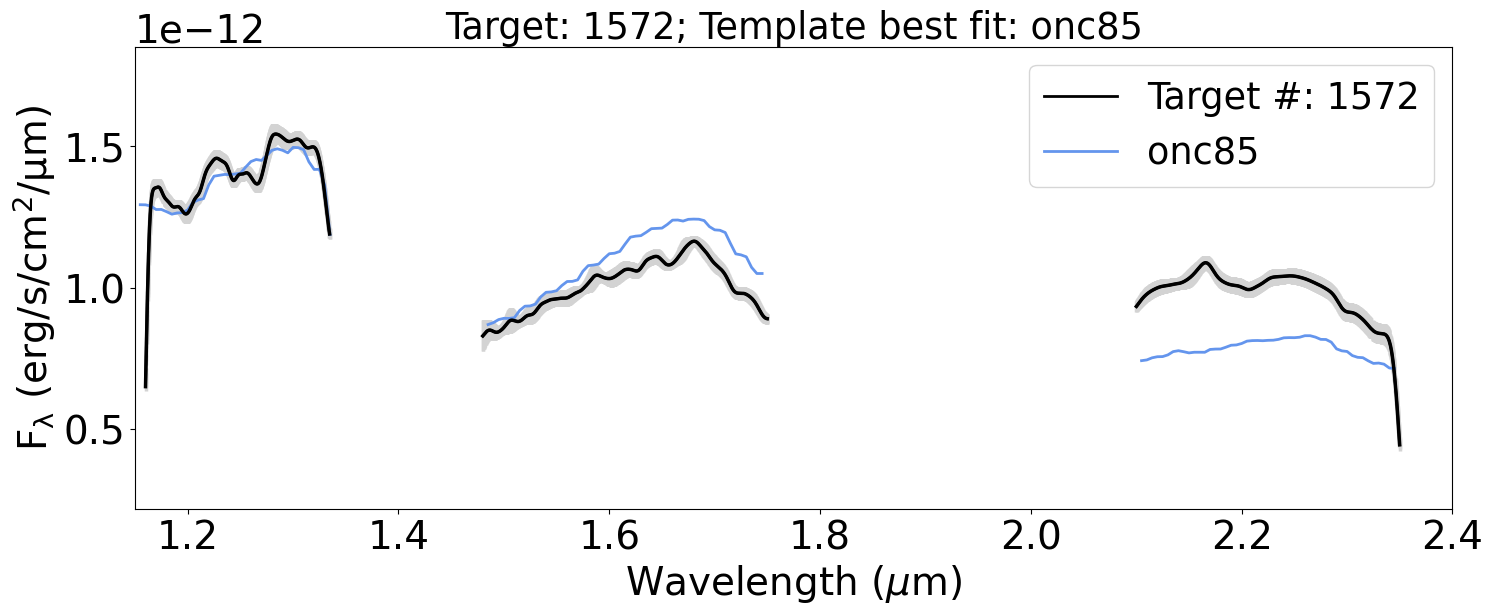}
    \includegraphics[width=0.49\linewidth]{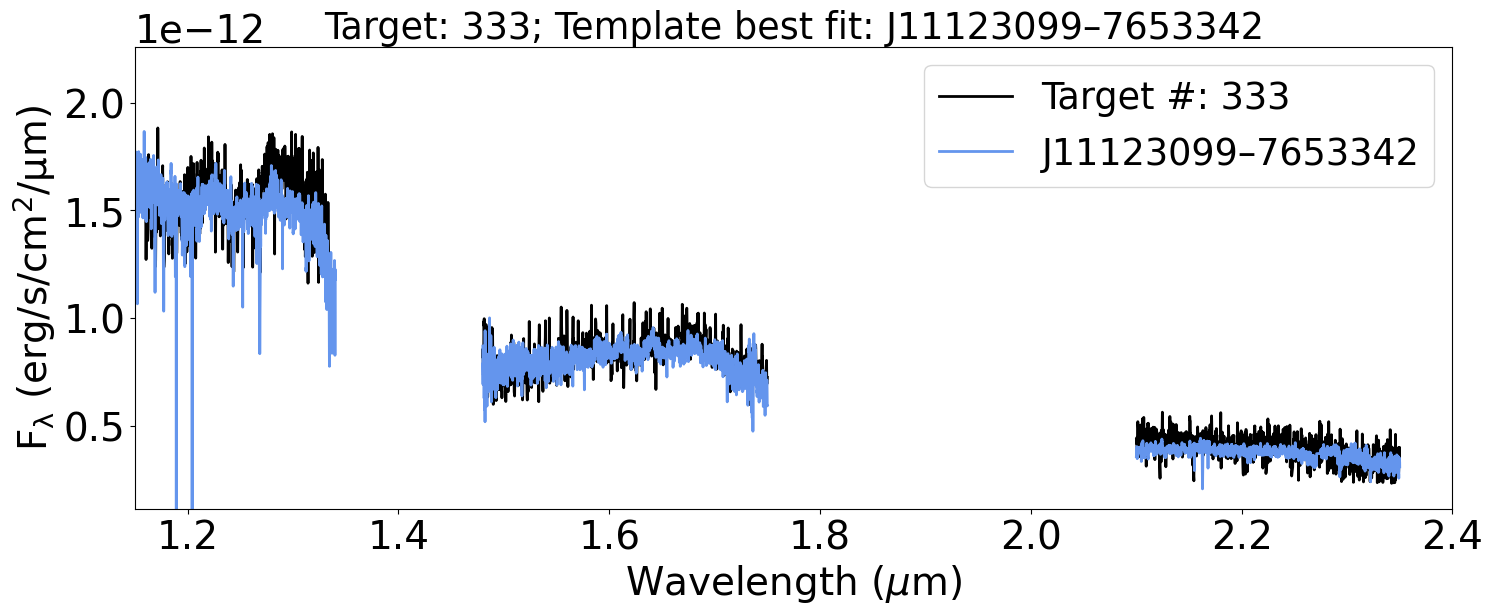}
    \includegraphics[width=0.49\linewidth]{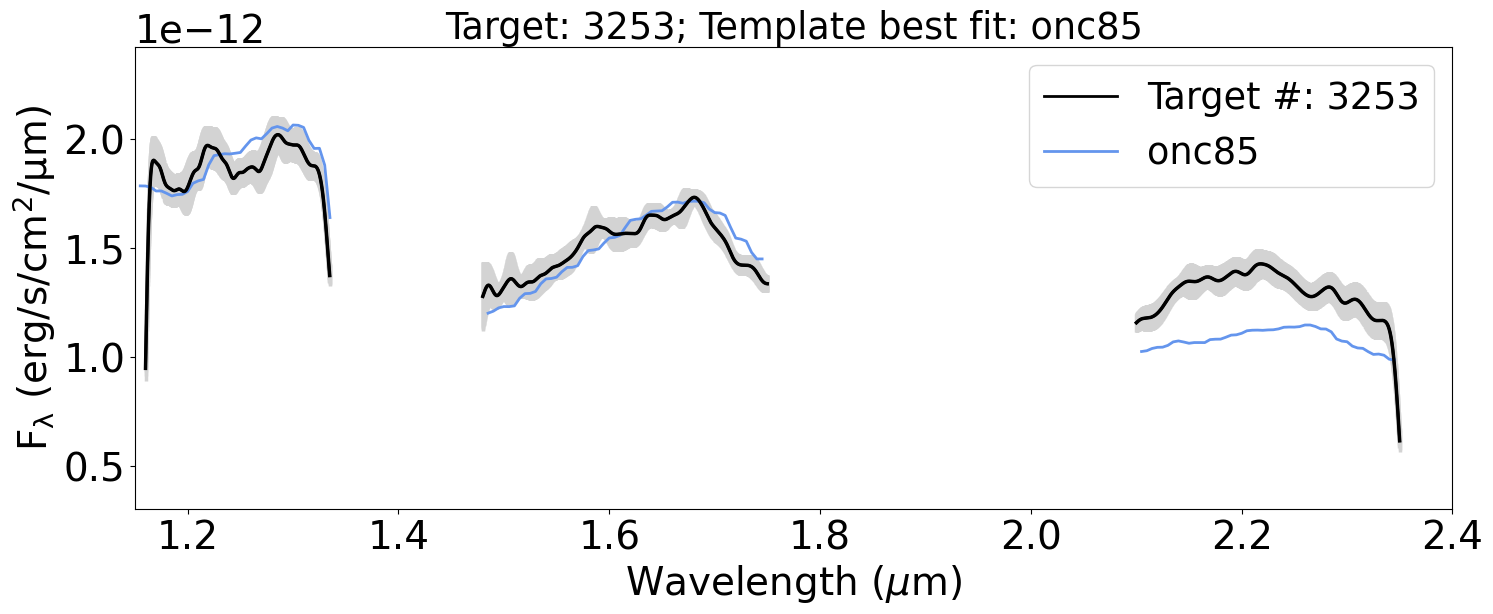}
    \includegraphics[width=0.49\linewidth]{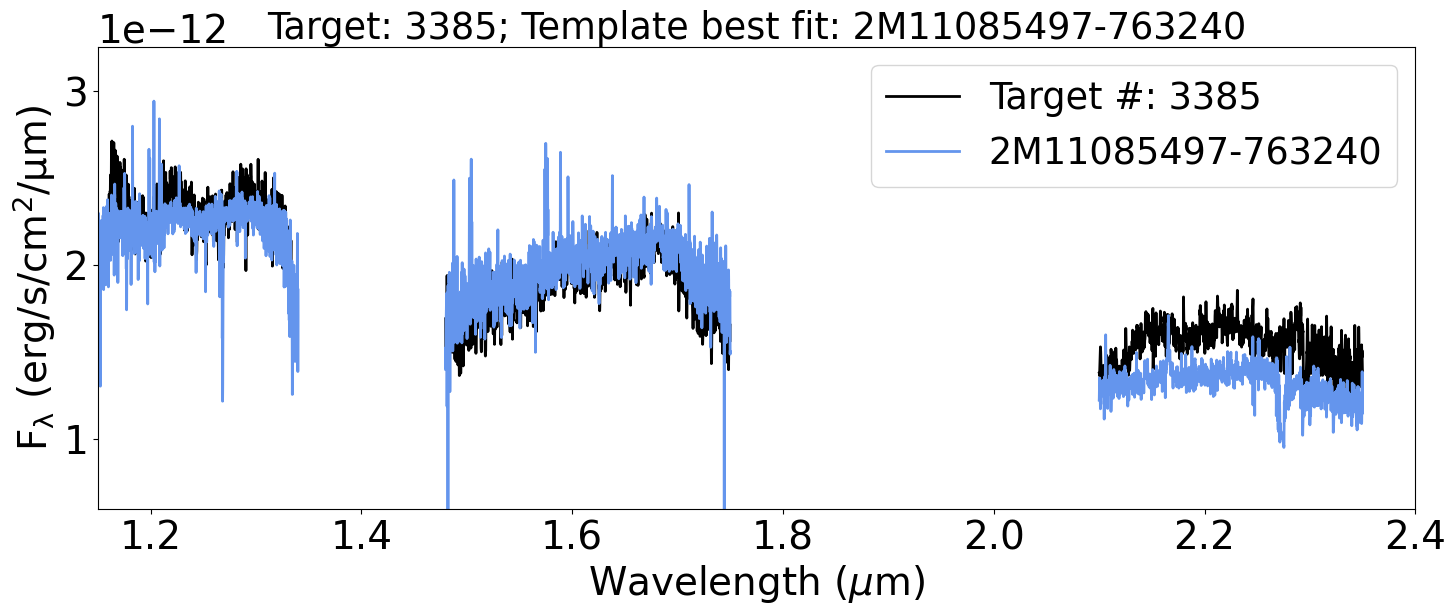}
    \includegraphics[width=0.49\linewidth]{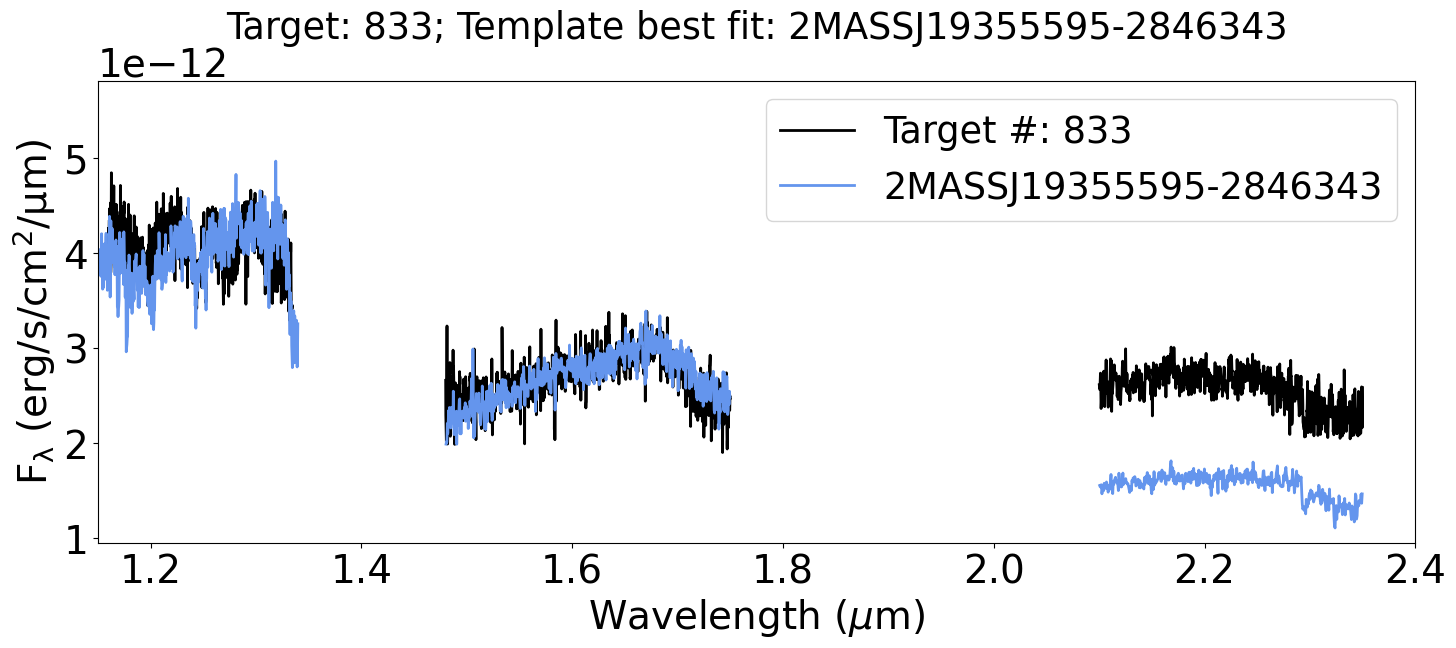}
    \includegraphics[width=0.49\linewidth]{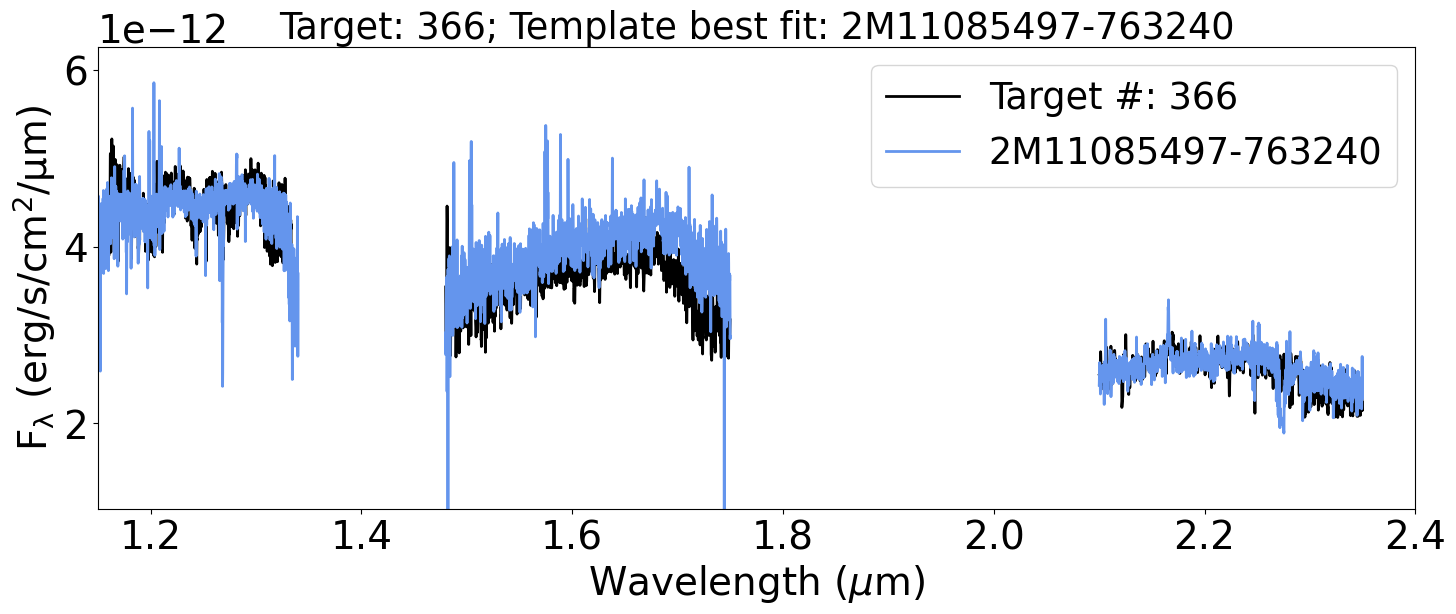}
    \includegraphics[width=0.49\linewidth]{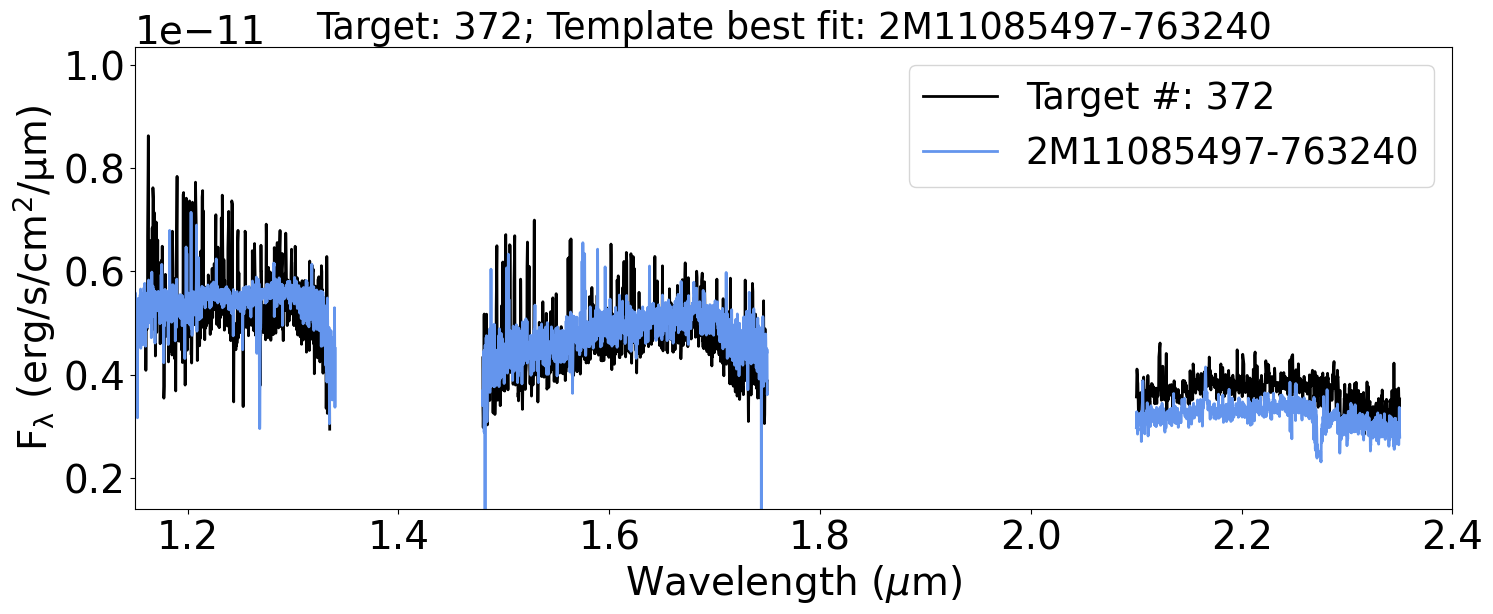}
    \includegraphics[width=0.49\linewidth]{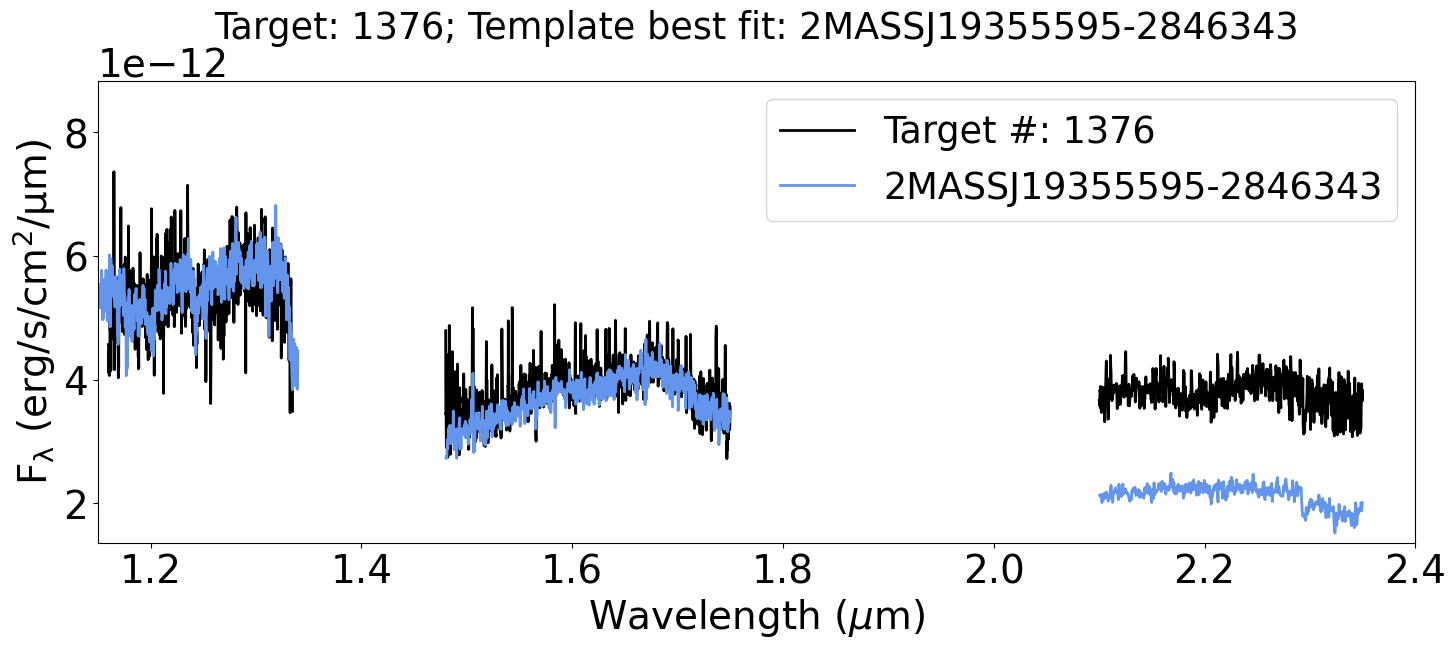}
    \includegraphics[width=0.49\linewidth]{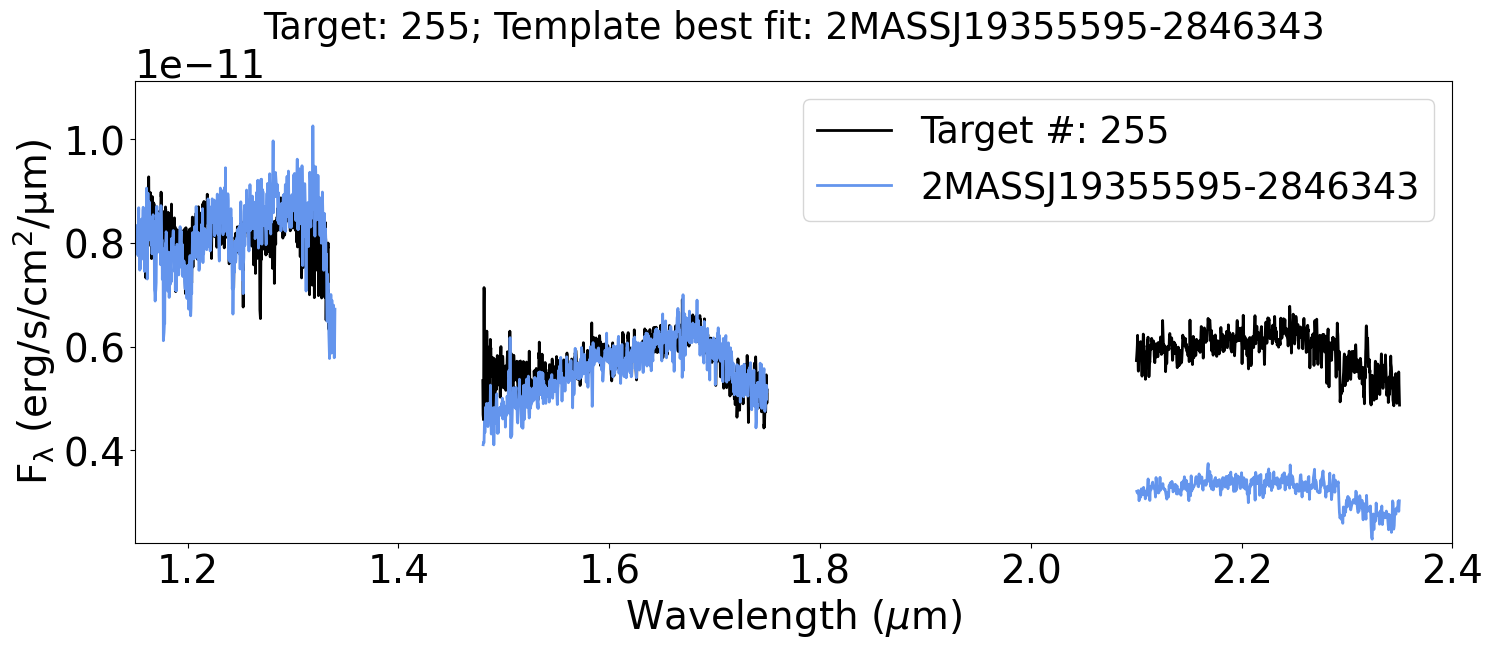}
    \includegraphics[width=0.49\linewidth]{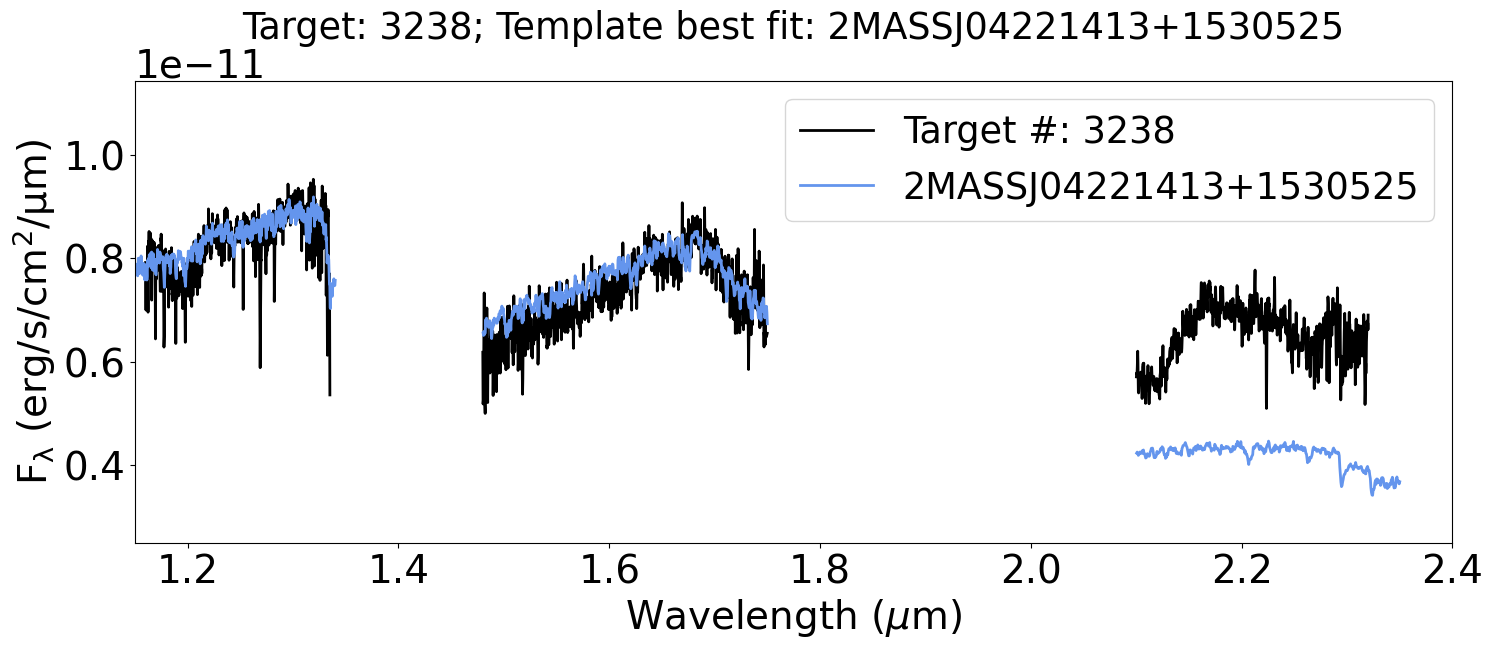}
    \includegraphics[width=0.49\linewidth]{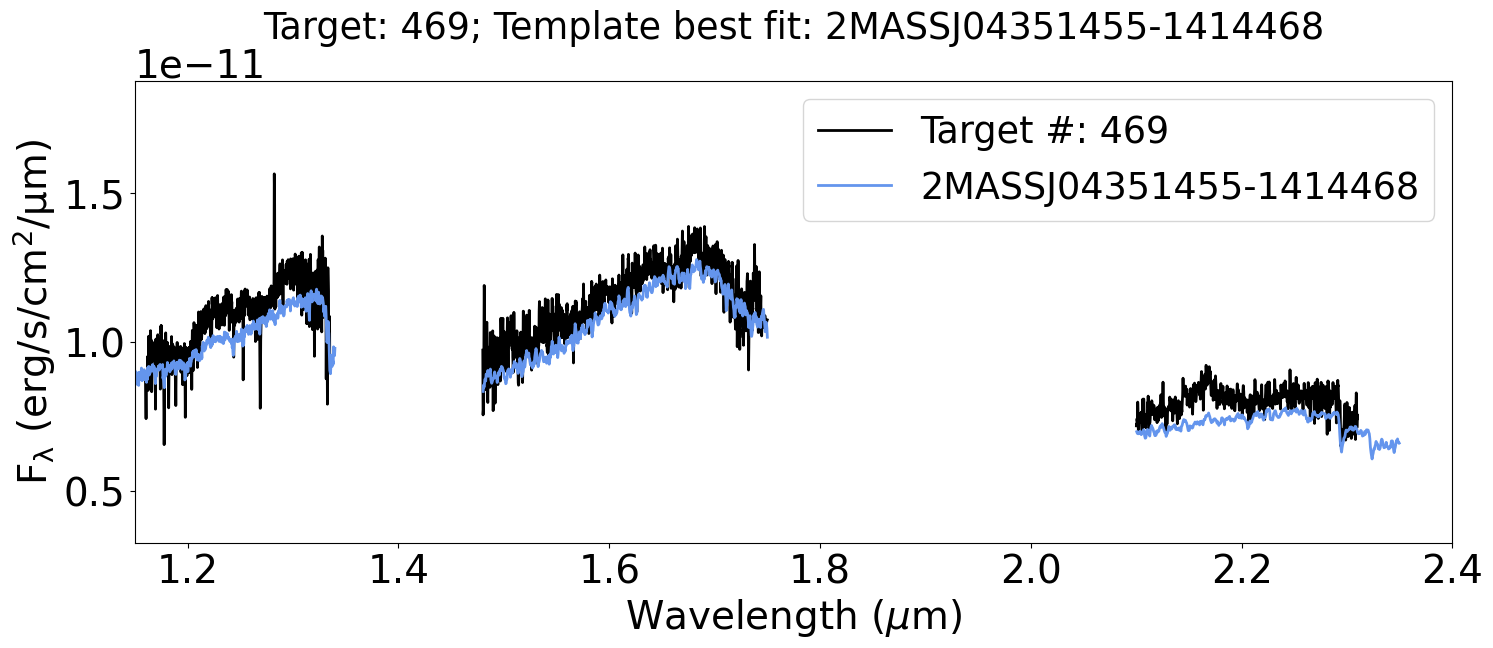}
    
    \caption{Best matches of our targets with templates from the literature.}
    \label{fig:best_matches_templates2}
\end{figure*}

\begin{figure*}
    \centering

    \includegraphics[width=0.49\linewidth]{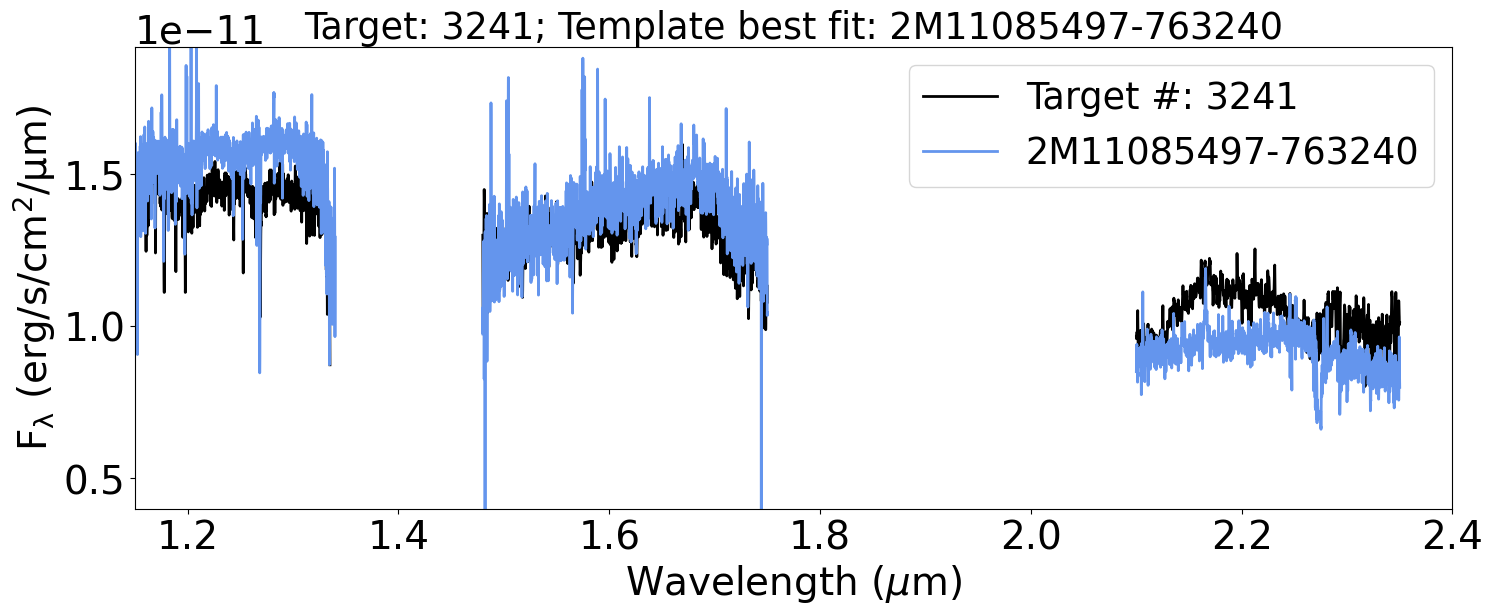}
    
    \caption{Best matches of our targets with templates from the literature.}
    \label{fig:best_matches_templates3}
\end{figure*}




\end{document}